\documentclass[10pt]{article} % For LaTeX2e
\usepackage[table]{xcolor}
\usepackage[accepted]{tmlr}
\usepackage{amsmath,amsfonts,bm}

\def\eqref#1{equation~\ref{#1}}
\def\1{\bm{1}}

\DeclareMathAlphabet{\mathsfit}{\encodingdefault}{\sfdefault}{m}{sl}
\SetMathAlphabet{\mathsfit}{bold}{\encodingdefault}{\sfdefault}{bx}{n}

\usepackage{hyperref}
\hypersetup{hidelinks} % To hide links
\usepackage{url}

\usepackage{latexsym}
\usepackage{booktabs}
\usepackage{multirow}
\usepackage{soul}
\usepackage{hhline}
\usepackage{adjustbox}
\usepackage{makecell}
\usepackage{colortbl}
\usepackage{amssymb}
\usepackage{hyperref}
\usepackage{pifont}    
\usepackage{array}
\usepackage[compact]{titlesec}
\usepackage{pgfplots}
\pgfplotsset{compat=1.18}

\usepackage[most]{tcolorbox}   % or \usepackage{tcolorbox}
\usepackage[normalem]{ulem}  % Allows for dashed underlining

\usepackage{tabularx}
\usepackage{longtable}
\usepackage{fontawesome5} % For better checkmarks
\usepackage{colortbl} % For row coloring
\usepackage[edges]{forest}

\usepackage{xspace}

\title{Code Reasoning for Software Engineering Tasks: \\A Survey and A Call to Action}

\author{\name Saurabh Pujar \email saurabh.pujar@ibm.com \\
      \addr IBM
      \AND
      \name Ira Ceka \email iceka@cs.columbia.edu \\
      \addr Columbia University
      \AND
      \name Irene Manotas \email irene.manotas@ibm.com\\
      \addr IBM
      \AND
      \name Gail Kaiser \email kaiser@cs.columbia.edu\\
      \addr Columbia University
      \AND
      \name Baishakhi Ray \email rayb@cs.columbia.edu\\
      \addr Columbia University
      \AND
      \name Shyam Ramji \email ramji@us.ibm.com\\
      \addr IBM}

\newcommand{\llms}{\text{LLMs}\xspace}

\newcommand{\cotshort}{\text{CoT}\xspace}
\newcommand{\swe}{SWE\xspace}

\newcommand{\sweagents}{SWE agents\xspace}

\newcommand{\cmark}{{\color{teal}\faCheckCircle}}
\newcommand{\xmark}{{\color{red}\faTimesCircle}}
\newcommand{\ignore}[1]{}

\def\month{06}  % Insert correct month for camera-ready version
\def\year{2026} % Insert correct year for camera-ready version
\def\openreview{\url{https://openreview.net/forum?id=zZa3u6LKwO}} % Insert correct link to OpenReview for camera-ready version

\begin{document}

\maketitle

\begin{abstract}
The rise of large language models (\llms) has led to dramatic improvements across a wide range of natural language tasks.
Their performance on certain tasks can be further enhanced by incorporating test-time reasoning techniques.
These inference-time advances have been adopted into the code domain, enabling complex software engineering (\swe) tasks such as code generation, test generation and issue resolution. 
However, the impact of different reasoning techniques on code-centric \swe tasks has not been systematically explored.
In this work, we survey code reasoning techniques that underpin these capabilities, with a focus on test-time compute and inference-time reasoning paradigms.
We examine a variety of code-specific reasoning methods and progressively build up to \sweagents, which combine planning, tool use, and multi-step interaction.
We also compare the impact of different techniques on coding tasks, highlighting their relative importance and outlining open challenges and future research directions.
Across commonly used models and benchmarks, we find that approaches exploiting code-specific signals (e.g., structure and execution feedback) are frequently associated with improved performance, motivating a dedicated study of code reasoning beyond natural-language reasoning.
Our contributions are: (1) to the best of our knowledge, the first dedicated survey of code reasoning for \swe{} tasks, highlighting overarching reasoning strategies, hybrid methods, and agentic approaches; (2) a taxonomy of inference-time techniques used to drive code reasoning, accompanied by a curated set of under-explored benchmarks with high potential for \swe{} evaluation; (3) a comparative analysis of reasoning design patterns across commonly used models and benchmarks; and (4) a synthesis of gaps in current methods and evaluation practices, identifying under-explored areas and concrete opportunities for future research.
\end{abstract}

\section{Introduction}
\label{sec:intro}

\definecolor{SelfrefineYellow}{RGB}{255,215,0}
\definecolor{CoTBlue}{RGB}{135,206,250}
\definecolor{InferenceGreen}{RGB}{193,240,193}
\definecolor{lightred}{RGB}{255,204,204}

\begin{figure*}[htbp]
\centering
  \noindent
  %\hspace*{-1.66cm} % Negative indent shifts figure to the left (tweak as needed)
  \includegraphics[width=0.97\textwidth, trim=0 0 0 0, clip]{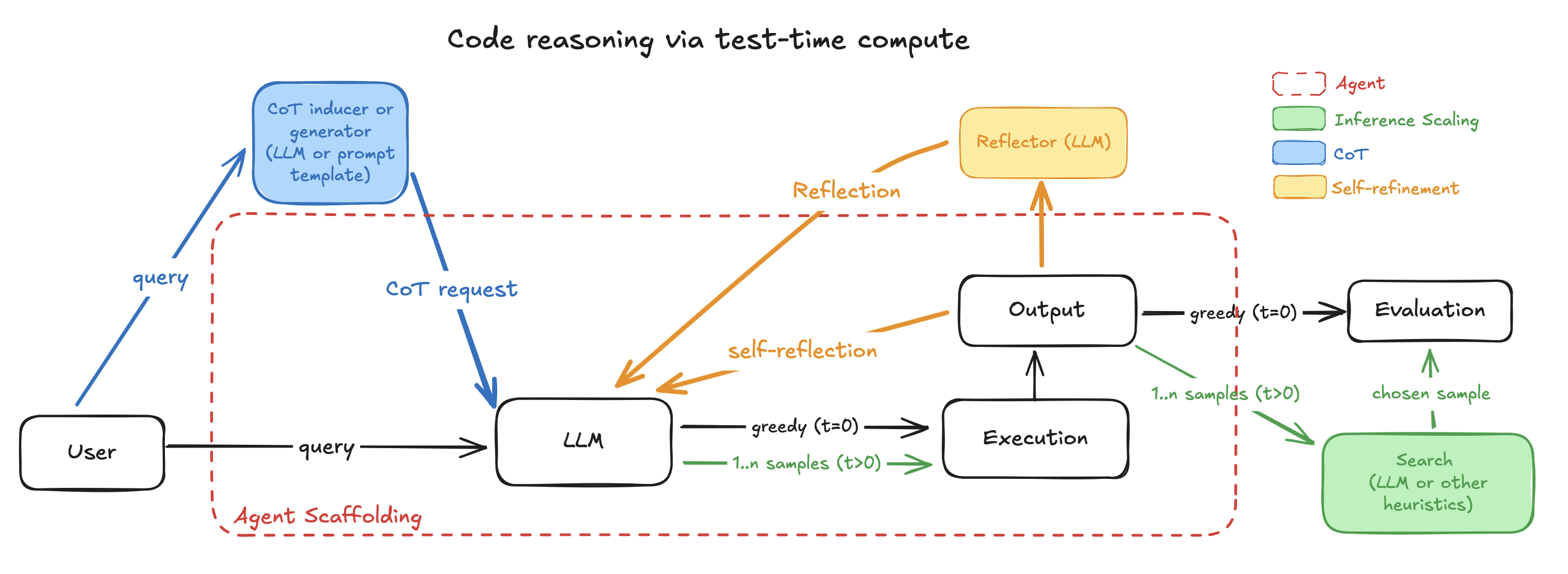}
  \caption{A simplified view of LLM inference for code tasks, illustrating both standard decoding and test-time compute-based reasoning techniques. The different colored regions indicate the core components of the reasoning methods covered in this survey. In standard inference, the user sends a query to an LLM, which produces a single greedy output (typically with temperature $t = 0$); for coding tasks, this output is executed and the resulting behavior is evaluated. \colorbox{CoTBlue}{\cotshort} (Sec.~\ref{sec:cot}) is induced by augmenting the user query with a reasoning-oriented prompt template or an auxiliary LLM. \colorbox{InferenceGreen}{Inference Scaling} (Sec.~\ref{sec:inferencescaling}) generates multiple outputs by sampling with temperature $t > 0$, and uses search over these candidates before evaluation. \colorbox{SelfrefineYellow}{Self-refine} (Sec.~\ref{sec:execution}) iteratively improves candidate outputs by having an LLM critique and revise them; the critic may be the same model (self-reflection) or a separate reflector model. Finally, an \colorbox{lightred}{Agent} (Sec.~\ref{sec:agentssec}) orchestrates LLM calls and manages the execution environment, including tool use for interacting with code and external systems.}
  \label{fig:sys-diag}
\end{figure*}

\citealp{naturalness} show that software is repetitive and predictable like natural language, and hence can be modeled using statistical techniques like LLMs.
Subsequently, \llms have been used effectively for a wide variety of Software Engineering (\swe) tasks\footnote{We use \swe tasks, Code tasks and Software engineering tasks interchangeably.}, including code generation~\citep{humaneval2021}, 
test generation~\citep{mündler2025swtbenchtestingvalidatingrealworld}, 
issue resolution~\citep{jimenez2024swebench} and others.
%language translation~\cite{unsuptrans} code summarization~\cite{codesumm} and others.
Many code specific datasets~\citep{codenet, xcodeeval}, models~\citep{li2023starcoder,nijkamp2023codegen} and benchmarks~\citep{apps2021,zhuo2025bigcodebenchbenchmarkingcodegeneration} have also been developed.
Despite this progress, LLMs have been shown to be limited in their capacity to solve real-world \swe tasks, like GitHub issue resolution~\cite{jimenez2024swebench}.
Recent development of Large Reasoning Models (LRMs)~\cite{guo2025deepseek,claudecode,jaech2024openai} and \sweagents have resulted in tremendous improvement on code tasks, including GitHub issue resolution.

In a recent survey, \citealp{yang2025code} explore how code and reasoning reinforce each other.
They compile works showing how incorporating code data improves reasoning, and how better reasoning leads to improvement on \swe tasks.
Reasoning is induced in \llms with test-time compute techniques that enable models to "think".
These underlying techniques that contribute to reasoning models, include Chain-of-Thought or CoT \citep{wei2022chain} which elicits reasoning, learning from environment feedback~\citep{chen2023teaching} and exploring multiple execution paths~\citep{yao2023tree}.
These techniques are primarily inference time, but can include some training with model generated synthetic reasoning data as well. 
Many recent surveys explore reasoning techniques, \swe \llms, benchmarks and agents, and we discuss them in Sec.~\ref{sec:related} and summarize the topics covered in Tab.~\ref{tab:surveys}.
We did not find any survey that explores the impact of reasoning, and specifically code-based reasoning techniques on \swe tasks.
%~\bray{what are SWE tasks?}.

Figure~\ref{fig:sys-diag} illustrates a simplified implementation of test-time compute based reasoning techniques, many of which have recently been adapted to code-centric tasks. 
However, the literature is fragmented, with inconsistent terminology and heterogeneous evaluation protocols, making it difficult to compare approaches and identify reliable design patterns.
At a high level, code reasoning methods can be grouped into chain-of-thought (Sec.~\ref{sec:cot}), self-refinement (Sec.~\ref{sec:execution}), and inference scaling (Sec.~\ref{sec:inferencescaling}). 
Beyond these components, software engineering agents (\sweagents{}; Sec.~\ref{sec:agentssec}) provide the scaffolding for \llms{} to iteratively apply test-time compute, such as planning, tool use, execution, and revision, to solve challenging code tasks such as GitHub issue resolution. 
In practice, many surveyed systems combine multiple techniques, as summarized in Table~\ref{tab:llm-hybrid}.

To better understand the impact of these design choices, we synthesize reported results under a common set of models and benchmarks (Sec.~\ref{sec:comparison}). This analysis suggests that approaches that explicitly leverage properties of code, such as its syntactic structure and executability, tend to achieve stronger performance, highlighting an important distinction from reasoning in the natural language domain. We also observe that evaluation is concentrated on a small set of popular code-generation benchmarks, which limits conclusions about broader \swe{} capabilities. Accordingly, we catalog additional tasks and benchmarks that better reflect diverse \swe{} workflows (Sec.~\ref{sec:benchmetrics}) and use these findings to motivate concrete directions for future research (Sec.~\ref{sec:gaps}).

\swe is one of the most interesting applications areas of Artificial Intelligence (AI) and there is growing research in this space.
As different reasoning techniques mature and agents become more robust, it is reasonable to expect more and more \swe tasks will be automated.
With our survey on code reasoning for code tasks, we hope to address this gap by making the following contributions:

\textbf{(1)} The first survey specific to reasoning for coding tasks, emphasizing reasoning techniques which borrow ideas from coding principles (Sec.~\ref{sec:codereason}). 
SWE Agents are given a special focus (Sec.~\ref{sec:agentssec}) since they often rely on multiple reasoning techniques.

\textbf{(2)} A Taxonomy covering different reasoning approaches and benchmarks for code Fig.~\ref{fig:taxonomy}. We also highlight approaches employing multiple reasoning techniques for LLMs and \sweagents (Tab.~\ref{tab:llm-hybrid}).

\textbf{(3)} Showcase benchmarks used to study the impact of reasoning on SWE tasks. We 
    % put together comprehensive 
    compiled comparison tables (Tab.~\ref{tab:apps-results},~\ref{tab:lcb-results},~\ref{tab:swe-bench},~\ref{tab:mbpp-family},~\ref{tab:mbpp-family-continued},~\ref{tab:he-family},~\ref{tab:he-family-continued}) showing the performance of different code reasoning and agentic approaches 
    % for the most commonly used benchmarks 
    (Sec.~\ref{subsec:taskbench}). We also highlight 
    % some uncommon but 
    promising benchmarks specific to code reasoning (Sec.~\ref{sec:evalreason}), and surface some new agent-specific benchmarks with potential for furthering SWE research.

\textbf{(4)} Comparison and discussion of results from different reasoning techniques examined in the survey (Sec.~\ref{sec:comparison}). In Sec.~\ref{sec:gaps}, we use this discussion to motivate future work.

A categorized list of all the works covered by our survey is also available on GitHub: \url{https://github.com/AI4Code-IBM-Columbia/code-reasoning-for-swe-tasks}

\section{Survey Methodology}
\label{app:survey-method}

We conducted a structured literature search using arXiv and Google Scholar to obtain broad coverage of relevant work on code reasoning for code-centric and \swe{} tasks. 
The literature search began in February 2025, with references collected from early February through May 2025. The search was updated in December 2025 to capture additional relevant work.
We used advanced search queries combining terms such as \textit{``code reasoning''}, \textit{``reasoning''} + \textit{``LLM''}, and \textit{``agents''} + \textit{``software engineering'', }\textit{``inference scaling''}, \textit{``reflection''},
\textit{``refinement''}, \textit{``code execution''} and \textit{``code benchmarks''}. 
We include works that evaluate on code/\swe{} tasks and employ test-time (inference-time) reasoning or agentic scaffolding. 
Purely training-time techniques are out of scope, unless the training data includes model generated test-time reasoning data like \cotshort{}, self-reflection or agent trajectories.

To capture recent work that builds on established foundations, we also followed citation links in Google Scholar and manually screened papers that cite widely used reference works in this area. We prioritized publications from major venues (e.g., ICSE, ASE, FSE, ACL, EMNLP, ICLR, NeurIPS, and TMLR), while also including relevant arXiv preprints to reflect fast-moving developments that may not yet be represented in archival proceedings.

Code reasoning technique records range from 2022 through 2025. Records for benchmarks and tasks range from 2021 through 2025. The search queries usually returned thousands of hits, so we went through the first few pages of the search results and screened for relevant titles. Among the records with relevant titles, we reviewed the abstract to verify whether the material was indeed relevant. If a paper was relevant, we checked the methodology to confirm that the reasoning was indeed test-time reasoning, then we checked that at least one of the tasks was related to code. From these records, we extended the search by using Google Scholar citations to include more relevant records.

To support transparency and reproducibility, we maintain an up-to-date, categorized list of surveyed papers and pointers to resources on GitHub\footnote{\url{https://github.com/AI4Code-IBM-Columbia/code-reasoning-for-swe-tasks}}.
New papers included will be categorized according to the Code Reasoning Taxonomy in Fig.~\ref{fig:taxonomy}.

\begin{figure}[htbp]
\centering
  \noindent
  %\hspace*{-1.66cm} % Negative indent shifts figure to the left (tweak as needed)
  \includegraphics[width=0.8\textwidth, trim=0 0 0 0, clip]{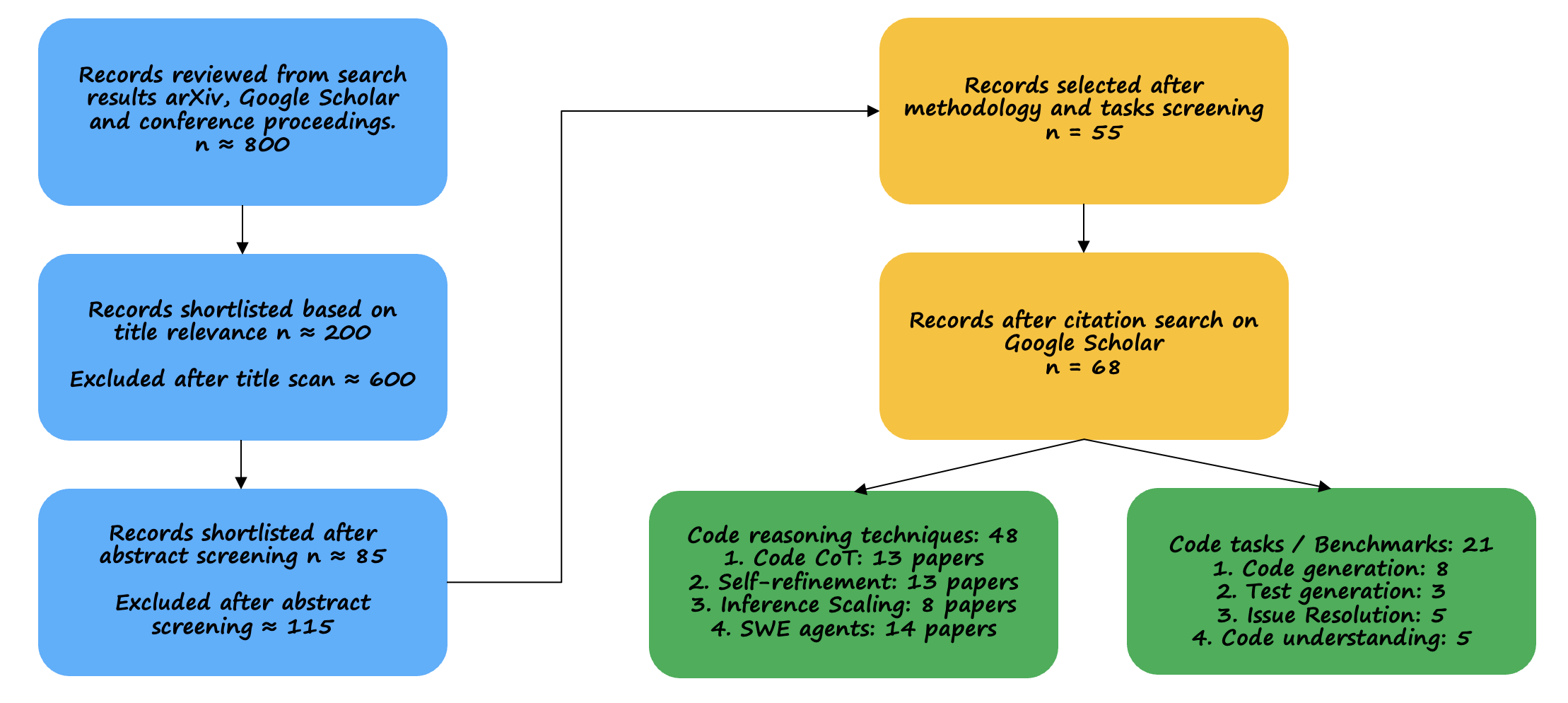}
  \caption{PRISMA-style flow of study selection and categorization. In total, 68 unique records were included. The category counts sum to 69 because one record, \cite{wang2024executable}, is counted in both code reasoning techniques (CodeAct) and code tasks (M3ToolEval).}
  \label{fig:prisma}
\end{figure}

\section{Related Surveys}
\label{sec:related}

\citealp{wei2022chain} introduce CoT as a form of in-context learning which elicits reasoning in LLMs.
In the same year, \citealp{dong2022survey} survey in-context learning techniques and reference CoT reasoning but do not expand on it.
\citealp{qiao2022reasoning} and \citealp{huang2022towards} survey methods and tasks for reasoning and extensively study CoT and other prompting approaches, but do not include software engineering tasks.
\citealp{chu2023navigate} also cover CoT reasoning extensively in a recent work.
They define a more general concept of XoT or X-of-Thought, which covers concepts like Program-of-Thought~\cite{chen2022program}, Tree-of-Thought~\cite{yao2023tree} etc. apart from CoT.
However, they focus on the impact of these techniques on reasoning benchmarks while we are more interested in how reasoning impacts code specific or software engineering benchmarks.
Other recent surveys also cover different types of reasoning techniques for LLMs.
\citealp{xu2025towards} discuss reinforcement learning based reasoning techniques, but they don't discuss code specific reasoning strategies. 
\citealp{plaat2024reasoning} classify the in-context reasoning approaches into prompting, evaluating and control (inference scaling and search) based strategies, but they don't focus on coding tasks.

In their work titled "Code to Think, Think to Code", \citealp{yang2025code} highlight the interplay between code properties and reasoning capabilities and how one enhances the other.
This survey makes the case that training with code related data improves performance on Math and reasoning benchmarks, while incorporating reasoning improves performance on coding benchmarks because some code properties reinforce reasoning capabilities and vice versa. 
Compared to this work, we dive deeper into reasoning techniques used for coding tasks and provide a taxonomy covering different strategies.
\cite{mei2025survey} provide a comprehensive survey on context engineering, which involves context retrieval, processing and management.
Because of the broad scope of their topic, they cover many aspects of \swe reasoning and tasks.
However they have a much more general focus because of which they are unable to cover many aspects of code specific reasoning and different code related benchmarks and tasks.

A lot of surveys do cover impact of LLMs and Agents on Software Engineering tasks but none so far have focused on reasoning based strategies. 
\citealp{zan2022large} survey 27 LLMs for natural language to code generation task.
\citealp{jiang2024survey} undertake an extensive survey for code generation covering not just LLMs but also LLM architectures, many different research topics, benchmarks and datasets, encompassing a total of 235 papers.
\cite{sun2024survey} also do a wide ranging survey covering 50 different models and their variants along with 20 different code-related task categories.
\cite{huynh2025large} survey many topics in this space including challenges and applications.
Apart from surveys covering multiple topics from the domain of AI for code/software engineering, there are also surveys that are more topic specific.
\citealp{wang2024enhancing} focus exclusively on reinforcement learning in code generation.
\citealp{chen2024survey} survey different evaluation techniques for coding tasks.
\citealp{yehudai2025survey} also focus on evaluation, but of LLM-agents and including Software Engineering (SWE) Agents.

We did not find any survey specific to code based reasoning techniques for software engineering tasks, covering agents and benchmarks and including a taxonomy.

\begin{table*}[t!]
\hspace*{1.6cm} % adjust value to shift left
{\scriptsize
\rowcolors{2}{gray!15}{white} % alternate row colors (gray/white)
\begin{tabular}{lccccc}
\toprule
\textbf{Survey} &
\makecell{\textbf{Covers}\\ \textbf{Reasoning}} &
\makecell{\textbf{Covers}\\ \textbf{SWE Tasks}} &
\makecell{\textbf{Covers}\\ \textbf{Agents}} &
\makecell{\textbf{Provides}\\ \textbf{Taxonomy}} &
\makecell{\textbf{Benchmarks}\\ \textbf{Coverage}} \\
\midrule
\citealp{dong2022survey} & \checkmark & \xmark & \xmark & \xmark & \xmark \\
\citealp{qiao2022reasoning} & \cmark & \xmark & \xmark & \xmark & \xmark \\
\citealp{huang2022towards} & \cmark & \xmark & \xmark & \xmark & \xmark \\
\citealp{zan2022large} & \xmark & \checkmark & \xmark & \xmark & \xmark \\
\citealp{chu2023navigate} & \checkmark & \checkmark & \xmark & \xmark & \xmark \\
\citealp{jiang2024survey} & \checkmark & \checkmark & \xmark & \xmark & \xmark \\
\cite{sun2024survey} & \cmark & \cmark & \xmark & \xmark & \checkmark \\
\citealp{plaat2024reasoning} & \cmark & \xmark & \xmark & \xmark & \xmark \\
\cite{wang2024enhancing} & \xmark & \checkmark & \xmark & \xmark & \checkmark \\
\cite{chen2024survey} & \xmark & \cmark & \xmark & \xmark & \checkmark \\
\cite{yehudai2025survey} & \xmark & \cmark & \cmark & \xmark & \cmark \\
\citealp{xu2025towards} & \cmark & \xmark & \xmark & \xmark & \xmark \\
\citealp{huynh2025large} & \checkmark & \checkmark & \xmark & \xmark & \xmark \\
\citealp{yang2025code} & \checkmark & \checkmark & \xmark & \xmark & \xmark \\
\citealp{mei2025survey} & \cmark & \checkmark & \checkmark & \checkmark & \checkmark \\
\textbf{Our Survey} & \cmark & \cmark & \cmark & \cmark & \cmark \\
\bottomrule
\end{tabular}
}
\caption{Comparison of existing surveys along five dimensions: reasoning, \swe tasks, agents, taxonomy, and benchmark coverage. \xmark~indicates that the topic is not covered at all. \checkmark~indicates partial coverage; for example, in the Reasoning column this may mean that only a single technique such as \cotshort{} is discussed, and in the SWE tasks column that only code generation is considered. \cmark~indicates comprehensive, in-depth coverage, specifically for code and \swe{} tasks. Our survey provides in-depth coverage across all these dimensions for test-time compute-based code reasoning on SWE tasks.
}
\label{tab:surveys}
\end{table*}

\definecolor{paired-light-blue}{RGB}{198, 219, 239}
\definecolor{paired-dark-blue}{RGB}{49, 130, 188}
\definecolor{paired-light-orange}{RGB}{251, 208, 162}
\definecolor{paired-dark-orange}{RGB}{230, 85, 12}
\definecolor{paired-light-green}{RGB}{199, 233, 193}
\definecolor{paired-dark-green}{RGB}{49, 163, 83}
\definecolor{paired-light-purple}{RGB}{218, 218, 235}
\definecolor{paired-dark-purple}{RGB}{117, 107, 176}
\definecolor{paired-light-gray}{RGB}{217, 217, 217}
\definecolor{paired-dark-gray}{RGB}{99, 99, 99}
\definecolor{paired-light-pink}{RGB}{222, 158, 214}
\definecolor{paired-dark-pink}{RGB}{123, 65, 115}
\definecolor{paired-light-red}{RGB}{231, 150, 156}
\definecolor{paired-dark-red}{RGB}{131, 60, 56}
\definecolor{paired-light-yellow}{RGB}{231, 204, 149}
\definecolor{paired-dark-yellow}{RGB}{141, 109, 49}
\definecolor{light-green}{RGB}{118, 207, 180}
\definecolor{raspberry}{RGB}{228, 24, 99}

\tikzset{%
    root/.style =          {align=center,text width=3cm,rounded corners=3pt, line width=0.5mm, fill=paired-light-gray!50,draw=paired-dark-gray!90},
    % root/.style =          {align=center,text width=3cm,rounded corners=3pt, line width=0.5mm, fill=paired-light-gray!50,draw=paired-dark-gray!90, yshift=10cm}, % Added yshift here
    data_section/.style =  {align=center,text width=4cm,rounded corners=3pt, fill=paired-light-blue!50,draw=paired-dark-blue!80,line width=0.4mm},
    model_section/.style = {align=center,text width=4cm,rounded corners=3pt, fill=paired-light-orange!50,draw=paired-dark-orange!80,line width=0.4mm},
    training_section/.style = {align=center,text width=4cm,rounded corners=3pt, fill=paired-light-green!50,draw=paired-dark-green!80, line width=0.4mm},
    inference_section/.style = {align=center,text width=4cm,rounded corners=3pt, fill=paired-light-red!35,draw=paired-light-red!90, line width=0.4mm},
    discussion_section/.style = {align=center,text width=4cm,rounded corners=3pt, fill=paired-light-purple!35,draw=paired-dark-purple!90, line width=0.4mm},% Using light red for Inference as in the example
    subsection/.style =    {align=center,text width=3.5cm,rounded corners=3pt}, % General subsection style, NO COLOR DEFINITION NOW
}

\begin{figure*}[!htb]
    \centering
    \resizebox{1\textwidth}{!}{
    \begin{forest}
        for tree={
            forked edges,
            grow'=0,
            draw,
            rounded corners,
            node options={align=center},
            text width=4cm,
            s sep=6pt,
            calign=child edge,
            calign child=(n_children()+1)/2,
            l sep=12pt,
        },
        [Code Reasoning, root
            [Techniques, root
                [Code \cotshort (\S\ref{sec:cot}), data_section
                    [Plan CoT (\ref{subsec:cotplan}), data_section
                        [ \textit{PlanSearch}$^\gamma$ ~\citep{wang2024planning};
                        \textit{Self-Planning} ~\citep{jiang2024self};
                        \textit{ClarifyGPT} ~\citep{mu2023clarifygpt}
                        ,data_section, text width=12cm
                        ] 
                    ]
                    [Code-structure \cotshort (\S\ref{subsec:cotstruct}), data_section
                        [\textit{SCoT} \citep{li2025structured};
                        \textit{MoT}$^\gamma$ \citep{pan2025modularization};
                        \textit{CodeChain} \citep{le2023codechain};
                        % \textit{SemCoder} \cite{ding2024semcoder};
                        \textit{CGO}$^\gamma$ \citep{yeo2025chain};
                        \textit{VisualCoder} \citep{chi-etal-2025-visualcoder},
                        data_section, text width=12cm
                        ] 
                    ]
                    [Training with \cotshort (\S\ref{subsec:cotft}), data_section
                        [\textit{UniCoder} \citep{sun2024unicoder};
                        \textit{COTTON} \citep{yang2024chain};
                        \textit{MSCoT}$^\gamma$ \citep{jin2025mscot};
                         \textit{SemCoder} \citep{ding2024semcoder};
                        \textit{ChainCoder} \citep{zheng2023chaincoder},
                        data_section, text width=12cm
                        ] 
                    ]
                ]
                [Self-refinement (\S\ref{sec:execution}), model_section
                    [Execution reflection (\S\ref{subsec:selfeval}), 
                    model_section [\textit{Self-debugging} \citep{chen2023teaching}; \textit{CodeCOT}$^\gamma$ \citep{huang2023codecot}; \textit{AlphaCodium}$^\gamma$ \citep{ridnik2024code}; \textit{Revisiting Self-debugging} \citep{chen2025revisit}; \textit{\(\mu\)Fix} \citep{tian-fixing-spec-cg-2025}, model_section, text width=12cm]
                    ]
                    [Training with feedback (\S\ref{subsec:trainexec}), model_section
                      [\textit{LEVER} \citep{Ni-LEVER-icml-2023}; \textit{CYCLE} \citep{ding2024cycle}; \textit{LeDex} \citep{jiang2025ledextrainingllmsbetter} , model_section, text width=12cm]   
                    ]
                    [Automated Test Generation (\S\ref{subsec:trainexec}), model_section
                      [\textit{UTGEN} \citep{prasad2025learninggenerateunittests};
                    \textit{AceCoder} \citep{zeng2025acecoderacingcoderrl};
                    \textit{DSTC}$^\gamma$ \citep{liu2024dstcdirectpreferencelearning}; %\textit{ASTER} \citep{pan2025asternaturalmultilanguageunit}; 
                    \textit{SWT-Bench} \citep{mündler2025swtbenchtestingvalidatingrealworld}; \textit{Otter}~\citep{ahmed2025otter}, model_section, text width=12cm]   
                    ]
                ]
                [Inference Scaling (\S\ref{sec:inferencescaling}), training_section
                    [Sampling (\S\ref{sec:sampling}), training_section  
                    [\textit{AlphaCode} \citep{li2022competition};
                        \textit{REx} \citep{tang2024code}; 
                        \textit{S*: Test-time Scaling} \citep{li2025s},
                        training_section, text width=12cm
                        ] 
                    ]
                    [Search (\S\ref{sec:search}), training_section     [%\textit{ToT} \citep{yao2023tree};
                        \textit{GToT}$^\gamma$ \citep{long2023large}; \textit{SWE-Search} \citep{antoniades2024swe}; \textit{CodeTree} \citep{li2024codetree};
                        \textit{Tree-of-Code} \citep{ni2024tree};
                        \textit{ORPS} \citep{yu2024outcome},
                        training_section, text width=12cm
                        ] 
                    ]
                ]
                [\sweagents (\S\ref{sec:agentssec}), inference_section 
                    [Workflow (\S\ref{subsec:workflow}), inference_section  
                        [\textit{Agentless} \citep{xia2024agentless};
                        \textit{AutoCodeRover} \citep{zhang2024autocoderover},
                        inference_section, text width=12cm
                        ] 
                    ]
                    [Agent Optimization (\S\ref{subsec:agentoptim}), inference_section 
                        [\textit{SWE-Agent} \citep{yang2024sweagent};
                        \textit{CodeAct} \citep{wang2024executable};
                        \textit{MASAI} \citep{arora2024masai};
                        \textit{CodeR}$^\gamma$ \citep{chen2024coder};
                        \textit{PairCoder} \citep{zhang2024pair};
                        \textit{HyperAgent}$^\gamma$ \citep{phan2024hyperagent};
                        \textit{AgileCoder} \citep{nguyen2024agilecoder};
                            \textit{OpenHands} \citep{wang2024openhands},
                            inference_section, text width=12cm
                        ] 
                    ]
                    [Training with trajectory (\S\ref{subsec:reasoningmodel}), inference_section  
                        [\textit{Lingma} \citep{ma2024lingma}; \textit{SWE-Gym} \citep{pan2024training};\textit{SWE-Fixer} \citep{xie2025swe};
                            \textit{SWE-RL} \citep{wei2025swe}, 
                            inference_section, text width=12cm
                        ] 
                    ]
                ]
            ]
            [Tasks, root
                [  , discussion_section
                [Code Generation (\S\ref{subsec:taskbench}), discussion_section  
                    [\textit{HE}$^\gamma$ \citep{chen2021evaluatinglargelanguagemodels}; \textit{MBPP}$^\gamma$ \citep{austin2021programsynthesislargelanguage};  \textit{APPS} \citep{hendrycks2021measuringcodingchallengecompetence}; \textit{CodeContests}\citep{codecontests-science.abq1158}; \textit{LCB} \citep{jain2024livecodebenchholisticcontaminationfree}; \textit{BigCodeBench} \citep{zhuo2025bigcodebenchbenchmarkingcodegeneration}; \textit{HEPack} \citep{Muennighoff2023OctoPackIT}; \textit{Spider} \citep{yu-etal-2018-spider, lei2025spider20evaluatinglanguage},
                        discussion_section, text width=12cm
                    ] 
                ]
                [Test Generation (\S\ref{subsec:testgen}), discussion_section  
                    [\textit{TestEval}~\citep{wang-etal-2025-testeval};
                     TDD-Bench-Verified \citep{ahmed2025otter};
                     SWT-Bench \citep{mündler2025swtbenchtestingvalidatingrealworld},
                        discussion_section, text width=12cm
                    ] 
                ]
                [Issue Resolution (\S\ref{subsec:issueres}), discussion_section  
                    [\textit{SWE-Bench}~\citep{jimenez2024swebenchlanguagemodelsresolve};
                     \textit{SWE-Bench Multimodal} \citep{yang2024swebenchmultimodalaisystems}; \textit{Multi-SWE-Bench} \citep{zan2025multiswebenchmultilingualbenchmarkissue}; \textit{SWE-PolyBench}$^\gamma$ \citep{rashid2025swe};\textit{M\(^3\)ToolEval} \citep{wang2024executable},
                        discussion_section, text width=12cm
                    ] 
                ]
                [Reasoning and Understanding (\S\ref{sec:evalreason}), discussion_section 
                    [CRUXEval \cite{gu2024cruxevalbenchmarkcodereasoning};
                    \textit{CodeMind}$^\gamma$ \citep{liu2024codemindframeworkchallengelarge};
                    \textit{ReEval} \citep{junkai-runtime-behavior-reasoning-icse2025};
                    \textit{ExeRScope} \citep{liu2025toolindepthanalysiscode}; \textit{CodeMMLU}~\citep{manh2025codemmlumultitaskbenchmarkassessing},
                        discussion_section, text width=12cm
                    ] 
                ]
                ]
            ]
       ]
    \end{forest}
}
    \caption{Code Reasoning Taxonomy. We organize prior work on code reasoning along two axes: \emph{techniques} and \emph{tasks}. Techniques are methods that elicit, enhance, or exploit reasoning in \llms{}. Tasks (often instantiated as benchmarks) are used to evaluate an LLM’s code-reasoning capability. Techniques are frequently combined within a single system; in this taxonomy, we categorize each publication by its dominant technique (i.e., the primary mechanism emphasized by the method). A $^\gamma$ symbol right after the technique's name indicates the publication has not been published yet in a peer-reviewed venue or is under revision at the time of this publication. Table~\ref{tab:llm-hybrid} highlights works that explicitly integrate multiple techniques. 
}

 % Caption is now in the forest label
    \label{fig:taxonomy}
\end{figure*}

\section{Code Reasoning: Techniques}
\label{sec:codereason}
\citealp{brown2020language} show that LLMs are few-shot learners.
Performance of LLMs on reasoning tasks is further enhanced by a certain kind of prompting called Chain-of-thought or CoT~\citep{wei2022chain} prompting which elicits LLM reasoning.
\citealp{wei2022emergent} suggest that in-context learning ability of LLMs, including CoT reasoning, is an emergent property of LLMs.
Code CoT papers (\citealp{li2025structured, jiang2024self, pan2025modularization} and others) suggest that code reasoning is a specific kind of reasoning and CoT can be more impactful when induced with prompts that recognize this difference.   
We survey such techniques in Sec.~\ref{sec:cot}.

One way code output is different from natural language output is that it can be executed and tested to validate its correctness.
% \citealp{yao2023tree} highlight that execution can be a way to check if the reasoning is correct. %% removed this reference from ToT paper
Self-refinement techniques using code execution as a way to evaluate and improve LLMs' output, such as Self-debugging \citep{chen2023teaching}, CodeCoT \citep{huang2023codecot}, LEVER \citep{Ni-LEVER-icml-2023} and others, are covered in Sec.~\ref{sec:execution}.

\citealp{yao2023tree} state that "System 2" thinking should involve exploring diverse solution paths rather than greedily picking one.
They connect CoT with inference scaling to enable exploration of multiple reasoning paths.
Inference scaling involves setting a temperature greater than 0 to generate multiple candidate solutions or samples during inference or test time, and then picking the best candidate. 
\citealp{li2022competition} effectively leverage this technique to generate competition level code.
Sec.~\ref{sec:inferencescaling} covers inference scaling used to explore multiple reasoning paths for software engineering tasks.

Many approaches use a combination of these techniques, although one technique usually dominates.
The dominant technique is the core method, contribution or novelty of a publication.
Tab.~\ref{tab:llm-hybrid} shows approaches which rely on multiple techniques.

\newtcolorbox{gapsndirections}{
  colback=gray!10,
  colframe=black,
  arc=1pt,
  boxrule=0.2pt,
  left=2pt, right=2pt, top=1pt, bottom=1pt
}

\subsection{Code Chain-of-thought}
\label{sec:cot}

CoT prompts for code can be categorized as plan-based or structure based.
Plan-based CoT is a natural language articulation of steps that need to be taken to solve a coding problem.
Code-structure-based CoT utilizes some code structure or programming concepts.
Besides these two prompting only techniques, another approach used by many is fine-tuning or instruction tuning for software engineering tasks with code CoT data.

\phantomsection\label{subsec:cotplan}
\textbf{Plan \cotshort.}
Several recent approaches enhance code generation by explicitly modeling intermediate reasoning or problem understanding steps.  
For instance, \texttt{PlanSearch}~\cite{wang2024planning} generates 3--6 problem observations, combines them into natural language plans, and translates these into pseudocode and then code.  
\texttt{Self-Planning}~\cite{jiang2024self} uses few-shot prompting to extract a high-level plan from the problem, which guides code generation. 
\texttt{ClarifyGPT}~\cite{mu2023clarifygpt} employs test generation to construct clarifying questions and answers that are appended to the prompt for code synthesis.

\phantomsection\label{subsec:cotstruct}
\textbf{Code-structure \cotshort.} 
In \texttt{SCoT}, \citealp{li2025structured} use programming structures, like sequence, branch and loop, as steps towards intermediate code, which is used to prompt the model to generate code.
\texttt{Chain of grounded objectives (CGO)}~\cite{yeo2025chain} embed appropriately structured functional objectives into the input prompts to enhance code generation. 
\citealp{pan2025modularization} propose a novel prompting technique, \texttt{Modularization-of-thought (MoT)}, which exploits modularization principles to decompose complex programming problems into smaller independent reasoning steps, via a multi-level reasoning graph. 
\citealp{le2023codechain} also elicit modularized code generation but in a multi-step technique called \texttt{CodeChain}, which is a chain of self-revisions applied by picking potentially correct representative sub-modules. 
Recently, VisualCoder \cite{chi-etal-2025-visualcoder} integrated multimodal CoT reasoning with a visual Control Flow Graph (CFG) to obtain deeper insights into execution flows and improve code execution reasoning and program repair generated solutions.

\label{subsec:cotft}
%\subsubsection{\cotshort{} fine-tuning}
\textbf{Training with \cotshort.} \citealp{sun2024unicoder} define \texttt{UniCoder}; they use an intermediate representation CoT based on PL conventions and use this to instruction-tune a model on a multi-task learning objective.
\citealp{yang2024chain} generate high-quality CoTs based on the \texttt{COTTON} framework, which trains light-LMs (< 10B parameters) to generate CoT comparable to those generated by strong teacher LLMs.
\texttt{ChainCoder}~\cite{zheng2023chaincoder} generates code iteratively in a "course-to-fine" approach and trains a model using an AST-based vocabulary.
\texttt{SemCoder}~\cite{ding2024semcoder} uses a monologue reasoning approach to train a model to learn program semantics, which is generated by asking the Code LLM to summarize the program functionalities, key properties and constraints, and reason about code execution step-by-step using a bi-directional monologue reasoning method. 
\texttt{MSCoT}~\cite{jin2025mscot} extends SCoT~\cite{li2025structured} to 11 more programming languages beyond Python; a trained MSCoT model generates structured-CoT before producing code in multiple languages.

\subsection{Self-refinement}
\label{sec:execution}

Self-refinement involves executing LLM-generated code in an environment and having the same or a different LLM reason about the execution output.
This reasoning can be fed back to the LLM to refine the code.

\label{subsec:selfeval} \
\textbf{Execution reflection.} 
These strategies utilize code execution feedback to select the final prediction from a LLM. 
In \citet{chen2023teaching}, the {\texttt{Self-debugging} approach, teaches the model to self-debug i.e., debug the model's predicted code, via few shot prompting and without additional model training. 
A similar approach was taken in Code Chain-of-Thought (\texttt{CodeCoT}) by \citet{huang2023codecot}, where CoT is used as a first step to generate the code, then a LLM generates test cases to validate whether the code has syntax errors during the execution. 
\texttt{AlphaCodium}, proposed by ~\citet{ridnik2024code}, is a flow with two key phases, (a) pre-processing to generate reflection and  (b) iterative code generation, to improve code LLM performance that does not require training a model. 
In \texttt{revisited self-debugging}~\cite{chen2025revisit} authors explored both post-execution and in-execution self-debugging, leveraging self-generated tests. 
More recently, \citet{tian-fixing-spec-cg-2025} proposed \(\mu\)Fix (Misunderstanding Fixing) where thought-eliciting prompting techniques are combined with feedback-based prompting to improve the code generation performance of LLMs. 

\label{subsec:trainexec}
\textbf{Training with feedback.} We pinpoint approaches that train an LLM, leveraging execution data, to improve model performance. 
LEarning to VERify \cite{Ni-LEVER-icml-2023} (\texttt{LEVER}) is an approach where verifiers are trained to check whether the generated code is correct or not based on three sources of information: the natural language input, the program itself, and its execution results.
CYCLE~\cite{ding2024cycle} trains code LLMs to self-refine using natural language specifications, generated code, and execution feedback, while avoiding repeated errors via a Past Generation Mask. 
Similarly, \cite{jiang2025ledextrainingllmsbetter} proposed \texttt{LEDEX}, a training framework to improve the self-debugging capability of LLMs using a chain of explanations on the wrong code followed by code refinement. 

\textbf{Automated Test Generation.} Unit Tests (UT) are one of the fundamental pieces to assess the correctness of code and give execution-based feedback to code generation models. 
UTGEN \cite{prasad2025learninggenerateunittests} is a data creation and training recipe that bootstraps training data for UT generation and works by perturbing code to simulate errors, generating failing tests and augmenting it with CoT rationales.

AceCoder \cite{zeng2025acecoderacingcoderrl} leverages automated large-scale test-case synthesis to enhance code model training. They proposed a pipeline that generates extensive \textit{(question, test-cases)} pairs from existing code data.
Similarly, \citet{liu2024dstcdirectpreferencelearning} propose Direct Preference Learning with Only Self-Generated Tests and Code (DSTC), using only self-generated code snippets and tests to construct preference pairs with direct preference learning to improve LM coding accuracy without external annotations.
ASTER \cite{pan2025asternaturalmultilanguageunit} is 
a multilingual UT-generator built with LLMs guided by lightweight program analysis.

\subsection{Inference Scaling}
\label{sec:inferencescaling}
Several approaches to code generation, code repair, and test-case generation use \textit{tree-based} strategies to guide decisions and explore reasoning paths, while others use sampling. \hspace{6em}

\phantomsection\label{sec:sampling} 
\textbf{Sampling.} In \texttt{AlphaCode}, \citet{li2022competition} filter and cluster samples according to program behavior on model-generated test inputs, selecting one candidate per cluster. 
The authors of \texttt{REx} \citep{tang2024code} frame iterative code repair, or \textit{refinement}, as a multi-armed bandit problem which is solved using Thompson sampling. 
In \texttt{S*}, \citet{li2025s} take a hybrid sampling approach, first generating N diverse programs in parallel then refining them using iterative debugging (informed by execution). 
CodeTree~\citep{li2024codetree} and ToC~\citep{ni2024tree} both model reasoning as tree search—CodeTree combines planning, execution-guided reasoning, and heuristics (test-pass rate, LM critique) via multi-agent roles, while ToC uses a binary pass/fail heuristic with reflective, multi-strategy execution for diverse solutions.

\phantomsection\label{sec:search} 
\textbf{Search.} 
\texttt{Tree-of-Thoughts (ToT)} \citep{yao2023tree} allows LMs to explore multiple reasoning paths over thoughts, where thoughts are language sequences that serve as intermediate steps towards problem solutions and represent the states or nodes of the tree. 
Similarly, \texttt{Guided tree-of-thought (GToT)} \citep{long2023large} uses tree-search guided by an LLM heuristic; it generates intermediate solutions through prompting, employs a checker to validate these solutions, and uses a controller to manage search and backtracking, enabling long-range reasoning. 
For test generation, \citet{ouedraogo2024large} show that \texttt{GToT} effectively produces syntactically-correct, compilable test suites with superior code coverage. \citet{yu2024outcome} propose \texttt{Outcome-Refining Process Supervision (ORPS)}, a beam-search approach for code generation over a "reasoning tree". 
\texttt{SWE-Search}~\citep{antoniades2024swe} is a moatless-tools~\citep{moatless} based multi-agent framework which integrates Monte-Carlo Tree Search with self-improvement for bug-fixing.  

\definecolor{lightgray}{gray}{0.95}
\definecolor{lightergray}{gray}{0.98}

\definecolor{lightblue}{RGB}{224,242,255}
\definecolor{lightorange}{RGB}{255,236,204}
\definecolor{lightgreen}{RGB}{217,240,211}
\definecolor{lightred}{RGB}{255,204,204}

\begin{table*}[htbp!]
\centering
\resizebox{\textwidth}{!}{
\begin{tabular}{l|ccc|ccc|cc|ccc}
\toprule
\textbf{Approach} & \multicolumn{3}{c|}{\colorbox{lightblue}{\textbf{Code \cotshort}}} & \multicolumn{3}{c|}{\colorbox{lightorange}{\textbf{Self-refinement}}} & \multicolumn{2}{c|}{\colorbox{lightgreen}{\textbf{Inference Scaling}}} & \multicolumn{3}{c}{\colorbox{lightred}{\textbf{\sweagents}}} \\
\hhline{~|---|---|--|---}
 & \makecell{Plan} & \makecell{Struct} & \makecell{Train\\\cotshort}  & \makecell{Exec.\\ref.} & \makecell{Train\\feedback} & \makecell{ATG} & \makecell{Samp.} & \makecell{Search} & \makecell{Work-\\flow} & \makecell{Agent\\Opt.} & \makecell{Train\\traj.} \\
\midrule
\rowcolor{gray!10} \colorbox{lightgreen}{AlphaCode} \citeyearpar{li2022competition} & & &  &  &  &  & \cmark & \cmark &  &  &  \\
\colorbox{lightblue}{ClarifyGPT} \citeyearpar{mu2023clarifygpt}, \colorbox{lightblue}{Self-Planning} \citeyearpar{jiang2024self} & \cmark & & &  &  &  &  &  &  &  &  \\
\rowcolor{gray!10} \colorbox{lightblue}{CodeChain} \citeyearpar{le2023codechain}, \colorbox{lightblue}{SCoT} \citeyearpar{li2025structured}, \colorbox{lightblue}{CGO} \citeyearpar{yeo2025chain}, & & \cmark & &  &  &  &  &  &  &  &  \\
\rowcolor{gray!10} \colorbox{lightblue}{MoT} \citeyearpar{pan2025modularization}, \colorbox{lightblue}{VisualCoder} \citeyearpar{chi-etal-2025-visualcoder} & & \cmark & &  &  &  &  &  &  &  &  \\
\colorbox{lightblue}{ChainCoder} \citeyearpar{zheng2023chaincoder}, \colorbox{lightblue}{UniCoder} \citeyearpar{sun2024unicoder}, \colorbox{lightblue}{MSCoT} \citeyearpar{jin2025mscot} & & \cmark & \cmark &  &  &  &  &  &  &  &  \\
\rowcolor{gray!10} \colorbox{lightorange}{LEVER} \citeyearpar{Ni-LEVER-icml-2023}, \colorbox{lightorange}{Self-Debugging} \citeyearpar{chen2023teaching}, & & & & \cmark &  &  & \cmark & \cmark &  &  &  \\
\rowcolor{gray!10} \colorbox{lightgreen}{ORPS} \citeyearpar{yu2024outcome}, \colorbox{lightgreen}{REx} \citeyearpar{tang2024code} & & & & \cmark &  &  & \cmark & \cmark &  &  &  \\
\colorbox{lightorange}{CodeCoT} \citeyearpar{huang2023codecot}, \colorbox{lightorange}{AlphaCodium} \citeyearpar{ridnik2024code} & \cmark &  &  & \cmark &   & \cmark &  &  &  &  &  \\
\rowcolor{gray!10} \colorbox{lightgreen}{GToT} \citeyearpar{long2023large} & & &  &  &  &  & \cmark & \cmark &  &  & \cmark \\
\colorbox{lightblue}{PlanSearch} \citeyearpar{wang2024planning} & \cmark & & & \cmark &  &  & \cmark &  &  &  &  \\
\rowcolor{gray!10} \colorbox{lightblue}{COTTON} \citeyearpar{yang2024chain} & \cmark &  & \cmark &  &  &  &  &  &  &  &  \\
\colorbox{lightblue}{SemCoder} \citeyearpar{ding2024semcoder} & & \cmark & \cmark &  &  & \cmark &  &  &  &  &  \\
\rowcolor{gray!10} \colorbox{lightorange}{CYCLE} \citeyearpar{ding2024cycle} & & &  & \cmark & \cmark &  & \cmark &  &  &  &  \\
\colorbox{lightorange}{DSTC} \citeyearpar{liu2024dstcdirectpreferencelearning} &  &  &  &  & \cmark & \cmark &  &  &  &  &   \\
\rowcolor{gray!10} \colorbox{lightgreen}{CodeTree} \citeyearpar{li2024codetree}, \colorbox{lightgreen}{Tree-of-Code} \citeyearpar{ni2024tree}, \colorbox{lightgreen}{SWE-Search} \citeyearpar{antoniades2024swe} &  &  &  &  &  &  & \cmark & \cmark &  & \cmark & \\
\colorbox{lightred}{Agentless} \citeyearpar{xia2024agentless} &  &  &  & \cmark &  & & &  & \cmark &  & \\
\rowcolor{gray!10} \colorbox{lightred}{AutoCodeRover} \citeyearpar{zhang2024autocoderover}, \colorbox{lightred}{PairCoder} \citeyearpar{zhang2024pair} &  &  &  & \cmark &   &  & &  & \cmark & \cmark & \\
\colorbox{lightred}{CodeAct} \citeyearpar{wang2024executable} &  &  &  & \cmark &  &  & &  &  & \cmark & \cmark \\
\rowcolor{gray!10} \colorbox{lightred}{OpenHands} \citeyearpar{wang2024openhands}, \colorbox{lightred}{MASAI} \citeyearpar{arora2024masai}, \colorbox{lightred}{CodeR} \citeyearpar{chen2024coder} &  &  &  & \cmark &  &  &  &  &  & \cmark & \\
\colorbox{lightred}{AgileCoder} \citeyearpar{nguyen2024agilecoder}, \colorbox{lightred}{HyperAgent} \citeyearpar{phan2024hyperagent}, \colorbox{lightred}{SWE-Agent} \citeyearpar{yang2024sweagent} &  &  &  & \cmark &  &  &  &  &  & \cmark & \\
\rowcolor{gray!10} \colorbox{lightred}{Lingma} \citeyearpar{ma2024lingma}, \colorbox{lightred}{SWE-Fixer} \citeyearpar{xie2025swe} & &  &  & \cmark &  &  &  &  & \cmark &  & \cmark \\
\colorbox{lightred}{SWE-Gym} \citeyearpar{pan2024training} &  &  &  & \cmark &  &  & \cmark & \cmark &  & \cmark & \cmark \\
\rowcolor{gray!10} \colorbox{lightorange}{Rev.\ Self-Debugging} \citeyearpar{chen2025revisit} & & &  & \cmark & \cmark &  &  &  & \cmark &  &  \\
\colorbox{lightgreen}{S*} \citeyearpar{li2025s} & & &  & \cmark & \cmark & \cmark &  &  &  & \cmark &  \\
\rowcolor{gray!10} \colorbox{lightorange}{$\mu$Fix} \citeyearpar{tian-fixing-spec-cg-2025} & \cmark &  &  & \cmark & &  &  &  &  &  &  \\
\colorbox{lightorange}{LeDex} \citeyearpar{jiang2025ledextrainingllmsbetter} & \cmark &  &  & \cmark & \cmark &  & \cmark &  &  &  &   \\
\rowcolor{gray!10} \colorbox{lightorange}{UTGEN} \citeyearpar{prasad2025learninggenerateunittests}, \colorbox{lightorange}{AceCoder} \citeyearpar{zeng2025acecoderacingcoderrl} &  &  &  &  & \cmark & \cmark & \cmark &  &  &  &   \\
\colorbox{lightorange}{SWT-Bench} \citeyearpar{mündler2025swtbenchtestingvalidatingrealworld}, \colorbox{lightorange}{Otter} \citeyearpar{ahmed2025otter} & \cmark &  &  &  &  & \cmark & \cmark &  &  & \cmark &   \\
\rowcolor{gray!10} \colorbox{lightred}{SWE-RL} \citeyearpar{wei2025swe} &  &  &  &  &  &  &  &  &  &  & \cmark \\
\bottomrule
\end{tabular}
}
\caption{Summary of the test-time compute-based reasoning techniques used in the papers surveyed. For each work, we assign a dominant technique as shown in the taxonomy Fig.~\ref{fig:taxonomy}. Additional techniques used by the same approach are marked with \cmark. For example, PlanSearch is categorized under \colorbox{lightblue}{Code \cotshort} as its dominant technique, but it also incorporates elements of \colorbox{lightorange}{Self-refinement} and \colorbox{lightgreen}{Inference Scaling}.}
\label{tab:llm-hybrid}
\end{table*}

\subsection{\sweagents}
\label{sec:agentssec}
Agentic systems use many of the reasoning techniques described in Sec.~\ref{sec:codereason} for different tasks.
Software Engineering (SWE) agents take a programming problem and iteratively solve it by self-debugging based on the feedback provided by the environment.
The self-debugging is enabled by CoT style natural language reflection~\citep{shinn2023reflexion} on environment feedback.
The reasoning is done by an LLM which interacts with the agent execution environment with tool calls~\citep{yao2023react}. 

\label{subsec:workflow}
\textbf{Workflow.} \citealp{effectagents} draw a distinction between Agents and LLM-based workflows stating that the latter are simpler, have a fixed path and do not require an LLM to make a decision.
Agentless \citep{xia2024agentless} is a three step process for Github issue resolution involving localization, repair and patch validation. AutoCodeRover~\citep{zhang2024autocoderover} uses program structure, in the form of an Abstract Syntax Tree (AST), to enhance code search and look at a software as classes and functions, rather than a set of files.

\label{subsec:agentoptim}
\textbf{Agent Optimization} can often lead to performance gains.
There can be many ways to improve an SE agent, including but not limited to, better environment management or agent-environment interface, improved workflow or architecture, and incorporating more tools. SWE-Agent~\citep{yang2024sweagent} is an agent capable of editing repository-level code by generating a thought and a command, and subsequently incorporating the feedback from the command’s execution into the environment. 
In CodeAct, \cite{wang2024executable} propose to use executable Python code to consolidate LLM agents' actions into a unified action space.
OpenHands~\citep{wang2024openhands} is a platform for developing flexible AI agents that interact with the digital world by writing code, interacting with the command line or browsing the web.
This platform allows for integration of other specialist agents, like CodeAct~\citep{wang2024executable} for software engineering.
There are other multi-agent techniques like MASAI~\cite{arora2024masai}, CodeR~\cite{chen2024coder}, PairCoder~\cite{zhang2024pair}, HyperAgent~\cite{phan2024hyperagent} and AgileCoder~\cite{nguyen2024agilecoder}. 

\label{subsec:reasoningmodel}
\textbf{Training with trajectory.} 
Some agentic systems improve the underlying reasoning model by training on agent trajectories, which include steps like CoT, tool calls, and patches. 
\citet{ma2024lingma} note that software evolution spans code, reasoning, tools, and cross-role interactions, and fine-tune their \texttt{Lingma} SWE-GPT models (7B, 72B) on repository understanding, bug localization, patching, and rejection sampling from pull requests.
\citealp{pan2024training} build \texttt{SWE-Gym} from 2,438 real-world Python tasks—each with a runnable codebase, unit tests, and an NL spec. Using OpenHands scaffolding~\citep{wang2024openhands}, they fine-tune Qwen2.5-Coder-32B~\citep{hui2024qwen2} on 491 agent–environment trajectories and train a verifier on the same data for scalable inference.
\texttt{SWE-Fixer}~\citep{xie2025swe} uses a fine-tuned Qwen2.5-7B retriever, boosted with BM25, to identify relevant files, while a fine-tuned Qwen2.5-72B editor generates patches for GitHub issues.
In \texttt{SWE-RL}~\citep{wei2025swe}, Llama 3~\citep{grattafiori2024llama} is trained with lightweight rule rewards and GRPO~\citep{shao2024deepseekmath} on 11M filtered PRs, producing Llama3-SWE-RL-70B, the top medium-sized model on SWE-bench Verified~\citep{swebenchver} upon release. \hspace{2em}

\section{Code Reasoning: Tasks}
\label{sec:benchmetrics}

In this section we discuss the different tasks and benchmarks which are used to evaluate code reasoning techniques described in Sec.~\ref{sec:codereason}.

\subsection{Code Generation}
\label{subsec:taskbench}

For code generation, a popular task, most common benchmarks include \textit{HumanEval (HE)} ~\citep{chen2021evaluatinglargelanguagemodels}, \texttt{HumanEvalPack} ~\citep{Muennighoff2023OctoPackIT}, \textit{MBPP} ~\citep{austin2021programsynthesislargelanguage},  \textit{APPS} ~\citep{hendrycks2021measuringcodingchallengecompetence}, and \textit{CodeContests}~\citep{codecontests-science.abq1158}.

More recently, \textit{LiveCodeBench (LCB)}~\citep{jain2024livecodebenchholisticcontaminationfree} collected new problems over time from contests platforms including LeetCode, AtCoder, and CodeForces. \textit{BigCodeBench}~\citep{zhuo2025bigcodebenchbenchmarkingcodegeneration} challenges LLMs to invoke multiple function calls as tools from multiple libraries and domains for different fine-grained tasks. 
% Each task has \(5.6\) test cases, on average, with an average branch coverage of \(99\%\).
CRUXEval~\citep{gu2024cruxevalbenchmarkcodereasoning} includes both input and output predictions to evaluate code reasoning and code execution, respectively.
ConvCodeBench ~\citep{han2025convcodeworldbenchmarkingconversationalcode} is a benchmark for interactive code generation, it uses pre-generated feedback logs, avoiding costly LLM calls for verbal feedback while maintaining strong correlation with live results; \textit{Spider}~\citep{yu-etal-2018-spider, lei2025spider20evaluatinglanguage} is a benchmark to evaluate the generation of SQL queries from natural language. 

\subsection{Test Generation}
\label{subsec:testgen}
For test generation, benchmarks like \textit{TestEval}~\citep{wang-etal-2025-testeval} can help on three different aspects: overall coverage, targeted line/branch coverage, and targeted path coverage. 
\textit{SWT-Bench} \citep{mündler2025swtbenchtestingvalidatingrealworld} is another github based test-generation benchmark; Otter  \citep{ahmed2025otter} too, proposed an LLM-based solution to generate test cases from issues. \footnote{Appendix \ref{ap:codeeval} lists metrics that can be used to assess code LLM performance.}

\subsection{Issue Resolution}
\label{subsec:issueres}
For GitHub issue resolution, \textit{SWE-Bench}~\citep{jimenez2024swebenchlanguagemodelsresolve} is a popular benchmark. Other variations of SWE-Bench include: \textit{SWE-Bench Multimodal} \citep{yang2024swebenchmultimodalaisystems} for visual and user-facing components, and \textit{Multi-SWE-Bench}~\citep{zan2025multiswebenchmultilingualbenchmarkissue} and \textit{SWE-PolyBench}~\citep{rashid2025swe} for more programming languages besides Python. \textit{M\(^3\)ToolEval} \citep{wang2024executable} is used for multi-turn, multi-tool complex tasks.

\subsection{Reasoning and Understanding}
\label{sec:evalreason}

Evaluating the ability of LLMs to both correctly and soundly reason about runtime behavior of code can help understand and verify whether the generated code aligns with the intended goal. 
\textit{ReEval}~\citep{junkai-runtime-behavior-reasoning-icse2025} helps to analyze how Code LLMs reason about runtime behaviors (e.g., program state, execution paths) of programs. 
ExeRScope \citep{liu2025toolindepthanalysiscode} analyzes the code execution reasoning output from LLMs for different code reasoning benchmarks and helps understand the impact of code properties (program constructs, complexity, dynamic program properties, and variable types) for such  benchmarks. 
\textit{CodeMMLU}~\citep{manh2025codemmlumultitaskbenchmarkassessing} is a large benchmark to evaluate both code understanding and code reasoning through a multiple-choice question-answering approach. 
CodeMind \citep{liu2024codemindframeworkchallengelarge} is a code reasoning benchmark for LLMs, evaluating Independent Execution Reasoning (IER), Dependent Execution Reasoning (DER), and Specification Reasoning (SR) tasks and metrics.

\section{Comparison and Discussion}
\label{sec:comparison}

\newtcolorbox{findingsboxold}{
  colback=blue!10,
  colframe=black,
  arc=1pt,
  boxrule=0.2pt,
  left=2pt, right=2pt, top=1pt, bottom=1pt
}

\newtcolorbox{findingsboxold-tight}{
  colback=blue!10,
  colframe=black,
  arc=1pt,
  boxrule=0.2pt,
  left=2pt, right=2pt, top=1pt, bottom=1pt,
  before skip=2pt,  % Very little space above the box
  after skip=2pt    % Very little space below the box
}

\newtcolorbox{findingsbox}{
  colback=gray!10,
  colframe=black,
  arc=2pt,
  boxrule=0.5pt,
  left=2pt, right=2pt, top=2pt, bottom=2pt,
  boxsep=0pt, % Controls spacing around the text inside the box
  width=\linewidth, % Full line width
  before skip=6pt, % Space before the box
  after skip=6pt % Space after the box
}

The ideal way to understand the impact of different code reasoning techniques is an exhaustive comparative study under a common experimental setup. 
However, such a controlled evaluation is often impractical. 
To enable fairer comparisons across heterogeneous experimental settings, we synthesize reported results on coding benchmarks while conditioning on the same underlying \llms{} (i.e., comparing techniques evaluated with the same model). 
Most of the comparable results are on code generation benchmarks, with a few on issue resolution.
We summarize reported performance for the techniques described in Sec.~\ref{sec:codereason} on the benchmarks and tasks in Sec.~\ref{sec:benchmetrics}. 
We further visualize technique performance on the intersection of models and benchmarks for which results are available, making cross-technique trends easier to interpret. 
This analysis is not exhaustive; it is restricted to benchmarks and models shared by only a subset of techniques.
All comparisons are cross-paper; each result is taken from its respective original paper. 
We do not normalize for differences in prompts, execution harnesses, or evaluation details.
The complete set of results are in the appendix in Tables~\ref{tab:apps-results}, \ref{tab:lcb-results}, \ref{tab:swe-bench}, \ref{tab:mbpp-family}, \ref{tab:mbpp-family-continued}, \ref{tab:he-family}, and \ref{tab:he-family-continued}.
Observations 1--5 are scoped to code generation tasks, where the majority of comparable results are available. Observation 6 extends to issue resolution, supported by SWE-Bench results.
We report the margins of difference for all comparisons in Tab.~\ref{tab:margins}.
We discuss threats to the validity of this cross-paper comparison approach in Sec.~\ref{sec:threats}.

\subsection{Plan \cotshort vs. Code Structure \cotshort}
\label{subsec:comp-plan-struct-cot}
\input{sections/results-obs-1-plot}
Section \ref{sec:cot} surveys works on code-specific \cotshort prompting and categorizes them into plan-based and structure-based.
Fig.~\ref{fig:obs1_mbpp_comparison} shows the performance of different plan and structure based techniques on code generation benchmarks (Sec.~\ref{subsec:taskbench}).
Structure-aware CoT techniques tend to outperform plan-based approaches across multiple code generation benchmarks.
With gpt-3.5-turbo, CGO outperforms Self-Planning on MBPP-S, MBPP+, HE, and HE+, and surpasses ClarifyGPT on MBPP-S and HE.
This pattern extends to Llama-3-8B-Instr, where CGO achieves better performance than Self-Planning on MBPP-S and MBPP+.
With gpt-4o mini, SCoT demonstrates superior performance over Self-Planning on MBPP and MBPP+, and on APPS with gpt-3.5-turbo.
Similarly, CodeChain outperforms Self-Planning on APPS with gpt-3.5-turbo.
These results suggest that code structure-based techniques tend to outperform plan-based approaches in the reported code generation benchmarks.

\begin{findingsboxold}
\textbf{\dashuline{Observation 1:} Reported results across several studies suggest that structure-aware CoT strategies can outperform planning-based CoT strategies in some code generation benchmarks.}

\end{findingsboxold}

\subsection{Modular Code Structure \cotshort}
\label{subsec:comp-mod-cot}

Code structure aware \cotshort can be sub-categorized into modular approaches, as shown in Fig.~\ref{fig:obs1_mbpp_comparison}. 
Modularity improves upon structure-based CoT by providing ultra-localized scoping with more clearly defined and specific functionality, eliminating the chance of error propagating to subsequent steps.
MoT and CodeChain represent modular \cotshort techniques, more specific types of structure-aware prompting.
MoT demonstrates superior performance over SCoT and Self-Planning with DeepSeek-R1~\citep{guo2025deepseek} on MBPP and HE, with comparable results on MBPP and MBPP+ using gpt-4o mini.
CodeChain exhibits similar performance gains, outperforming SCoT and Self-Planning on APPS with gpt-3.5.
Building on Observation 1, these results suggest that modular formats can outperform other structure-aware Code \cotshort prompting techniques on the code generation benchmarks studied.

\begin{findingsboxold}
\textbf{\dashuline{Observation 2:} Reported results across several studies suggest that modular CoT techniques can outperform other structure-aware and plan-based CoT approaches in some code generation benchmarks.}

\end{findingsboxold}

\subsection{Self-refinement vs. Code \cotshort}
\label{subsec:comp-refine}
%\todo{Check DS label, update caption, add stripes to some similar colors}

\input{sections/results-obs-define-patterns}

\begin{figure}[htbp]
  \centering
\begin{tikzpicture}

\tikzset{
  every axis/.style={
    ybar=0pt,
    bar width=7pt,
    height=3.6cm,
    grid=major,
    clip=false,
    tick label style={font=\small},
    x tick label style={font=\small, rotate=45, anchor=east, xshift=-3pt},
    ylabel style={font=\large},
    title style={yshift=1pt, font=\bfseries\large}
  }
}

% ======================================================================================
%                                   SINGLE ROW
% ======================================================================================

% (1) HE+
\begin{axis}[
  name=plot1,
  width=0.22\textwidth,
  ylabel={Pass@1 (\%)},
  title={HE+},
  symbolic x coords={Cl-3.5, gpt-3.5},
  xtick={Cl-3.5, gpt-3.5},
  enlarge x limits=0.55,
  ymin=65, ymax=95,
  ytick={65,70,75,80,85,90,95}
]
\addplot[ybar, bar shift=-5pt, bluePattern4]     coordinates { (Cl-3.5,81.6) };
\addplot[ybar, bar shift= 5pt, orangePattern7]   coordinates { (Cl-3.5,89.0) };
\addplot[ybar, bar shift=-10pt, bluePattern6]    coordinates { (gpt-3.5,67.3) };
\addplot[ybar, bar shift=  0pt, bluePatternwhitelinesleft] coordinates { (gpt-3.5,68.5) };
\addplot[ybar, bar shift= 10pt, orangePattern3]  coordinates { (gpt-3.5,80.5) };
\end{axis}

% (2) HE
\begin{axis}[
  at={(plot1.east)}, anchor=west, xshift=0.8cm,
  width=0.24\textwidth,
  name=plot2,
  title={HE},
  symbolic x coords={gpt-3.5},
  xtick={gpt-3.5},
  enlarge x limits=0.7,
  ymin=55, ymax=95,
  ytick={55,65,75,85,95}
]
\addplot[ybar, bar shift=-18pt, fill=blue!60]  coordinates { (gpt-3.5,60.6) };
\addplot[ybar, bar shift=-10pt, bluePattern6]  coordinates { (gpt-3.5,72.7) };
\addplot[ybar, bar shift=-2pt,  bluePattern1]  coordinates { (gpt-3.5,74.4) };
\addplot[ybar, bar shift= 6pt,  bluePatternwhitelinesleft] coordinates { (gpt-3.5,74.6) };
\addplot[ybar, bar shift=14pt,  orangePattern3]  coordinates { (gpt-3.5,90.2) };
\addplot[ybar, bar shift=22pt,  orangePattern2]  coordinates { (gpt-3.5,77.4) };
\end{axis}

% (3) MBPP-ET
\begin{axis}[
  at={(plot2.east)}, anchor=west, xshift=0.8cm,
  width=0.18\textwidth,
  name=plot3,
  title={MBPP-ET},
  symbolic x coords={gpt-3.5},
  xtick={gpt-3.5},
  enlarge x limits=0.7,
  ymin=50, ymax=75,
  ytick={50,55,60,65,70,75},
  bar width=6pt
]
\addplot[ybar, bar shift=-7pt, bluePattern1]    coordinates { (gpt-3.5,55.6) };
\addplot[ybar, bar shift= 0pt, orangePattern2]  coordinates { (gpt-3.5,60.4) };
\addplot[ybar, bar shift= 7pt, orangePattern3]  coordinates { (gpt-3.5,69.1) };
\end{axis}

% (4) MBPP
\begin{axis}[
  at={(plot3.east)}, anchor=west, xshift=1.0cm,
  width=0.14\textwidth,
  name=plot4,
  title={MBPP},
  symbolic x coords={gpt-3.5},
  xtick={gpt-3.5},
  enlarge x limits=0.7,
  ymin=40, ymax=80,
  ytick={40,50,60,70,80}
]
\addplot[ybar, bar shift=-5pt, fill=blue!60]    coordinates { (gpt-3.5,47.0) };
\addplot[ybar, bar shift= 5pt, orangePattern2]  coordinates { (gpt-3.5,74.2) };
\end{axis}

% (5) APPS
\begin{axis}[
  at={(plot4.east)}, anchor=west, xshift=0.8cm,
  width=0.28\textwidth,
  name=plot5,
  title={APPS},
  symbolic x coords={gpt-3.5, DS},
  xtick={gpt-3.5, DS},
  enlarge x limits=0.55,
  ymin=0, ymax=40,
  ytick={0,10,20,30,40}
]
\addplot[ybar, bar shift=-18pt, bluePattern6]        coordinates { (gpt-3.5,21.3) };
\addplot[ybar, bar shift=-10pt, fill=blue!60]        coordinates { (gpt-3.5,22.0) };
\addplot[ybar, bar shift=-2pt,  bluePattern2Struct]  coordinates { (gpt-3.5,26.4) };
\addplot[ybar, bar shift= 6pt,  orangePattern3]      coordinates { (gpt-3.5,35.7) };
\addplot[ybar, bar shift=-18pt, fill=blue!60]        coordinates { (DS,4.3) };
\addplot[ybar, bar shift=-10pt, orangePattern2]      coordinates { (DS,4.7) };
\addplot[ybar, bar shift=-2pt,  orangePattern5light] coordinates { (DS,11.3) };
\addplot[ybar, bar shift= 6pt,  orangePattern3]      coordinates { (DS,14.0) };
\end{axis}

% (6) CodeContests
\begin{axis}[
  at={(plot5.east)}, anchor=west, xshift=1.0cm,
  width=0.16\textwidth,
  name=plot6,
  ylabel={Pass@5 (\%)},
  ylabel style={at={(axis description cs:-0.33,0.5)}},
  title={CodeContests},
  symbolic x coords={gpt-3.5},
  xtick={gpt-3.5},
  enlarge x limits=0.7,
  ymin=0, ymax=20,
  ytick={0,5,10,15,20}
]
\addplot[ybar, bar shift=-5pt, bluePattern2Struct]   coordinates { (gpt-3.5,14.1) };
\addplot[ybar, bar shift= 5pt, orangePattern5]       coordinates { (gpt-3.5,17.0) };
\end{axis}

% ======================================================================================
%                            CENTERED LEGENDS
% ======================================================================================

% Center position under entire row:
\coordinate (midpoint) at ($(plot1.south)!0.5!(plot6.south)$);
% Global x-axis label
\node[font=\small, anchor=north] at ([yshift=-1.3cm]midpoint) {Models};

% ========== Plan-based CoT ==========
\node[draw=gray!40, dashed, anchor=north, inner sep=4pt]
at ([yshift=-2.0cm, xshift=-4.8cm]midpoint) {   % << MOVED LEFT
  \normalsize
  \begin{tikzpicture}[baseline]
    \node[anchor=west, font=\small] at (0,0) {\textbf{Plan-based CoT:}};
    \fill[bluePattern6] (0.00,-0.73) rectangle (0.25,-0.29);
    \fill[bluePattern6] (0.37,-0.71) rectangle (0.60,-0.31);
    \node[anchor=west, font=\small] at (0.68,-0.51) {Self-Planning};
    \fill[bluePattern4] (2.95,-0.73) rectangle (3.20,-0.29);
    \fill[bluePattern4] (3.32,-0.71) rectangle (3.55,-0.31);
    \node[anchor=west, font=\small] at (3.63,-0.51) {PlanSearch};
    \fill[bluePattern1] (5.85,-0.73) rectangle (6.10,-0.29);
    \fill[bluePattern1] (6.22,-0.71) rectangle (6.45,-0.31);
    \node[anchor=west, font=\small] at (6.53,-0.51) {ClarifyGPT};
  \end{tikzpicture}
};

% ========== Self-refinement ==========
\node[draw=gray!40, dashed, anchor=north, inner sep=4pt]
at ([yshift=-3.5cm, xshift=-1.0cm]midpoint) {
  \normalsize
\begin{tikzpicture}[baseline]
  \node[anchor=west, font=\small] at (0,0) {\textbf{Self-refinement:}};
  \fill[orangePattern3] (-0.10,-0.73) rectangle (0.15,-0.29);
  \fill[orangePattern3] (0.27,-0.71) rectangle (0.50,-0.31);
  \node[anchor=west, font=\small] at (0.52,-0.51) {$\mu$Fix};
  \fill[orangePattern2] (1.56,-0.73) rectangle (1.81,-0.29);
  \fill[orangePattern2] (1.95,-0.71) rectangle (2.18,-0.31);
  \node[anchor=west, font=\small] at (2.20,-0.51) {Self-Debugging};
  \fill[orangePattern7] (4.70,-0.73) rectangle (4.95,-0.29);
  \fill[orangePattern7] (5.07,-0.71) rectangle (5.30,-0.31);
  \node[anchor=west, font=\small] at (5.32,-0.51) {Revisiting Self-Debugging};
  \fill[orangePattern5] (9.80,-0.73) rectangle (10.05,-0.29);
  \fill[orangePattern5] (10.17,-0.71) rectangle (10.40,-0.31);
  \node[anchor=west, font=\small] at (10.48,-0.51) {AlphaCodium};
  \fill[orangePattern5light] (12.90,-0.73) rectangle (13.15,-0.29);
  \fill[orangePattern5light] (13.27,-0.71) rectangle (13.50,-0.31);
  \node[anchor=west, font=\small] at (13.58,-0.51) {CYCLE};
\end{tikzpicture}
};

% ========== Structure-aware ==========
\node[draw=gray!40, dashed, anchor=north, inner sep=4pt]
at ([yshift=-2.0cm, xshift=3.4cm]midpoint) {
  \normalsize
  \begin{tikzpicture}[baseline]
    \node[anchor=west, font=\small] at (0,0) {\textbf{Structure-aware CoT:}};
    \fill[blue!60] (0.00,-0.73) rectangle (0.25,-0.29);
    \fill[blue!60] (0.37,-0.71) rectangle (0.60,-0.31);
    \node[anchor=west, font=\small] at (0.68,-0.51) {SCoT};
    \fill[bluePattern2Struct] (2.30,-0.73) rectangle (2.55,-0.29);
    \fill[bluePattern2Struct] (2.67,-0.71) rectangle (2.90,-0.31);
    \node[anchor=west, font=\small] at (2.98,-0.51) {CodeChain};
    \fill[bluePatternwhitelinesleft] (5.20,-0.73) rectangle (5.45,-0.29);
    \fill[bluePatternwhitelinesleft] (5.57,-0.71) rectangle (5.80,-0.31);
    \node[anchor=west, font=\small] at (5.88,-0.51) {CGO};
  \end{tikzpicture}
};

% ======================================================================================

\end{tikzpicture}
 \renewcommand{\thefootnote}{$\S$}
 \caption{Comparison between code \cotshort and self-refinement techniques on code generation benchmarks HE+, HE, MBPP-ET, MBPP, APPS, and CodeContests with models Claude 3.5 Sonnet (Cl-3.5), GPT-3.5-turbo (GPT-3.5), and DeepSeek-Coder (DS). Code \cotshort is sub-categorized into plan-based and structure-aware \cotshort. CodeContests results are reported as pass@5. Note: Y-axis scales differ across subplots.}
 %\protect\footnotemark}
\vspace{-6pt}
\label{fig:obs3_five_benchmarks}
\end{figure}

Self-refinement (Sec.~\ref{sec:execution}) uses the execution output as feedback to iteratively improve the model generated code.
Incorrect intermediate code is discarded and the model can improve upon it.
As can be seen in Fig.~\ref{fig:obs3_five_benchmarks}, using self-refinement has a bigger impact on the studied code generation benchmarks than any code \cotshort based technique.

Across multiple code generation benchmarks, self-refinement methods tend to outperform CoT approaches.
With gpt-3.5, $\mu$Fix and Self-Debugging surpass other CoT baselines (CGO, SCoT, Self-Planning, ClarifyGPT) and outperform UniCoder on HumanEval.
On MBPP-ET with gpt-3.5, these methods also outperform ClarifyGPT, while Self-Debugging achieves particularly strong gains on MBPP with gpt-3.5, surpassing SCoT by a large margin.
With Claude-3.5~\citep{anthropic2024claude35sonnet}, Revisiting Self-Debugging beats PlanSearch on HE+.
Similar patterns emerge on APPS with gpt-3.5, where $\mu$Fix outperforms CodeChain, SCoT, and Self-Planning.
These findings extend to DeepSeek-Coder~\citep{guo2024deepseek}, where $\mu$Fix and Self-Debugging outperform SCoT, and CYCLE models, which are smaller finetuned models, also surpass SCoT.

\begin{findingsboxold}
\textbf{\dashuline{Observation 3:} Reported results across several studies suggest that execution-aware strategies can outperform Code \cotshort based methods in some code generation benchmarks.}

\end{findingsboxold}

\subsection{Inference Scaling}

\begin{figure}[htbp]
\centering
\hspace{-1.9cm}
\begin{tikzpicture}

% Global axis style to match other plots
\tikzset{
  every axis/.style={
    ybar=0pt,
    bar width=7pt,
    height=3.6cm,
    grid=major,
    clip=false,
    tick label style={font=\small},
    x tick label style={font=\small, rotate=45, anchor=east, xshift=-3pt},
    ylabel style={font=\large},
    title style={yshift=3pt, font=\bfseries\large}
  }
}

% ============================== 1) MBPP+ ==============================
\begin{axis}[
      name=plot1,
      width=0.30\textwidth,
      enlarge x limits=0.35,
      ylabel={Pass@1 (\%)},
      symbolic x coords={gpt-4o, 4o-mini},
      xtick={gpt-4o, 4o-mini},
      ymin=35, ymax=85,
      ytick={35,45,55,65,75,85},
      title={MBPP+}
]
  \addplot[ybar, bar shift=-5pt,  orangePattern7]           coordinates { (gpt-4o,76.5) };
  \addplot[ybar, bar shift=5pt,   darkGreenLightWest]       coordinates { (gpt-4o,80.7) };
  \addplot[ybar, bar shift=-20pt, bluePattern6]             coordinates { (4o-mini,42.4) };
  \addplot[ybar, bar shift=-12pt, fill=blue!60]             coordinates { (4o-mini,51.4) };
  \addplot[ybar, bar shift=-4pt,  bluePurpleStripesDark]    coordinates { (4o-mini,58.1) };
  \addplot[ybar, bar shift=4pt,   bluePattern4]             coordinates { (4o-mini,73.5) };
  \addplot[ybar, bar shift=12pt,  darkGreenLightWest]       coordinates { (4o-mini,77.0) };
\end{axis}

% ============================== 2) MBPP ==============================
\begin{axis}[
      at={(plot1.east)}, anchor=west, xshift=0.8cm,
      name=plot2,
      width=0.30\textwidth,
      enlarge x limits=0.35,
      symbolic x coords={gpt-4o, 4o-mini},
      xtick={gpt-4o, 4o-mini},
      ymin=45, ymax=100,
      ytick={45,55,65,75,85,95},
      title={MBPP}
]
  \addplot[ybar, bar shift=-5pt,  orangePattern7]       coordinates { (gpt-4o,91.5) };
  \addplot[ybar, bar shift=5pt,   darkGreenLightWest]   coordinates { (gpt-4o,98.7) };
  \addplot[ybar, bar shift=-14pt, bluePattern6]         coordinates { (4o-mini,52.1) };
  \addplot[ybar, bar shift=-6pt,  fill=blue!60]         coordinates { (4o-mini,63.9) };
  \addplot[ybar, bar shift=2pt,   bluePurpleStripesDark] coordinates { (4o-mini,73.9) };
  \addplot[ybar, bar shift=10pt,  lightGreenDarkEast]   coordinates { (4o-mini,95.7) };
\end{axis}

% ============================== 3) APPS ==============================
\begin{axis}[
      at={(plot2.east)}, anchor=west, xshift=0.8cm,
      name=plot3,
      width=0.16\textwidth,
      enlarge x limits=0.6,
      symbolic x coords={gpt-4},
      xtick={gpt-4},
      ymin=55, ymax=75,
      ytick={55,60,65,70,75},
      xlabel={Models},
      xlabel style={font=\small, xshift=-40pt},
      title={APPS}
]
  \addplot[ybar, bar shift=-5pt, bluePattern2Struct]   coordinates { (gpt-4,61.5) };
  \addplot[ybar, bar shift=5pt,  lightGreenSolid]      coordinates { (gpt-4,70.0) };
\end{axis}

% ============================== 4) LCB ==============================
\begin{axis}[
      at={(plot3.east)}, anchor=west, xshift=0.8cm,
      name=plot4,
      width=0.21\textwidth,
      enlarge x limits=0.35,
      symbolic x coords={o1-mini, 4o-mini},
      xtick={o1-mini, 4o-mini},
      ymin=35, ymax=90,
      ytick={35,45,55,65,75,85},
      title={LCB}
]
  \addplot[ybar, bar shift=-5pt,  bluePattern4]
    coordinates { (o1-mini,69.5) (4o-mini,39.0) };
  \addplot[ybar, bar shift=5pt,   limeGreenCrossDark]
    coordinates { (o1-mini,85.3) (4o-mini,61.3) };
\end{axis}

% ============================== 5) HumanEval+ ==============================
\begin{axis}[
      at={(plot4.east)}, anchor=west, xshift=0.8cm,
      name=plot5,
      width=0.16\textwidth,
      enlarge x limits=0.6,
      symbolic x coords={gpt-4o},
      xtick={gpt-4o},
      ymin=80, ymax=92,
      ytick={80,84,88,92},
      title={HE+}
]
  \addplot[ybar, bar shift=-5pt, orangePattern7]       coordinates { (gpt-4o,87.8) };
  \addplot[ybar, bar shift=5pt,  darkGreenLightWest]   coordinates { (gpt-4o,86.0) };
\end{axis}

% ============================== 6) HumanEval ==============================
\begin{axis}[
      at={(plot5.east)}, anchor=west, xshift=0.8cm,
      name=plot6,
      width=0.16\textwidth,
      enlarge x limits=0.6,
      symbolic x coords={gpt-4o},
      xtick={gpt-4o},
      ymin=88, ymax=98,
      ytick={88,90,92,94,96,98},
      title={HE}
]
  \addplot[ybar, bar shift=-5pt, orangePattern7]       coordinates { (gpt-4o,92.1) };
  \addplot[ybar, bar shift=5pt,  darkGreenLightWest]   coordinates { (gpt-4o,94.5) };
\end{axis}

% ============================== CENTERING HELPERS ==============================

% Midpoint under all six plots
\coordinate (midpoint) at ($(plot1.south)!0.5!(plot6.south)$);

\end{tikzpicture}

\vspace{0.3cm}

\begin{tikzpicture}

% ================= LEFT LEGEND =================
\node[draw=gray!40, dashed, anchor=west, inner sep=4pt] at (0,0.3) {
  \normalsize
  \begin{tikzpicture}[baseline]
    \node[anchor=west, font=\small] at (0,0) {\textbf{Code CoT:}};

    \fill[bluePurpleStripesDark]  (0.00,-0.73) rectangle (0.25,-0.29);
    \fill[bluePurpleStripesDark]  (0.37,-0.71) rectangle (0.60,-0.31);
    \node[anchor=west, font=\small] at (0.72,-0.51) {MoT};

    \fill[blue!60]    (2.00,-0.73) rectangle (2.25,-0.29);
    \fill[blue!60]    (2.37,-0.71) rectangle (2.60,-0.31);
    \node[anchor=west, font=\small] at (2.72,-0.51) {SCoT};

    \fill[bluePattern4]     (4.00,-0.73) rectangle (4.25,-0.29);
    \fill[bluePattern4]     (4.37,-0.71) rectangle (4.60,-0.31);
    \node[anchor=west, font=\small] at (4.72,-0.51) {PlanSearch};

    \fill[bluePattern2Struct]    (0.00,-1.35) rectangle (0.25,-0.91);
    \fill[bluePattern2Struct]    (0.37,-1.33) rectangle (0.60,-0.93);
    \node[anchor=west, font=\small] at (0.72,-1.13) {CodeChain};

    \fill[bluePattern6]    (2.90,-1.35) rectangle (3.15,-0.91);
    \fill[bluePattern6]    (3.27,-1.33) rectangle (3.50,-0.93);
    \node[anchor=west, font=\small] at (3.62,-1.13) {Self-Planning};
  \end{tikzpicture}
};

% ================= MIDDLE LEGEND =================
\node[draw=gray!40, dashed, anchor=west, inner sep=4pt] at (7.5,0.31) {
  \normalsize
  \begin{tikzpicture}[baseline]
    \node[anchor=west, font=\small] at (0,0) {\textbf{Self-refinement:}};
    \fill[orangePattern7] (0.00,-0.93) rectangle (0.25,-0.49);
    \fill[orangePattern7] (0.37,-0.91) rectangle (0.60,-0.51);
    \node[anchor=west, align=left] at (0.72,-0.71) {Revisiting\\Self-Debugging};
  \end{tikzpicture}
};

% ================= RIGHT LEGEND =================
\node[draw=gray!40, dashed, anchor=west, inner sep=4pt] at (12.0,0.3) {
  \normalsize
  \begin{tikzpicture}[baseline]
    \node[anchor=west, font=\small] at (0,0) {\textbf{Inference Scaling:}};

    \fill[darkGreenLightWest] (0.00,-0.73) rectangle (0.25,-0.29);
    \fill[darkGreenLightWest] (0.37,-0.71) rectangle (0.60,-0.31);
    \node[anchor=west, font=\small] at (0.68,-0.51) {CodeTree};

    \fill[lightGreenDarkEast] (2.70,-0.73) rectangle (2.95,-0.29);
    \fill[lightGreenDarkEast] (3.07,-0.71) rectangle (3.30,-0.31);
    \node[anchor=west, font=\small] at (3.42,-0.51) {ORPs};

    \fill[limeGreenCrossDark] (0.00,-1.35) rectangle (0.25,-0.91);
    \fill[limeGreenCrossDark] (0.37,-1.33) rectangle (0.60,-0.93);
    \node[anchor=west, font=\small] at (0.70,-1.13) {S*};

    \fill[lightGreenSolid] (2.70,-1.35) rectangle (2.95,-0.91);
    \fill[lightGreenSolid] (3.07,-1.33) rectangle (3.30,-0.93);
    \node[anchor=west, font=\small] at (3.42,-1.13) {REx};
  \end{tikzpicture}
};

\end{tikzpicture}

\renewcommand{\thefootnote}{$\S$}
\caption{Performance comparison between Inference Scaling, Code \cotshort and Self-refinement on Code Generation benchmarks of MBPP+, MBPP, APPS, LCB, HumanEval+ (HE+) and HumanEval (HE) with models GPT-4o, GPT-4o mini (4o-mini), GPT-4 and OpenAI O1-mini (o1-mini). Note: Y-axis scales differ across subplots.}
%\protect\footnotemark}
\label{fig:obs4_four_benchmarks_2}
\end{figure}

Inference Scaling (Sec.~\ref{sec:inferencescaling}) involves generating multiple LLM outputs and then selecting the best candidate among them for evaluation.
In Fig.~\ref{fig:obs4_four_benchmarks_2}, we can see that inference scaling outperforms code \cotshort on the code generation benchmarks studied.
ORPS outperforms MoT, SCoT, and Self-Planning on MBPP with GPT-4o mini~\citep{openai_gpt4omini_blog}.
REx with GPT-4~\citep{achiam2023gpt} also claims to achieve the state-of-the-art on APPS, with roughly 70\%, beating CodeChain.
S* also beats PlanSearch on LCB with o1-mini~\citep{jaech2024openai} and GPT-4o mini.
Interesting to note PlanSearch, which incorporates inference scaling techniques as highlighted in Tab.~\ref{tab:llm-hybrid}, outperforms other Code \cotshort methods by a big margin on MBPP+ with GPT-4o mini.
CodeTree also does better than multiple Code \cotshort methods on MBPP+ with GPT-4o mini.

While there is some evidence that inference scaling could outperform self-refinement, as CodeTree outperforms Revisiting Self-Debugging on MBPP+ (80.7 vs.\ 76.5), MBPP (98.7 vs.\ 91.5), and HumanEval (94.5 vs.\ 92.1), Revisiting Self-Debugging performs better on HumanEval+ (87.8 vs.\ 86.0). Since these comparisons are limited to a single inference scaling and self-refinement approach pair, this remains an open question.

\begin{findingsboxold}

\textbf{\dashuline{Observation 4:} Reported results across several studies suggest that approaches  integrating inference scaling can outperform CoT-dominant strategies in some code generation benchmarks.}

\end{findingsboxold}

\subsection{\sweagents}
\label{subsec:comp-agent}
\begin{figure}[htbp]
  \centering
  \begin{tikzpicture}

    % ==========================
    % Global axis style (match Fig. 1)
    % ==========================
    \tikzset{
      every axis/.style={
        ybar=0pt,
        bar width=7pt,
        height=3.6cm,
        grid=major,
        clip=false,
        tick label style={font=\small},
        x tick label style={font=\small, rotate=45, anchor=east, xshift=-3pt},
        ylabel style={font=\large},
        title style={yshift=1pt, font=\bfseries\large}
      }
    }

    % ==========================
    % Agentic colors (new, distinct)
    % ==========================
    \definecolor{redPatternMediumDotsDark}{rgb}{0.9,0.0,0.5} % PairCoder
    \definecolor{redPatternDarkLinesWest}{rgb}{0.6,0.4,0.2} % AgileCoder

    % ==========================
    % Plot 1: HE (gpt-4, DS-C)
    % ==========================
    \begin{axis}[
      name=plot1,
      width=0.26\textwidth,
      ylabel={Pass@1 (\%)},
      title={HE},
      symbolic x coords={gpt-4, DS-C},
      xtick={gpt-4, DS-C},
      enlarge x limits=0.55,
      ymin=65, ymax=100,
      ytick={65,75,85,95},
      bar width=6pt
    ]
      % gpt-4 data
      \addplot[ybar, bar shift=-15pt, redPatternMediumDotsDark]         coordinates {(gpt-4,93.9)}; % PairCoder
      \addplot[ybar, bar shift= -5pt, redPatternDarkLinesWest]         coordinates {(gpt-4,90.9)}; % AgileCoder
      \addplot[ybar, bar shift=  5pt, bluePattern1]           coordinates {(gpt-4,87.8)}; % ClarifyGPT (match Fig. 1)

      % DS-C data
      \addplot[ybar, bar shift=-15pt, redPatternMediumDotsDark]         coordinates {(DS-C,85.4)}; % PairCoder
      \addplot[ybar, bar shift= -5pt, orangePattern3]         coordinates {(DS-C,83.5)}; % μFix (match Fig. 1)
      \addplot[ybar, bar shift=  5pt, orangePattern2]         coordinates {(DS-C,77.4)}; % Self-Debugging (match Fig. 1)
      \addplot[ybar, bar shift= 15pt, fill=teal!60]           coordinates {(DS-C,70.6)}; % UniCoder (new color)
    \end{axis}

    % ==========================
    % Plot 2: MBPP+ (gpt-3.5)
    % ==========================
    \begin{axis}[
      at={(plot1.east)}, anchor=west, xshift=1.0cm,
      name=plot2,
      width=0.16\textwidth,
      title={MBPP+},
      symbolic x coords={gpt-3.5},
      xtick={gpt-3.5},
      enlarge x limits=0.7,
      ymin=65, ymax=80,
      ytick={65,70,75,80}
    ]
      \addplot[ybar, bar shift=-10pt, redPatternMediumDotsDark]         coordinates {(gpt-3.5,77.7)}; % PairCoder
      \addplot[ybar, bar shift=  0pt,  bluePatternwhitelinesleft] coordinates {(gpt-3.5,73.7)}; % CGO (match Fig. 1)
      \addplot[ybar, bar shift= 10pt,  bluePattern6]          coordinates {(gpt-3.5,68.1)}; % Self-Planning (match Fig. 1)
    \end{axis}

    % ==========================
    % Plot 3: MBPP (gpt-3.5, DS-C)
    % ==========================
    \begin{axis}[
      at={(plot2.east)}, anchor=west, xshift=1.0cm,
      name=plot3,
      width=0.26\textwidth,
      title={MBPP},
      symbolic x coords={gpt-3.5, DS-C},
      xtick={gpt-3.5, DS-C},
      enlarge x limits=0.55,
      ymin=45, ymax=85,
      ytick={45,55,65,75,85},
      bar width=6pt
    ]
      % gpt-3.5 data
      \addplot[ybar, bar shift=-15pt, redPatternDarkLinesWest]         coordinates {(gpt-3.5,80.9)}; % AgileCoder
      \addplot[ybar, bar shift= -5pt, redPatternMediumDotsDark]         coordinates {(gpt-3.5,80.6)}; % PairCoder
      \addplot[ybar, bar shift=  5pt, orangePattern2]         coordinates {(gpt-3.5,74.2)}; % Self-Debugging (match Fig. 1)
      \addplot[ybar, bar shift= 15pt, fill=blue!60]           coordinates {(gpt-3.5,47.0)}; % SCoT (match Fig. 1)

      % DS-C data
      \addplot[ybar, bar shift= -5pt, redPatternMediumDotsDark]         coordinates {(DS-C,78.8)}; % PairCoder
      \addplot[ybar, bar shift=  5pt, fill=teal!60]           coordinates {(DS-C,64.3)}; % UniCoder (new color)
    \end{axis}

    % ==========================
    % Plot 4: MBPP-S (gpt-4)
    % ==========================
    \begin{axis}[
      at={(plot3.east)}, anchor=west, xshift=1.0cm,
      name=plot4,
      width=0.16\textwidth,
      title={MBPP-S},
      symbolic x coords={gpt-4},
      xtick={gpt-4},
      enlarge x limits=0.7,
      ymin=75, ymax=95,
      ytick={75,80,85,90,95}
    ]
      \addplot[ybar, bar shift=-5pt, redPatternMediumDotsDark]          coordinates {(gpt-4,91.2)}; % PairCoder
      \addplot[ybar, bar shift= 5pt, bluePattern1]            coordinates {(gpt-4,78.7)}; % ClarifyGPT (match Fig. 1)
    \end{axis}

    % ==========================
    % Centered global x-label
    % ==========================
    \coordinate (midpoint) at ($(plot1.south)!0.5!(plot4.south)$);
    \node[font=\small, anchor=north] at ([yshift=-1.3cm]midpoint) {Models};

  \end{tikzpicture}

\centering
\begin{tikzpicture}
  % --------- \sweagents (left box) ---------
  \node[draw=gray!40, dashed, anchor=west, inner sep=4pt] at (0,0) {
    \normalsize
    \begin{tikzpicture}[baseline]
      \node[anchor=west, font=\small] at (0,0) {\textbf{\sweagents:}};
      \fill[redPatternMediumDotsDark] (0.00,-0.73) rectangle (0.25,-0.29);
      \fill[redPatternMediumDotsDark] (0.37,-0.71) rectangle (0.60,-0.31);
      \node[anchor=west, font=\small] at (0.68,-0.51) {PairCoder};
      \fill[redPatternDarkLinesWest] (0.00,-1.33) rectangle (0.25,-0.89);
      \fill[redPatternDarkLinesWest] (0.37,-1.31) rectangle (0.60,-0.91);
      \node[anchor=west, font=\small] at (0.68,-1.11) {AgileCoder};
    \end{tikzpicture}
  };
  
  % --------- Self-refinement (center box) ---------
  \node[draw=gray!40, dashed, anchor=west, inner sep=2pt] at (3.5,0) {
    \normalsize
    \begin{tikzpicture}[baseline]
      \node[anchor=west, font=\small] at (0,0) {\textbf{Self-refinement:}};
      \fill[orangePattern2] (0.00,-0.73) rectangle (0.25,-0.29);
      \fill[orangePattern2] (0.37,-0.71) rectangle (0.60,-0.31);
      \node[anchor=west, font=\small] at (0.68,-0.51) {Self-Debugging};
      \fill[orangePattern3] (0.00,-1.33) rectangle (0.25,-0.89);
      \fill[orangePattern3] (0.37,-1.31) rectangle (0.60,-0.91);
      \node[anchor=west, font=\small] at (0.68,-1.11) {$\mu$Fix};
    \end{tikzpicture}
  };
      
  % --------- CoT / Code-based (right box) ---------
  \node[draw=gray!40, dashed, anchor=west, inner sep=4pt] at (7.5,0) {
    \normalsize
    \begin{tikzpicture}[baseline]
      \node[anchor=west, font=\small] at (0,0) {\textbf{CoT:}};
      % ClarifyGPT
      \fill[bluePattern1] (0.00,-0.73) rectangle (0.25,-0.29);
      \fill[bluePattern1] (0.37,-0.71) rectangle (0.60,-0.31);
      \node[anchor=west, font=\small] at (0.68,-0.51) {ClarifyGPT};
      % UniCoder
      \fill[teal!60] (3.00,-0.73) rectangle (3.25,-0.29);
      \fill[teal!60] (3.37,-0.71) rectangle (3.60,-0.31);
      \node[anchor=west, font=\small] at (3.68,-0.51) {UniCoder};
      % SCoT
      \fill[blue!60] (5.60,-0.73) rectangle (5.85,-0.29);
      \fill[blue!60] (5.97,-0.71) rectangle (6.20,-0.31);
      \node[anchor=west, font=\small] at (6.28,-0.51) {SCoT};
      % Self-Planning
      \fill[bluePattern6] (0.00,-1.21) rectangle (0.25,-0.77);
      \fill[bluePattern6] (0.37,-1.19) rectangle (0.60,-0.79);
      \node[anchor=west, font=\small] at (0.68,-0.99) {Self-Planning};
      % CGO
      \fill[bluePatternwhitelinesleft] (3.20,-1.21) rectangle (3.45,-0.77);
      \fill[bluePatternwhitelinesleft] (3.57,-1.19) rectangle (3.80,-0.79);
      \node[anchor=west, font=\small] at (3.88,-0.99) {CGO};
    \end{tikzpicture}
  };
\end{tikzpicture}
  \renewcommand{\thefootnote}{$\S$}
  \caption{Comparison of \sweagents-based approaches with self-refinement and code \cotshort-based approaches on code generation benchmarks of HE, MBPP+, MBPP, MBPP-S with models GPT-4, DeepSeek-Coder (DS-C) and GPT-3.5-turbo (GPT-3.5). Note: Y-axis scales differ across subplots.}
  %\protect\footnotemark}
  \label{fig:obs5_agent}
\label{fig:legend}
\end{figure}

\sweagents (Sec.~\ref{sec:agentssec}) succeed by integrating chain-of-thought reasoning, execution-based validation, and sampling into a unified framework--thus leveraging code’s structured syntax, executable semantics, and error feedback all in one.
Fig.~\ref{fig:obs5_agent} shows that agent-based approaches perform better than self-refinement and code \cotshort on the code generation benchmarks studied.

Agentic approaches tend to outperform both CoT and self-refinement methods across the benchmarks and models studied.
With GPT-4, PairCoder and AgileCoder outperform ClarifyGPT on HumanEval, and PairCoder surpasses ClarifyGPT on MBPP-S.
On GPT-3.5, PairCoder beats CGO and Self-Planning on MBPP+, while both PairCoder and AgileCoder surpass SCoT and Self-Debugging on MBPP.
With DeepSeek-Coder, PairCoder outperforms $\mu$Fix, Self-Debugging, and UniCoder on HumanEval, and also UniCoder on MBPP.

\begin{findingsboxold}
\textbf{\dashuline{Observation 5:} Reported results across several studies suggest that orchestrating multiple reasoning techniques through agentic scaffolding can outperform single-strategy approaches in some code generation benchmarks.}
\end{findingsboxold}

\subsection{\sweagents with Inference Scaling}
\label{subsec:comp-agent-inf}
\input{sections/results-obs-6}

\sweagents generally have some elements of \cotshort and self-refinement as part of the scaffolding.
Techniques that combine inference scaling with \sweagents have also been proposed.
Fig.~\ref{fig:obs6_agent_inf_scaling} shows that such techniques which combine inference scaling with \sweagents perform better than \sweagents, self-refinement and inference scaling based approaches on the code generation and issue resolution benchmarks studied.

Across multiple benchmarks shown, inference scaling integrated with agentic approaches demonstrates strong performance.
With GPT-4o mini, CodeTree outperforms MoT, SCoT, Self-Planning, and PlanSearch on HumanEval+ and MBPP+, and surpasses ORPS on MBPP.
These gains extend to GPT-4o, where CodeTree similarly outperforms these CoT-based strategies.
SWE-Search, which combines inference scaling within an agentic framework, dominates the leaderboard on SWE-Bench Lite.

\begin{findingsboxold}
\textbf{\dashuline{Observation 6:} Reported results across several studies suggest that integrating search-based inference scaling within agentic frameworks can achieve state-of-the-art performance on some code generation and issue resolution benchmarks.}

\end{findingsboxold}

\section{Gaps and Future Directions}
\label{sec:gaps}

Sec.~\ref{sec:codereason} gave an overview of various inference time code reasoning techniques and in Sec.~\ref{sec:comparison} we saw how this impacted code tasks, primarily code generation but also issue resolution. 
Despite the vast array of work in this domain covered by our survey, there still appear to be gaps which can be explored by the research community.
In this section we try to motivate why some of these gaps must be explored and identify a set of Call To Actions (CTA) as specific research directions.

\subsection{Structure-based \cotshort understanding and applications}

Section \ref{sec:cot} surveys works that formulate \cotshort in plan-based, structure-based, and modular methods.
As per Sec.~\ref{subsec:comp-plan-struct-cot} and Sec.~\ref{subsec:comp-mod-cot}, structure-based methods outperform plan-based methods and modular methods outperform other Code \cotshort prompting methods on the code generation benchmarks studied.
To further examine this result, first we must understand why chain-of-thought (CoT) prompting helps over direct prompting. 
One hypothesis from \citet{prystawski2023think}'s work provides theoretical and experimental evidence that intermediate steps (i.e., chain-of-thought reasoning) reduce bias in transformers. 
They show that when training data has local structure (as textual data does), intermediate variables (CoT) can outperform direct prediction (no CoT). 
This suggests that \textbf{CoT reasoning helps most when a model is asked to make inferences about concepts that do not co-occur in the training data, but which can be chained together through topics that do.}
Section \ref{sec:cot} surveys works that formulate CoT in plan-based, structure-based, and modular arrangements. 
The results suggest that structure-aware strategies can outperform plan-based approaches, and modular formats can outperform structure-aware ones.
We posit that because code has properties of \textit{structured syntax}, the primitive structures invoked within the CoT are highly local in the training data. 
Structures (such as idents, branches, loop invariants, functions, etc) are seen countless times in the training corpus. 
The model’s ability to estimate probabilities (and thus its ability to arrive at a correct solution) become sharper by eliciting these localities. 
Modular structures may push this same principle further.
Based on this thesis we present our first call-to-action below.

\begin{gapsndirections}
\textbf{CTA-1$^{\star}$: Deeper research on code structure and its impact on code tasks will help confirm hypotheses such as: (1) CoT helps by chaining concepts through intermediate topics; (2) code's structured syntax makes primitives (branches, loops, functions) highly local in training data, sharpening probability estimates; (3) modular structures amplify this effect. Validating these may lead to further improvements.}
\end{gapsndirections}
\begingroup\renewcommand{\thefootnote}{}\footnotetext{$^{\star}$This CTA stems from hypotheses we formulate based on general trends observed in our analysis, and should be treated as interpretive hypotheses rather than definitive conclusions.}\endgroup

Code structure improving \cotshort result should encourage research in exploiting other code properties.
Current Code \cotshort approaches consider a structured representation of the code and also the reasoning and generation of modular solutions. 
Code \cotshort approaches should also consider other mechanisms for generating code such as logic programming or event-driven programming (when it applies). 
Changing the programming paradigm will modify the way the LLM reasons about the solution and might help generate code more aligned to a given specification.

\begin{gapsndirections}
\textbf{CTA-2: Future work should investigate Code \cotshort approaches that consider other programming paradigms and software design principles for code tasks.}
\end{gapsndirections}

Inference scaling approaches (Sec.~\ref{sec:inferencescaling}) sample multiple solutions from \llms, selecting the best solutions using different strategies. 
Table~\ref{tab:llm-hybrid} shows many approaches use a combination of inference scaling, self-refinement and \sweagents.
However there is little to no exploration of inference scaling and Code \cotshort approaches.
The only technique that explores this is PlanSearch which does better than other code \cotshort techniques on MBPP+ with GPT-4o mini (Fig.~\ref{fig:obs4_four_benchmarks_2}).
Works on inference scaling can also leverage other strategies from CoT such as structure of code and modular code generation that have shown to increase the correctness of the generated solutions, compared to plan-based \cotshort approaches.

\begin{gapsndirections}
\textbf{CTA-3: Future work should explore combining inference scaling with code-structure based \cotshort approaches.}
\end{gapsndirections}

\subsection{Self-refinement}
Self-refinement (Sec.~\ref{sec:execution}) involves executing the model generated code and then feeding the execution output back to the model for iterative improvement.
This technique outperforms code \cotshort prompting techniques (Sec.~\ref{subsec:comp-refine}) on the code generation benchmarks studied.
We posit that execution may help because executing code can be used as a deterministic check. 
Any chain that violates the check can be discarded. 
Hence, bad chains are filtered out, so variance may collapse faster. 
However, even with reduced variance, LLMs can still exhibit issues, such as model rigidity. 
Because code is inherently deterministic (i.e. under certain assumptions, a given input consistently produces the same output), this can lead models to develop rigid generation patterns in training. 
For example, \citet{twist2025llms} show that LLMs exhibit a strong bias towards certain programming languages, like Python; \citet{liu2025code} document the pervasiveness of repetition in LLM-based code generation, where models often reproduce patterns observed in training. 
\citet{zhang2025unveiling} demonstrate that LLMs favor certain libraries and APIs by default, reflecting the distribution of their training corpora. 
Furthermore, \citet{pan2025code} show that LLMs struggle to generalize to the architectural design principles of given projects, leading to the generation of conflicting code. 
This phenomenon compels the integration of search in order to explore diverse trajectories, which may explain the recent success of inference scaling techniques.

\begin{gapsndirections}
\textbf{CTA-4$^{\star}$: Future work should investigate whether execution serves as a deterministic check that mitigates model rigidity by filtering bad chains and reducing variance. Understanding this mechanism could inform better training data construction and guide the development of more effective search-based inference scaling techniques.}
\end{gapsndirections}

So far self-refinement approaches have considered execution and unit tests as feedback to self-refine the generated code. 
To increase the robustness of the generated solutions, these approaches could also reason about other characteristics of the generated code such as efficiency, security, and usability, among others. 
When generating modularized code, besides reasoning about the functionality of the solution, approaches could also consider reasoning about other important aspects such as maintainability and scalability of the proposed code solution. Mechanisms to validate these properties exist in the generated code can also be used as feedback.
Besides unit tests, feedback from integration tests should be considered to further evaluate code modules that were tested as separate units (e.g., with unit testing) work correctly also when interacting with each other.

\begin{gapsndirections}
\textbf{CTA-5: Future work should extend self-refinement approaches beyond unit tests to incorporate software quality metrics (e.g., efficiency, security, maintainability, scalability) and integration testing as feedback mechanisms.}
\end{gapsndirections}

\subsection{\sweagents Improvements and Benchmarks}

\sweagents (Sec.~\ref{sec:agentssec}) orchestrate LLMs, tools and execution environments resulting in state-of-the-art performance (Sec.~\ref{subsec:comp-agent} and Sec.~\ref{subsec:comp-agent-inf}) on code generation and relatively more challenging issue resolution task.
Many approaches covered in our survey also involve multi-agent solutions.
Although enabling good results, such agent and multi-agent approaches also increase the degree of complexity of AI systems.
Such systems will not be immune to errors, which can propagate through the system and stifle its performance.
It is possible that many errors are common across LLMs, agents and tasks.
A better understanding and development of techniques to overcome these errors can result in improved performance of multiple systems on multiple tasks.

\begin{gapsndirections}
\textbf{CTA-6: Incorporating a structured way to analyze errors in agents can help them overcome repetitions of the same error, find alternative methods to solve a given step, avoiding waste of resources and increasing their ability to find a solution.}
\end{gapsndirections}

Overall, patterns of errors made by agents have recently become a topic of growing interest. 
For instance, ~\citet{cuadron2025danger} examined error patterns in extended internal reasoning chains (i.e., overthinking) in reasoning models, while frameworks such as TRAIL~\citep{deshpande2025trail} have focused on categorizing the errors exhibited by agents, and judges the ability of agents to label existing errors in categories.
However, these efforts have largely remained at the level of descriptive taxonomies of existing errors.
What remains missing is a concrete error benchmark that evaluates agents’ ability to recover from their own mistakes or fix injected mistakes.
Such a benchmark could be constructed by injecting error instances or categories into a boilerplate agent performing clean tasks, with evaluation metrics including both the raw competence ceiling and the category-wise error recovery rate, alongside latency costs (i.e., the number of steps or seconds required for the agent to recover).

\begin{gapsndirections}
\textbf{CTA-7: Future work should develop error recovery benchmarks that evaluate agents' ability to recover from both self-induced and injected mistakes, measuring category-wise recovery rates and latency costs.}
\end{gapsndirections}

Beyond agents' errors, another critical frontier is benchmarking tool-use, particularly for SWE agents. 
While works such as GAIA ~\citep{mialon2023gaia} provide benchmarks for general tool use in generic AI agents, there remains a gap in specific tool-use benchmarks tailored to SWE agents which targets the challenge of system orchestration, specifically.
Furthermore, recent work built taxonomies of agentic decision-making pathways~\citep{ceka2026understanding} showing that agents continue to struggle with certain complexities, performing especially poorly on more advanced SWE-bench problems. They also highlight the presence of “hot nodes”(i.e., critical components in these pathways) that strongly influence whether an agent can successfully generate patches. Based on the aforementioned work, two additional benchmark directions could be: (i) benchmarks that explicitly target complex issues that agents consistently fail on, and 
(ii) benchmarks, and approaches, that emphasize the specific components of agent decision-making that are most predictive of successful outcomes. 
For example, TDD-Bench (Otter) ~\citep{ahmed2025otter} isolates reproduction test case generation as one such critical sub-component of SWE Agents.

\begin{gapsndirections}
\textbf{CTA-8: Despite progress on agents' benchmark development, there remains a gap in specific tool-use benchmarks tailored to SWE agents which targets the challenge of system orchestration; benchmarks that explicitly target complex issues that agents consistently fail on, and 
benchmarks, and techniques, that emphasize the specific components of agent decision-making that are most predictive of successful outcomes. These are areas where future research should focus on to advance the understanding and performance of Agentic solutions.}
\end{gapsndirections}

\section{Limitations}
\label{sec:limitations}
Code reasoning for software engineering is a relatively new and rapidly evolving area. As a result, our survey emphasizes test-time reasoning techniques for which we observed a substantial body of work on code-centric tasks. While we aimed for broad coverage, our taxonomy may not include emerging techniques that have limited adoption or evaluation in the code/\swe{} literature to date.

Our paper selection process (Sec.~\ref{app:survey-method}) relies on keyword- and citation-based search, which cannot guarantee complete recall. Consequently, some relevant papers may be missing. To support transparency and community feedback, we plan to maintain an up-to-date, categorized list of surveyed papers at \url{https://github.com/AI4Code-IBM-Columbia/code-reasoning-for-swe-tasks}.

Many systems combine multiple reasoning components (e.g., prompting, sampling/search, execution feedback, and refinement). Our taxonomy assigns each paper to a dominant technique based on the primary mechanism emphasized by the method.
The primary mechanism can be thought of as core contribution, method or novelty in the publication.
To provide a more nuanced view and mitigate potential misrepresentation, we explicitly highlight hybrid approaches in Tab.~\ref{tab:llm-hybrid}.

This survey focuses on published literature and therefore excludes some highly capable but closed systems. Although such systems may employ code-specific reasoning, their methods are often unclear and their results difficult to verify. In addition, we focus primarily on the effectiveness of different techniques. Cost is an important consideration, especially for test-time compute, but most surveyed papers do not report cost-related metrics. When such metrics are reported, they are not discussed consistently across papers, making comparison difficult.

Finally, due to space constraints and the complexity of many approaches, our descriptions focus on the most salient and broadly applicable ideas rather than all implementation details of every system.
\subsection{Threats to Validity of Cross-Paper Comparisons}
\label{sec:threats}

The comparisons synthesized in this survey (Sec.~\ref{sec:comparison}) rely on reported results from different papers, which vary in model choice, prompting strategy, inference budgets, benchmark versions (e.g., MBPP vs.\ MBPP-S vs.\ MBPP-ET), and evaluation harnesses. These factors introduce confounders that make direct attribution of performance differences to specific techniques difficult. Our comparative analysis does not control for differences in prompts, compute budgets, tool access, or evaluation harnesses, which may affect comparability. To mitigate some of these threats, we condition all comparisons on the same underlying base model and look for trends that hold consistently across multiple benchmarks, including different splits of the same benchmark (e.g., MBPP, MBPP-S, MBPP-ET) as well as entirely different benchmarks (e.g., HumanEval, APPS). A trend where one technique outperforms another across multiple controlled splits and independent benchmarks is stronger evidence than a single benchmark result. Where trends are supported by many such data points across these controlled same-model comparisons, we present suggestive claims rather than definitive conclusions. Since we try to control for base model across aggregated results, we are limited to the statistics reported by each approach's original paper. Most papers do not report confidence intervals or variance for their results. To help assess the strength of each comparison, we report the margin of difference between approaches where applicable. To facilitate this, we provide a comparison table in the appendix (Tab.~\ref{tab:margins}) that makes it easy to see which specific methods and categories of methods show trends across multiple splits of benchmarks and distinct benchmarks, along with their margins. Future work should aim to verify these trends in a controlled experimental setup with identical conditions across techniques.

\section{Conclusion}
\label{sec:future}

Test or inference-time reasoning techniques have driven major recent gains in AI and have been rapidly adopted in software engineering (SWE). 
In this survey, we focus on code reasoning for SWE tasks, providing a systematic overview of this emerging area.
We begin by proposing a taxonomy of reasoning techniques, including SWE agents (Sec.~\ref{sec:codereason}), which defines the scope of code reasoning and clarifies the common design patterns that underlie diverse methods.

Our analysis reveals that many state-of-the-art systems combine multiple reasoning strategies, a trend we summarize in Table~\ref{tab:llm-hybrid}.
Despite this diversity of approaches, we observed that evaluation is concentrated on a small set of code-generation benchmarks, leaving other critical SWE tasks comparatively underexplored.
To address this imbalance, we catalog a broader set of tasks and benchmarks for code reasoning (Sec.~\ref{sec:benchmetrics}).

To assess the impact of existing techniques, we perform a comparative analysis over commonly used models and benchmarks (Sec.~\ref{sec:comparison}), highlighting relative strengths of different code reasoning methods.
These observations motivate several future research directions, including the design of more realistic and comprehensive SWE benchmarks and exploration of reasoning and agent architectures that better capture the constraints and real-world engineering workflows, while taking advantage of features that formal languages afford (Sec.~\ref{sec:gaps}).
We hope this survey serves as a foundation for the growing field of code reasoning for SWE and supports the community in building more capable, reliable, and practical AI tools.

\subsection*{Acknowledgments}

This work was supported in part by the National Science
Foundation Graduate Research Fellowship Program (NSF
GRFP), IBM, NSF CNS-2247370, NSF CCF-2313055, DARPA/NIWC
Pacific N66001-21-C-4018. Any opinions, findings, conclusions or recommendations expressed herein are those of
the authors and do not necessarily reflect those of the US
Government, NSF, IBM, or DARPA.

\bibliography{main}

@inproceedings{chi-etal-2025-visualcoder,
    title = "{V}isual{C}oder: Guiding Large Language Models in Code Execution with Fine-grained Multimodal Chain-of-Thought Reasoning",
    author = "Chi, Cuong Le  and
      Hoang, Chau Truong Vinh  and
      Huy, Phan Nhat  and
      Le, Dung D.  and
      Nguyen, Tien N  and
      Bui, Nghi D. Q.",
    editor = "Chiruzzo, Luis  and
      Ritter, Alan  and
      Wang, Lu",
    booktitle = "Findings of the Association for Computational Linguistics: NAACL 2025",
    month = apr,
    year = "2025",
    address = "Albuquerque, New Mexico",
    publisher = "Association for Computational Linguistics"
}

@misc{yang2024sweagent,
      title={SWE-agent: Agent-Computer Interfaces Enable Automated Software Engineering}, 
      author={John Yang and Carlos E. Jimenez and Alexander Wettig and Kilian Lieret and Shunyu Yao and Karthik Narasimhan and Ofir Press},
      year={2024},
      eprint={2405.15793},
      archivePrefix={arXiv},
      primaryClass={cs.SE},
      url={https://arxiv.org/abs/2405.15793}, 
}

@article{sun2024survey,
  title={A survey of neural code intelligence: Paradigms, advances and beyond},
  author={Sun, Qiushi and Chen, Zhirui and Xu, Fangzhi and Cheng, Kanzhi and Ma, Chang and Yin, Zhangyue and Wang, Jianing and Han, Chengcheng and Zhu, Renyu and Yuan, Shuai and others},
  journal={arXiv preprint arXiv:2403.14734},
  year={2024}
}

@article{yang2025code,
  title={Code to Think, Think to Code: A Survey on Code-Enhanced Reasoning and Reasoning-Driven Code Intelligence in LLMs},
  author={Yang, Dayu and Liu, Tianyang and Zhang, Daoan and Simoulin, Antoine and Liu, Xiaoyi and Cao, Yuwei and Teng, Zhaopu and Qian, Xin and Yang, Grey and Luo, Jiebo and others},
  journal={arXiv preprint arXiv:2502.19411},
  year={2025}
}

@article{wang2024planning,
  title={Planning in natural language improves llm search for code generation},
  author={Wang, Evan and Cassano, Federico and Wu, Catherine and Bai, Yunfeng and Song, Will and Nath, Vaskar and Han, Ziwen and Hendryx, Sean and Yue, Summer and Zhang, Hugh},
  journal={arXiv preprint arXiv:2409.03733},
  year={2024}
}

@article{jiang2024self,
  title={Self-planning code generation with large language models},
  author={Jiang, Xue and Dong, Yihong and Wang, Lecheng and Fang, Zheng and Shang, Qiwei and Li, Ge and Jin, Zhi and Jiao, Wenpin},
  journal={ACM Transactions on Software Engineering and Methodology},
  volume={33},
  number={7},
  pages={1--30},
  year={2024},
  publisher={ACM New York, NY}
}

@article{mu2023clarifygpt,
  title={Clarifygpt: Empowering llm-based code generation with intention clarification},
  author={Mu, Fangwen and Shi, Lin and Wang, Song and Yu, Zhuohao and Zhang, Binquan and Wang, Chenxue and Liu, Shichao and Wang, Qing},
  journal={arXiv preprint arXiv:2310.10996},
  year={2023}
}

@article{yang2024chain,
  title={Chain-of-thought in neural code generation: From and for lightweight language models},
  author={Yang, Guang and Zhou, Yu and Chen, Xiang and Zhang, Xiangyu and Zhuo, Terry Yue and Chen, Taolue},
  journal={IEEE Transactions on Software Engineering},
  year={2024},
  publisher={IEEE}
}

@article{li2025structured,
  title={Structured chain-of-thought prompting for code generation},
  author={Li, Jia and Li, Ge and Li, Yongmin and Jin, Zhi},
  journal={ACM Transactions on Software Engineering and Methodology},
  volume={34},
  number={2},
  pages={1--23},
  year={2025},
  publisher={ACM New York, NY}
}

@article{le2023codechain,
  title={Codechain: Towards modular code generation through chain of self-revisions with representative sub-modules},
  author={Le, Hung and Chen, Hailin and Saha, Amrita and Gokul, Akash and Sahoo, Doyen and Joty, Shafiq},
  journal={arXiv preprint arXiv:2310.08992},
  year={2023}
}

@misc{zheng2023chaincoder,
      title={Outline, Then Details: Syntactically Guided Coarse-To-Fine Code Generation}, 
      author={Wenqing Zheng and S P Sharan and Ajay Kumar Jaiswal and Kevin Wang and Yihan Xi and Dejia Xu and Zhangyang Wang},
      year={2023},
      eprint={2305.00909},
      archivePrefix={arXiv},
      primaryClass={cs.PL},
      url={https://arxiv.org/abs/2305.00909}, 
}

@article{yeo2025chain,
  title={Chain of Grounded Objectives: Bridging Process and Goal-oriented Prompting for Code Generation},
  author={Yeo, Sangyeop and Hwang, Seung-won and Ma, Yu-Seung},
  journal={arXiv preprint arXiv:2501.13978},
  year={2025}
}

@inproceedings{
chen2023teaching,
title={Teaching Large Language Models to Self-Debug},
author={Xinyun Chen and Maxwell Lin and Nathanael Sch{\"a}rli and Denny Zhou},
booktitle={The Twelfth International Conference on Learning Representations},
year={2024},
url={https://openreview.net/forum?id=KuPixIqPiq}
}

@article{ding2024semcoder,
  title={Semcoder: Training code language models with comprehensive semantics reasoning},
  author={Ding, Yangruibo and Peng, Jinjun and Min, Marcus and Kaiser, Gail and Yang, Junfeng and Ray, Baishakhi},
  journal={Advances in Neural Information Processing Systems},
  volume={37},
  pages={60275--60308},
  year={2024}
}

@article{huang2023codecot,
  title={Codecot: Tackling code syntax errors in cot reasoning for code generation},
  author={Huang, Dong and Bu, Qingwen and Qing, Yuhao and Cui, Heming},
  journal={arXiv preprint arXiv:2308.08784},
  year={2023}
}

@article{ding2024cycle,
  title={Cycle: Learning to self-refine the code generation},
  author={Ding, Yangruibo and Min, Marcus J and Kaiser, Gail and Ray, Baishakhi},
  journal={Proceedings of the ACM on Programming Languages},
  volume={8},
  number={OOPSLA1},
  pages={392--418},
  year={2024},
  publisher={ACM New York, NY, USA}
}

@article{chen2025revisit,
  title={Revisit Self-Debugging with Self-Generated Tests for Code Generation},
  author={Chen, Xiancai and Tao, Zhengwei and Zhang, Kechi and Zhou, Changzhi and Gu, Wanli and He, Yuanpeng and Zhang, Mengdi and Cai, Xunliang and Zhao, Haiyan and Jin, Zhi},
  journal={arXiv preprint arXiv:2501.12793},
  year={2025}
}

@article{li2022competition,
  title={Competition-level code generation with alphacode},
  author={Li, Yujia and Choi, David and Chung, Junyoung and Kushman, Nate and Schrittwieser, Julian and Leblond, R{\'e}mi and Eccles, Tom and Keeling, James and Gimeno, Felix and Dal Lago, Agustin and others},
  journal={Science},
  volume={378},
  number={6624},
  pages={1092--1097},
  year={2022},
  publisher={American Association for the Advancement of Science}
}

@article{yao2023tree,
  title={Tree of thoughts: Deliberate problem solving with large language models},
  author={Yao, Shunyu and Yu, Dian and Zhao, Jeffrey and Shafran, Izhak and Griffiths, Tom and Cao, Yuan and Narasimhan, Karthik},
  journal={Advances in neural information processing systems},
  volume={36},
  pages={11809--11822},
  year={2023}
}

@article{ridnik2024code,
  title={Code generation with alphacodium: From prompt engineering to flow engineering},
  author={Ridnik, Tal and Kredo, Dedy and Friedman, Itamar},
  journal={arXiv preprint arXiv:2401.08500},
  year={2024}
}

@article{yu2024outcome,
  title={Outcome-Refining Process Supervision for Code Generation},
  author={Yu, Zhuohao and Gu, Weizheng and Wang, Yidong and Zeng, Zhengran and Wang, Jindong and Ye, Wei and Zhang, Shikun},
  journal={arXiv preprint arXiv:2412.15118},
  year={2024}
}

@article{sun2024unicoder,
  title={Unicoder: Scaling code large language model via universal code},
  author={Sun, Tao and Chai, Linzheng and Yang, Jian and Yin, Yuwei and Guo, Hongcheng and Liu, Jiaheng and Wang, Bing and Yang, Liqun and Li, Zhoujun},
  journal={arXiv preprint arXiv:2406.16441},
  year={2024}
}

@inproceedings{Ni-LEVER-icml-2023,
author = {Ni, Ansong and Iyer, Srini and Radev, Dragomir and Stoyanov, Ves and Yih, Wen-tau and Wang, Sida I. and Lin, Xi Victoria},
title = {LEVER: learning to verify language-to-code generation with execution},
year = {2023},
publisher = {JMLR.org},
booktitle = {Proceedings of the 40th International Conference on Machine Learning},
articleno = {1086},
series = {ICML'23}
}

@INPROCEEDINGS{junkai-runtime-behavior-reasoning-icse2025,
author = { Chen, Junkai and Pan, Zhiyuan and Hu, Xing and Li, Zhenhao and Li, Ge and Xia, Xin },
booktitle = { 2025 IEEE/ACM 47th International Conference on Software Engineering (ICSE) },
title = {{ Reasoning Runtime Behavior of a Program with LLM: How Far Are We? }},
year = {2025},
volume = {},
ISSN = {1558-1225},
pages = {140-152},
doi = {10.1109/ICSE55347.2025.00012},
url = {https://doi.ieeecomputersociety.org/10.1109/ICSE55347.2025.00012},
publisher = {IEEE Computer Society},
address = {Los Alamitos, CA, USA},
month =May}

@INPROCEEDINGS{liu2025toolindepthanalysiscode,
      title={A Tool for In-depth Analysis of Code Execution Reasoning of Large Language Models}, 
      author={Changshu Liu and Reyhaneh Jabbarvand},
      primaryClass={cs.SE},
      url={https://arxiv.org/abs/2501.18482}, 
     booktitle = { Companion Proceedings of the 33rd ACM International Conference on the Foundations of Software Engineering },
    year = {2025},
    publisher = {Association for Computing Machinery},
    address = {New York, NY, USA},
    series = {FSE 2025}
}

@misc{jiang2025ledextrainingllmsbetter,
      title={LeDex: Training LLMs to Better Self-Debug and Explain Code}, 
      author={Nan Jiang and Xiaopeng Li and Shiqi Wang and Qiang Zhou and Soneya Binta Hossain and Baishakhi Ray and Varun Kumar and Xiaofei Ma and Anoop Deoras},
      year={2025},
      eprint={2405.18649},
      archivePrefix={arXiv},
      primaryClass={cs.CL},
      url={https://arxiv.org/abs/2405.18649}, 
}

@article{dong2022survey,
  title={A survey on in-context learning},
  author={Dong, Qingxiu and Li, Lei and Dai, Damai and Zheng, Ce and Ma, Jingyuan and Li, Rui and Xia, Heming and Xu, Jingjing and Wu, Zhiyong and Liu, Tianyu and others},
  journal={arXiv preprint arXiv:2301.00234},
  year={2022}
}

@article{qiao2022reasoning,
  title={Reasoning with language model prompting: A survey},
  author={Qiao, Shuofei and Ou, Yixin and Zhang, Ningyu and Chen, Xiang and Yao, Yunzhi and Deng, Shumin and Tan, Chuanqi and Huang, Fei and Chen, Huajun},
  journal={arXiv preprint arXiv:2212.09597},
  year={2022}
}

@article{zan2022large,
  title={Large language models meet nl2code: A survey},
  author={Zan, Daoguang and Chen, Bei and Zhang, Fengji and Lu, Dianjie and Wu, Bingchao and Guan, Bei and Wang, Yongji and Lou, Jian-Guang},
  journal={arXiv preprint arXiv:2212.09420},
  year={2022}
}

@article{huang2022towards,
  title={Towards reasoning in large language models: A survey},
  author={Huang, Jie and Chang, Kevin Chen-Chuan},
  journal={arXiv preprint arXiv:2212.10403},
  year={2022}
}

@article{chu2023navigate,
  title={Navigate through enigmatic labyrinth a survey of chain of thought reasoning: Advances, frontiers and future},
  author={Chu, Zheng and Chen, Jingchang and Chen, Qianglong and Yu, Weijiang and He, Tao and Wang, Haotian and Peng, Weihua and Liu, Ming and Qin, Bing and Liu, Ting},
  journal={arXiv preprint arXiv:2309.15402},
  year={2023}
}

@article{xu2025towards,
  title={Towards Large Reasoning Models: A Survey of Reinforced Reasoning with Large Language Models},
  author={Xu, Fengli and Hao, Qianyue and Zong, Zefang and Wang, Jingwei and Zhang, Yunke and Wang, Jingyi and Lan, Xiaochong and Gong, Jiahui and Ouyang, Tianjian and Meng, Fanjin and others},
  journal={arXiv preprint arXiv:2501.09686},
  year={2025}
}

@article{plaat2024reasoning,
  title={Reasoning with large language models, a survey},
  author={Plaat, Aske and Wong, Annie and Verberne, Suzan and Broekens, Joost and van Stein, Niki and Back, Thomas},
  journal={arXiv preprint arXiv:2407.11511},
  year={2024}
}

@article{wang2024enhancing,
  title={Enhancing Code LLMs with Reinforcement Learning in Code Generation},
  author={Wang, Junqiao and Zhang, Zeng and He, Yangfan and Song, Yuyang and Shi, Tianyu and Li, Yuchen and Xu, Hengyuan and Wu, Kunyu and Qian, Guangwu and Chen, Qiuwu and others},
  journal={arXiv preprint arXiv:2412.20367},
  year={2024}
}

@article{chen2024survey,
  title={A survey on evaluating large language models in code generation tasks},
  author={Chen, Liguo and Guo, Qi and Jia, Hongrui and Zeng, Zhengran and Wang, Xin and Xu, Yijiang and Wu, Jian and Wang, Yidong and Gao, Qing and Wang, Jindong and others},
  journal={arXiv preprint arXiv:2408.16498},
  year={2024}
}

@article{huynh2025large,
  title={Large Language Models for Code Generation: A Comprehensive Survey of Challenges, Techniques, Evaluation, and Applications},
  author={Huynh, Nam and Lin, Beiyu},
  journal={arXiv preprint arXiv:2503.01245},
  year={2025}
}

@article{jiang2024survey,
  title={A survey on large language models for code generation},
  author={Jiang, Juyong and Wang, Fan and Shen, Jiasi and Kim, Sungju and Kim, Sunghun},
  journal={arXiv preprint arXiv:2406.00515},
  year={2024}
}

@article{yehudai2025survey,
  title={Survey on Evaluation of LLM-based Agents},
  author={Yehudai, Asaf and Eden, Lilach and Li, Alan and Uziel, Guy and Zhao, Yilun and Bar-Haim, Roy and Cohan, Arman and Shmueli-Scheuer, Michal},
  journal={arXiv preprint arXiv:2503.16416},
  year={2025}
}

@article{wei2022chain,
  title={Chain-of-thought prompting elicits reasoning in large language models},
  author={Wei, Jason and Wang, Xuezhi and Schuurmans, Dale and Bosma, Maarten and Xia, Fei and Chi, Ed and Le, Quoc V and Zhou, Denny and others},
  journal={Advances in neural information processing systems},
  volume={35},
  pages={24824--24837},
  year={2022}
}

@article{chen2022program,
  title={Program of thoughts prompting: Disentangling computation from reasoning for numerical reasoning tasks},
  author={Chen, Wenhu and Ma, Xueguang and Wang, Xinyi and Cohen, William W},
  journal={arXiv preprint arXiv:2211.12588},
  year={2022}
}

@article{pan2025modularization,
  title={Modularization is Better: Effective Code Generation with Modular Prompting},
  author={Pan, Ruwei and Zhang, Hongyu},
  journal={arXiv preprint arXiv:2503.12483},
  year={2025}
}

@article{jin2025mscot,
  title={MSCoT: Structured Chain-of-Thought Generation for Multiple Programming Languages},
  author={Jin, Naizhu and Li, Zhong and Zhang, Tian and Zeng, Qingkai},
  journal={arXiv preprint arXiv:2504.10178},
  year={2025}
}

@article{humaneval2021,
  title={Evaluating large language models trained on code},
  author={Chen, Mark and Tworek, Jerry and Jun, Heewoo and Yuan, Qiming and Pinto, Henrique Ponde De Oliveira and Kaplan, Jared and Edwards, Harri and Burda, Yuri and Joseph, Nicholas and Brockman, Greg and others},
  journal={arXiv preprint arXiv:2107.03374},
  year={2021}
}

@misc{apps2021,
      title={Measuring Coding Challenge Competence With APPS}, 
      author={Dan Hendrycks and Steven Basart and Saurav Kadavath and Mantas Mazeika and Akul Arora and Ethan Guo and Collin Burns and Samir Puranik and Horace He and Dawn Song and Jacob Steinhardt},
      year={2021},
      eprint={2105.09938},
      archivePrefix={arXiv},
      primaryClass={cs.SE},
      url={https://arxiv.org/abs/2105.09938}, 
}

@article{tang2024code,
  title={Code repair with llms gives an exploration-exploitation tradeoff},
  author={Tang, Hao and Hu, Keya and Zhou, Jin and Zhong, Si Cheng and Zheng, Wei-Long and Si, Xujie and Ellis, Kevin},
  journal={Advances in Neural Information Processing Systems},
  volume={37},
  pages={117954--117996},
  year={2024}
}

@article{li2025s,
  title={S*: Test time scaling for code generation},
  author={Li, Dacheng and Cao, Shiyi and Cao, Chengkun and Li, Xiuyu and Tan, Shangyin and Keutzer, Kurt and Xing, Jiarong and Gonzalez, Joseph E and Stoica, Ion},
  journal={arXiv preprint arXiv:2502.14382},
  year={2025}
}

@article{long2023large,
  title={Large language model guided tree-of-thought},
  author={Long, Jieyi},
  journal={arXiv preprint arXiv:2305.08291},
  year={2023}
}

@article{ouedraogo2024large,
  title={Large-scale, Independent and Comprehensive study of the power of LLMs for test case generation},
  author={Ou{\'e}draogo, Wendk{\^u}uni C and Kabor{\'e}, Kader and Tian, Haoye and Song, Yewei and Koyuncu, Anil and Klein, Jacques and Lo, David and Bissyand{\'e}, Tegawend{\'e} F},
  journal={arXiv preprint arXiv:2407.00225},
  year={2024}
}

@article{li2024codetree,
  title={Codetree: Agent-guided tree search for code generation with large language models},
  author={Li, Jierui and Le, Hung and Zhou, Yingbo and Xiong, Caiming and Savarese, Silvio and Sahoo, Doyen},
  journal={arXiv preprint arXiv:2411.04329},
  year={2024}
}

@article{ni2024tree,
  title={Tree-of-Code: A Hybrid Approach for Robust Complex Task Planning and Execution},
  author={Ni, Ziyi and Li, Yifan and Dong, Daxiang},
  journal={arXiv preprint arXiv:2412.14212},
  year={2024}
}

@INPROCEEDINGS {tian-fixing-spec-cg-2025,
author = { Tian, Zhao and Chen, Junjie and Zhang, Xiangyu },
booktitle = { 2025 IEEE/ACM 47th International Conference on Software Engineering (ICSE) },
title = {{ Fixing Large Language Models' Specification Misunderstanding for Better Code Generation }},
year = {2025},
volume = {},
ISSN = {1558-1225},
pages = {645-645},
url = {https://doi.ieeecomputersociety.org/10.1109/ICSE55347.2025.00108},
publisher = {IEEE Computer Society},
address = {Los Alamitos, CA, USA},
month =May}

@article{brown2020language,
  title={Language models are few-shot learners},
  author={Brown, Tom and Mann, Benjamin and Ryder, Nick and Subbiah, Melanie and Kaplan, Jared D and Dhariwal, Prafulla and Neelakantan, Arvind and Shyam, Pranav and Sastry, Girish and Askell, Amanda and others},
  journal={Advances in neural information processing systems},
  volume={33},
  pages={1877--1901},
  year={2020}
}

@article{wei2022emergent,
  title={Emergent abilities of large language models},
  author={Wei, Jason and Tay, Yi and Bommasani, Rishi and Raffel, Colin and Zoph, Barret and Borgeaud, Sebastian and Yogatama, Dani and Bosma, Maarten and Zhou, Denny and Metzler, Donald and others},
  journal={arXiv preprint arXiv:2206.07682},
  year={2022}
}

@inproceedings{wang2024executable,
  title={Executable code actions elicit better llm agents},
  author={Wang, Xingyao and Chen, Yangyi and Yuan, Lifan and Zhang, Yizhe and Li, Yunzhu and Peng, Hao and Ji, Heng},
  booktitle={Forty-first International Conference on Machine Learning},
  year={2024}
}

@article{antoniades2024swe,
  title={SWE-Search: Enhancing Software Agents with Monte Carlo Tree Search and Iterative Refinement},
  author={Antoniades, Antonis and {\"O}rwall, Albert and Zhang, Kexun and Xie, Yuxi and Goyal, Anirudh and Wang, William},
  journal={arXiv preprint arXiv:2410.20285},
  year={2024}
}

@inproceedings{yao2023react,
  title={React: Synergizing reasoning and acting in language models},
  author={Yao, Shunyu and Zhao, Jeffrey and Yu, Dian and Du, Nan and Shafran, Izhak and Narasimhan, Karthik and Cao, Yuan},
  booktitle={International Conference on Learning Representations (ICLR)},
  year={2023}
}

@article{shinn2023reflexion,
  title={Reflexion: Language agents with verbal reinforcement learning},
  author={Shinn, Noah and Cassano, Federico and Gopinath, Ashwin and Narasimhan, Karthik and Yao, Shunyu},
  journal={Advances in Neural Information Processing Systems},
  volume={36},
  pages={8634--8652},
  year={2023}
}

@misc{effectagents,
	Author = {Erik Schluntz and Barry Zhang},
	url = {https://www.anthropic.com/engineering/building-effective-agents},
	Title = {Building effective agents},
	Year = {2024}
}

@misc{claudecode,
	Author = {Anthropic},
	url = {https://www.anthropic.com/news/claude-3-7-sonnet},
	Title = {Claude 3.7 Sonnet and Claude Code},
	Year = {2025}
}

@article{xia2024agentless,
  title={Agentless: Demystifying llm-based software engineering agents},
  author={Xia, Chunqiu Steven and Deng, Yinlin and Dunn, Soren and Zhang, Lingming},
  journal={arXiv preprint arXiv:2407.01489},
  year={2024}
}

@inproceedings{wang2024openhands,
  title={Openhands: An open platform for ai software developers as generalist agents},
  author={Wang, Xingyao and Li, Boxuan and Song, Yufan and Xu, Frank F and Tang, Xiangru and Zhuge, Mingchen and Pan, Jiayi and Song, Yueqi and Li, Bowen and Singh, Jaskirat and others},
  booktitle={The Thirteenth International Conference on Learning Representations},
  year={2025}
}

@article{arora2024masai,
  title={Masai: Modular architecture for software-engineering ai agents},
  author={Arora, Daman and Sonwane, Atharv and Wadhwa, Nalin and Mehrotra, Abhav and Utpala, Saiteja and Bairi, Ramakrishna and Kanade, Aditya and Natarajan, Nagarajan},
  journal={arXiv preprint arXiv:2406.11638},
  year={2024}
}

@inproceedings{zhang2024autocoderover,
  title={Autocoderover: Autonomous program improvement},
  author={Zhang, Yuntong and Ruan, Haifeng and Fan, Zhiyu and Roychoudhury, Abhik},
  booktitle={Proceedings of the 33rd ACM SIGSOFT International Symposium on Software Testing and Analysis},
  pages={1592--1604},
  year={2024}
}

@article{chen2024coder,
  title={Coder: Issue resolving with multi-agent and task graphs},
  author={Chen, Dong and Lin, Shaoxin and Zeng, Muhan and Zan, Daoguang and Wang, Jian-Gang and Cheshkov, Anton and Sun, Jun and Yu, Hao and Dong, Guoliang and Aliev, Artem and others},
  journal={arXiv preprint arXiv:2406.01304},
  year={2024}
}

@inproceedings{zhang2024pair,
  title={A Pair Programming Framework for Code Generation via Multi-Plan Exploration and Feedback-Driven Refinement},
  author={Zhang, Huan and Cheng, Wei and Wu, Yuhan and Hu, Wei},
  booktitle={Proceedings of the 39th IEEE/ACM International Conference on Automated Software Engineering},
  pages={1319--1331},
  year={2024}
}

@article{phan2024hyperagent,
  title={Hyperagent: Generalist software engineering agents to solve coding tasks at scale},
  author={Phan, Huy Nhat and Nguyen, Tien N and Nguyen, Phong X and Bui, Nghi DQ},
  journal={arXiv preprint arXiv:2409.16299},
  year={2024}
}

@article{ma2024lingma,
  title={Lingma swe-gpt: An open development-process-centric language model for automated software improvement},
  author={Ma, Yingwei and Cao, Rongyu and Cao, Yongchang and Zhang, Yue and Chen, Jue and Liu, Yibo and Liu, Yuchen and Li, Binhua and Huang, Fei and Li, Yongbin},
  journal={arXiv preprint arXiv:2411.00622},
  year={2024}
}

@article{hui2024qwen2,
  title={Qwen2. 5-coder technical report},
  author={Hui, Binyuan and Yang, Jian and Cui, Zeyu and Yang, Jiaxi and Liu, Dayiheng and Zhang, Lei and Liu, Tianyu and Zhang, Jiajun and Yu, Bowen and Lu, Keming and others},
  journal={arXiv preprint arXiv:2409.12186},
  year={2024}
}

@article{pan2024training,
  title={Training Software Engineering Agents and Verifiers with SWE-Gym},
  author={Pan, Jiayi and Wang, Xingyao and Neubig, Graham and Jaitly, Navdeep and Ji, Heng and Suhr, Alane and Zhang, Yizhe},
  journal={arXiv preprint arXiv:2412.21139},
  year={2024}
}

@article{xie2025swe,
  title={SWE-Fixer: Training Open-Source LLMs for Effective and Efficient GitHub Issue Resolution},
  author={Xie, Chengxing and Li, Bowen and Gao, Chang and Du, He and Lam, Wai and Zou, Difan and Chen, Kai},
  journal={arXiv preprint arXiv:2501.05040},
  year={2025}
}

@article{wei2025swe,
  title={Swe-rl: Advancing llm reasoning via reinforcement learning on open software evolution},
  author={Wei, Yuxiang and Duchenne, Olivier and Copet, Jade and Carbonneaux, Quentin and Zhang, Lingming and Fried, Daniel and Synnaeve, Gabriel and Singh, Rishabh and Wang, Sida I},
  journal={arXiv preprint arXiv:2502.18449},
  year={2025}
}

@article{grattafiori2024llama,
  title={The llama 3 herd of models},
  author={Grattafiori, Aaron and Dubey, Abhimanyu and Jauhri, Abhinav and Pandey, Abhinav and Kadian, Abhishek and Al-Dahle, Ahmad and Letman, Aiesha and Mathur, Akhil and Schelten, Alan and Vaughan, Alex and others},
  journal={arXiv preprint arXiv:2407.21783},
  year={2024}
}

@misc{swebenchver,
	Author = {OpenAI},
	url = {https://openai.com/index/introducing-swe-bench-verified},
	Title = {Introducing swe-bench verified},
	Year = {2024}
}

@article{shao2024deepseekmath,
  title={Deepseekmath: Pushing the limits of mathematical reasoning in open language models},
  author={Shao, Zhihong and Wang, Peiyi and Zhu, Qihao and Xu, Runxin and Song, Junxiao and Bi, Xiao and Zhang, Haowei and Zhang, Mingchuan and Li, YK and Wu, Y and others},
  journal={arXiv preprint arXiv:2402.03300},
  year={2024}
}

@article{nguyen2024agilecoder,
  title={Agilecoder: Dynamic collaborative agents for software development based on agile methodology},
  author={Nguyen, Minh Huynh and Chau, Thang Phan and Nguyen, Phong X and Bui, Nghi DQ},
  journal={arXiv preprint arXiv:2406.11912},
  year={2024}
}

@misc{moatless,
	Author = {Albert Orwall},
	url = {https://github.com/aorwall/moatless-tools},
	Title = {Moatless tools},
	Year = {2024}
}

@inproceedings{naturalness,
author = {Hindle, Abram and Barr, Earl T. and Su, Zhendong and Gabel, Mark and Devanbu, Premkumar},
title = {On the naturalness of software},
year = {2012},
isbn = {9781467310673},
publisher = {IEEE Press},
booktitle = {Proceedings of the 34th International Conference on Software Engineering},
pages = {837–847},
numpages = {11},
location = {Zurich, Switzerland},
series = {ICSE '12}
}

@inproceedings{codenet,
 author = {Puri, Ruchir and Kung, David and Janssen, Geert and Zhang, Wei and Domeniconi, Giacomo and Zolotov, Vladimir and Dolby, Julian T and Chen, Jie and Choudhury, Mihir and Decker, Lindsey and Thost, Veronika and Thost, Veronika and Buratti, Luca and Pujar, Saurabh and Ramji, Shyam and Finkler, Ulrich and Malaika, Susan and Reiss, Frederick},
 booktitle = {Proceedings of the Neural Information Processing Systems Track on Datasets and Benchmarks},
 editor = {J. Vanschoren and S. Yeung},
 pages = {},
 title = {CodeNet: A Large-Scale AI for Code Dataset for Learning a Diversity of Coding Tasks},
 url = {https://datasets-benchmarks-proceedings.neurips.cc/paper_files/paper/2021/file/a5bfc9e07964f8dddeb95fc584cd965d-Paper-round2.pdf},
 volume = {1},
 year = {2021}
}

@inproceedings{xcodeeval,
    title = "{XC}ode{E}val: An Execution-based Large Scale Multilingual Multitask Benchmark for Code Understanding, Generation, Translation and Retrieval",
    author = "Khan, Mohammad Abdullah Matin  and
      Bari, M Saiful  and
      Do, Xuan Long  and
      Wang, Weishi  and
      Parvez, Md Rizwan  and
      Joty, Shafiq",
    booktitle = "Proceedings of the 62nd Annual Meeting of the Association for Computational Linguistics (Volume 1: Long Papers)",
    month = aug,
    year = "2024",
    address = "Bangkok, Thailand",
    publisher = "Association for Computational Linguistics",
    url = "https://aclanthology.org/2024.acl-long.367/",
    doi = "10.18653/v1/2024.acl-long.367",
    pages = "6766--6805",
}

@article{
li2023starcoder,
title={StarCoder: may the source be with you!},
author={Raymond Li and Loubna Ben allal and Yangtian Zi and Niklas Muennighoff and Denis Kocetkov and Chenghao Mou and Marc Marone and Christopher Akiki and Jia LI and Jenny Chim and Qian Liu and Evgenii Zheltonozhskii and Terry Yue Zhuo and Thomas Wang and Olivier Dehaene and Joel Lamy-Poirier and Joao Monteiro and Nicolas Gontier and Ming-Ho Yee and Logesh Kumar Umapathi and Jian Zhu and Ben Lipkin and Muhtasham Oblokulov and Zhiruo Wang and Rudra Murthy and Jason T Stillerman and Siva Sankalp Patel and Dmitry Abulkhanov and Marco Zocca and Manan Dey and Zhihan Zhang and Urvashi Bhattacharyya and Wenhao Yu and Sasha Luccioni and Paulo Villegas and Fedor Zhdanov and Tony Lee and Nadav Timor and Jennifer Ding and Claire S Schlesinger and Hailey Schoelkopf and Jan Ebert and Tri Dao and Mayank Mishra and Alex Gu and Carolyn Jane Anderson and Brendan Dolan-Gavitt and Danish Contractor and Siva Reddy and Daniel Fried and Dzmitry Bahdanau and Yacine Jernite and Carlos Mu{\~n}oz Ferrandis and Sean Hughes and Thomas Wolf and Arjun Guha and Leandro Von Werra and Harm de Vries},
journal={Transactions on Machine Learning Research},
issn={2835-8856},
year={2023},
url={https://openreview.net/forum?id=KoFOg41haE},
note={Reproducibility Certification}
}

@inproceedings{nijkamp2023codegen,
title={CodeGen: An Open Large Language Model for Code with Multi-Turn Program Synthesis},
author={Erik Nijkamp and Bo Pang and Hiroaki Hayashi and Lifu Tu and Huan Wang and Yingbo Zhou and Silvio Savarese and Caiming Xiong},
booktitle={The Eleventh International Conference on Learning Representations },
year={2023},
url={https://openreview.net/forum?id=iaYcJKpY2B_}
}

@article{guo2025deepseek,
  title={Deepseek-r1: Incentivizing reasoning capability in llms via reinforcement learning},
  author={Guo, Daya and Yang, Dejian and Zhang, Haowei and Song, Junxiao and Zhang, Ruoyu and Xu, Runxin and Zhu, Qihao and Ma, Shirong and Wang, Peiyi and Bi, Xiao and others},
  journal={arXiv preprint arXiv:2501.12948},
  year={2025}
}

@article{jaech2024openai,
  title={Openai o1 system card},
  author={Jaech, Aaron and Kalai, Adam and Lerer, Adam and Richardson, Adam and El-Kishky, Ahmed and Low, Aiden and Helyar, Alec and Madry, Aleksander and Beutel, Alex and Carney, Alex and others},
  journal={arXiv preprint arXiv:2412.16720},
  year={2024}
}

@inproceedings{
jimenez2024swebench,
title={{SWE}-bench: Can Language Models Resolve Real-world Github Issues?},
author={Carlos E Jimenez and John Yang and Alexander Wettig and Shunyu Yao and Kexin Pei and Ofir Press and Karthik R Narasimhan},
booktitle={The Twelfth International Conference on Learning Representations},
year={2024},
url={https://openreview.net/forum?id=VTF8yNQM66}
}

@article{codecontests-science.abq1158,
  author = {Yujia Li  and David Choi  and Junyoung Chung  and Nate Kushman  and Julian Schrittwieser  and R{\'e}mi Leblond  and Tom Eccles  and James Keeling  and Felix Gimeno  and Agustin Dal Lago  and Thomas Hubert  and Peter Choy  and Cyprien de Masson d’Autume  and Igor Babuschkin  and Xinyun Chen  and Po-Sen Huang  and Johannes Welbl  and Sven Gowal  and Alexey Cherepanov  and James Molloy  and Daniel J. Mankowitz  and Esme Sutherland Robson  and Pushmeet Kohli  and Nando de Freitas  and Koray Kavukcuoglu  and Oriol Vinyals },
  title = {Competition-level code generation with AlphaCode},
  journal = {Science},
  volume = {378},
  number = {6624},
  pages = {1092-1097},
  year = {2022},
  doi = {10.1126/science.abq1158},
  }

@misc{prasad2025learninggenerateunittests,
      title={Learning to Generate Unit Tests for Automated Debugging}, 
      author={Archiki Prasad and Elias Stengel-Eskin and Justin Chih-Yao Chen and Zaid Khan and Mohit Bansal},
      year={2025},
      eprint={2502.01619},
      archivePrefix={arXiv},
      primaryClass={cs.SE},
      url={https://arxiv.org/abs/2502.01619}, 
}

@misc{zeng2025acecoderacingcoderrl,
      title={ACECODER: Acing Coder RL via Automated Test-Case Synthesis}, 
      author={Huaye Zeng and Dongfu Jiang and Haozhe Wang and Ping Nie and Xiaotong Chen and Wenhu Chen},
      year={2025},
      eprint={2502.01718},
      archivePrefix={arXiv},
      primaryClass={cs.SE},
      url={https://arxiv.org/abs/2502.01718}, 
}

@misc{liu2024dstcdirectpreferencelearning,
      title={DSTC: Direct Preference Learning with Only Self-Generated Tests and Code to Improve Code LMs}, 
      author={Zhihan Liu and Shenao Zhang and Yongfei Liu and Boyi Liu and Yingxiang Yang and Zhaoran Wang},
      year={2024},
      eprint={2411.13611},
      archivePrefix={arXiv},
      primaryClass={cs.SE},
      url={https://arxiv.org/abs/2411.13611}, 
}

@misc{pan2025asternaturalmultilanguageunit,
      title={ASTER: Natural and Multi-language Unit Test Generation with LLMs}, 
      author={Rangeet Pan and Myeongsoo Kim and Rahul Krishna and Raju Pavuluri and Saurabh Sinha},
      year={2025},
      eprint={2409.03093},
      archivePrefix={arXiv},
      primaryClass={cs.SE},
      url={https://arxiv.org/abs/2409.03093}, 
}

@inproceedings{zhuo2024ice,
  title={ICE-Score: Instructing Large Language Models to Evaluate Code},
  author={Zhuo, Terry Yue},
  booktitle={Findings of the Association for Computational Linguistics: EACL 2024},
  pages={2232--2242},
  year={2024}
}

@inproceedings{wang-etal-2025-testeval,
    title = "{TESTEVAL}: Benchmarking Large Language Models for Test Case Generation",
    author = "Wang, Wenhan  and
      Yang, Chenyuan  and
      Wang, Zhijie  and
      Huang, Yuheng  and
      Chu, Zhaoyang  and
      Song, Da  and
      Zhang, Lingming  and
      Chen, An Ran  and
      Ma, Lei",
    booktitle = "Findings of the Association for Computational Linguistics: NAACL 2025",
    month = apr,
    year = "2025",
    publisher = "Association for Computational Linguistics",
    url = "https://aclanthology.org/2025.findings-naacl.197/",
    ISBN = "979-8-89176-195-7",
}

@misc{mündler2025swtbenchtestingvalidatingrealworld,
      title={SWT-Bench: Testing and Validating Real-World Bug-Fixes with Code Agents}, 
      author={Niels Mündler and Mark Niklas Müller and Jingxuan He and Martin Vechev},
      year={2025},
      eprint={2406.12952},
      archivePrefix={arXiv},
      primaryClass={cs.SE},
      url={https://arxiv.org/abs/2406.12952}, 
}

@misc{jimenez2024swebenchlanguagemodelsresolve,
      title={SWE-bench: Can Language Models Resolve Real-World GitHub Issues?}, 
      author={Carlos E. Jimenez and John Yang and Alexander Wettig and Shunyu Yao and Kexin Pei and Ofir Press and Karthik Narasimhan},
      year={2024},
      eprint={2310.06770},
      archivePrefix={arXiv},
      primaryClass={cs.CL},
      url={https://arxiv.org/abs/2310.06770}, 
}

@article{Muennighoff2023OctoPackIT,
  title={OctoPack: Instruction Tuning Code Large Language Models},
  author={Niklas Muennighoff and Qian Liu and Qi Liu and Armel Randy Zebaze and Qinkai Zheng and Binyuan Hui and Terry Yue Zhuo and Swayam Singh and Xiangru Tang and Leandro von Werra and S. Longpre},
  journal={ArXiv},
  year={2023},
  volume={abs/2308.07124},
  url={https://api.semanticscholar.org/CorpusID:260886874}
}

@misc{manh2025codemmlumultitaskbenchmarkassessing,
      title={CodeMMLU: A Multi-Task Benchmark for Assessing Code Understanding \& Reasoning Capabilities of CodeLLMs}, 
      author={Dung Nguyen Manh and Thang Phan Chau and Nam Le Hai and Thong T. Doan and Nam V. Nguyen and Quang Pham and Nghi D. Q. Bui},
      year={2025},
      eprint={2410.01999},
      archivePrefix={arXiv},
      primaryClass={cs.SE},
      url={https://arxiv.org/abs/2410.01999}, 
}

@misc{han2025convcodeworldbenchmarkingconversationalcode,
      title={ConvCodeWorld: Benchmarking Conversational Code Generation in Reproducible Feedback Environments}, 
      author={Hojae Han and Seung-won Hwang and Rajhans Samdani and Yuxiong He},
      year={2025},
      eprint={2502.19852},
      archivePrefix={arXiv},
      primaryClass={cs.SE},
      url={https://arxiv.org/abs/2502.19852}, 
}

@misc{chen2021evaluatinglargelanguagemodels,
      title={Evaluating Large Language Models Trained on Code}, 
      author={Mark Chen and Jerry Tworek and Heewoo Jun and Qiming Yuan and Henrique Ponde de Oliveira Pinto and Jared Kaplan and Harri Edwards and Yuri Burda and Nicholas Joseph and Greg Brockman and Alex Ray and Raul Puri and Gretchen Krueger and Michael Petrov and Heidy Khlaaf and Girish Sastry and Pamela Mishkin and Brooke Chan and Scott Gray and Nick Ryder and Mikhail Pavlov and Alethea Power and Lukasz Kaiser and Mohammad Bavarian and Clemens Winter and Philippe Tillet and Felipe Petroski Such and Dave Cummings and Matthias Plappert and Fotios Chantzis and Elizabeth Barnes and Ariel Herbert-Voss and William Hebgen Guss and Alex Nichol and Alex Paino and Nikolas Tezak and Jie Tang and Igor Babuschkin and Suchir Balaji and Shantanu Jain and William Saunders and Christopher Hesse and Andrew N. Carr and Jan Leike and Josh Achiam and Vedant Misra and Evan Morikawa and Alec Radford and Matthew Knight and Miles Brundage and Mira Murati and Katie Mayer and Peter Welinder and Bob McGrew and Dario Amodei and Sam McCandlish and Ilya Sutskever and Wojciech Zaremba},
      year={2021},
      eprint={2107.03374},
      archivePrefix={arXiv},
      primaryClass={cs.LG},
      url={https://arxiv.org/abs/2107.03374}, 
}

@article{codescore-2025,
author = {Dong, Yihong and Ding, Jiazheng and Jiang, Xue and Li, Ge and Li, Zhuo and Jin, Zhi},
title = {CodeScore: Evaluating Code Generation by Learning Code Execution},
year = {2025},
publisher = {Association for Computing Machinery},
address = {New York, NY, USA},
volume = {34},
number = {3},
issn = {1049-331X},
url = {https://doi.org/10.1145/3695991},
journal = {ACM Trans. Softw. Eng. Methodol.},
month = feb,
articleno = {77},
numpages = {22},
}

@misc{peng2024humanevalxlmultilingualcodegeneration,
      title={HumanEval-XL: A Multilingual Code Generation Benchmark for Cross-lingual Natural Language Generalization}, 
      author={Qiwei Peng and Yekun Chai and Xuhong Li},
      year={2024},
      eprint={2402.16694},
      archivePrefix={arXiv},
      primaryClass={cs.CL},
      url={https://arxiv.org/abs/2402.16694}, 
}

@article{prystawski2023think,
  title={Why think step by step? reasoning emerges from the locality of experience},
  author={Prystawski, Ben and Li, Michael and Goodman, Noah},
  journal={Advances in Neural Information Processing Systems},
  volume={36},
  pages={70926--70947},
  year={2023}
}

@inproceedings{evalplus-neurips-2023,
  title = {Is Your Code Generated by Chat{GPT} Really Correct? Rigorous Evaluation of Large Language Models for Code Generation},
  author = {Liu, Jiawei and Xia, Chunqiu Steven and Wang, Yuyao and Zhang, Lingming},
  booktitle = {Thirty-seventh Conference on Neural Information Processing Systems},
  year = {2023},
  url = {https://openreview.net/forum?id=1qvx610Cu7},
}

@misc{hendrycks2021measuringcodingchallengecompetence,
      title={Measuring Coding Challenge Competence With APPS}, 
      author={Dan Hendrycks and Steven Basart and Saurav Kadavath and Mantas Mazeika and Akul Arora and Ethan Guo and Collin Burns and Samir Puranik and Horace He and Dawn Song and Jacob Steinhardt},
      year={2021},
      eprint={2105.09938},
      archivePrefix={arXiv},
      primaryClass={cs.SE},
      url={https://arxiv.org/abs/2105.09938}, 
}

@misc{zhuo2025bigcodebenchbenchmarkingcodegeneration,
      title={BigCodeBench: Benchmarking Code Generation with Diverse Function Calls and Complex Instructions}, 
      author={Terry Yue Zhuo and Minh Chien Vu and Jenny Chim and Han Hu and Wenhao Yu and Ratnadira Widyasari and Imam Nur Bani Yusuf and Haolan Zhan and Junda He and Indraneil Paul and Simon Brunner and Chen Gong and Thong Hoang and Armel Randy Zebaze and Xiaoheng Hong and Wen-Ding Li and Jean Kaddour and Ming Xu and Zhihan Zhang and Prateek Yadav and Naman Jain and Alex Gu and Zhoujun Cheng and Jiawei Liu and Qian Liu and Zijian Wang and Binyuan Hui and Niklas Muennighoff and David Lo and Daniel Fried and Xiaoning Du and Harm de Vries and Leandro Von Werra},
      year={2025},
      eprint={2406.15877},
      archivePrefix={arXiv},
      primaryClass={cs.SE},
      url={https://arxiv.org/abs/2406.15877}, 
}

@misc{jain2024livecodebenchholisticcontaminationfree,
      title={LiveCodeBench: Holistic and Contamination Free Evaluation of Large Language Models for Code}, 
      author={Naman Jain and King Han and Alex Gu and Wen-Ding Li and Fanjia Yan and Tianjun Zhang and Sida Wang and Armando Solar-Lezama and Koushik Sen and Ion Stoica},
      year={2024},
      eprint={2403.07974},
      archivePrefix={arXiv},
      primaryClass={cs.SE},
      url={https://arxiv.org/abs/2403.07974}, 
}

@misc{austin2021programsynthesislargelanguage,
      title={Program Synthesis with Large Language Models}, 
      author={Jacob Austin and Augustus Odena and Maxwell Nye and Maarten Bosma and Henryk Michalewski and David Dohan and Ellen Jiang and Carrie Cai and Michael Terry and Quoc Le and Charles Sutton},
      year={2021},
      eprint={2108.07732},
      archivePrefix={arXiv},
      primaryClass={cs.PL},
      url={https://arxiv.org/abs/2108.07732}, 
}

@inproceedings{yu-etal-2018-spider,
    title = "{S}pider: A Large-Scale Human-Labeled Dataset for Complex and Cross-Domain Semantic Parsing and Text-to-{SQL} Task",
    author = "Yu, Tao  and
      Zhang, Rui  and
      Yang, Kai  and
      Yasunaga, Michihiro  and
      Wang, Dongxu  and
      Li, Zifan  and
      Ma, James  and
      Li, Irene  and
      Yao, Qingning  and
      Roman, Shanelle  and
      Zhang, Zilin  and
      Radev, Dragomir",
    editor = "Riloff, Ellen  and
      Chiang, David  and
      Hockenmaier, Julia  and
      Tsujii, Jun{'}ichi",
    booktitle = "Proceedings of the 2018 Conference on Empirical Methods in Natural Language Processing",
    month = oct # "-" # nov,
    year = "2018",
    publisher = "Association for Computational Linguistics",
    url = "https://aclanthology.org/D18-1425/",
    doi = "10.18653/v1/D18-1425",
}

@misc{lei2025spider20evaluatinglanguage,
      title={Spider 2.0: Evaluating Language Models on Real-World Enterprise Text-to-SQL Workflows}, 
      author={Fangyu Lei and Jixuan Chen and Yuxiao Ye and Ruisheng Cao and Dongchan Shin and Hongjin Su and Zhaoqing Suo and Hongcheng Gao and Wenjing Hu and Pengcheng Yin and Victor Zhong and Caiming Xiong and Ruoxi Sun and Qian Liu and Sida Wang and Tao Yu},
      year={2025},
      eprint={2411.07763},
      archivePrefix={arXiv},
      primaryClass={cs.CL},
      url={https://arxiv.org/abs/2411.07763}, 
}

@misc{yang2024swebenchmultimodalaisystems,
      title={SWE-bench Multimodal: Do AI Systems Generalize to Visual Software Domains?}, 
      author={John Yang and Carlos E. Jimenez and Alex L. Zhang and Kilian Lieret and Joyce Yang and Xindi Wu and Ori Press and Niklas Muennighoff and Gabriel Synnaeve and Karthik R. Narasimhan and Diyi Yang and Sida I. Wang and Ofir Press},
      year={2024},
      eprint={2410.03859},
      archivePrefix={arXiv},
      primaryClass={cs.CL},
      url={https://arxiv.org/abs/2410.03859}, 
}

@misc{zan2025multiswebenchmultilingualbenchmarkissue,
      title={Multi-SWE-bench: A Multilingual Benchmark for Issue Resolving}, 
      author={Daoguang Zan and Zhirong Huang and Wei Liu and Hanwu Chen and Linhao Zhang and Shulin Xin and Lu Chen and Qi Liu and Xiaojian Zhong and Aoyan Li and Siyao Liu and Yongsheng Xiao and Liangqiang Chen and Yuyu Zhang and Jing Su and Tianyu Liu and Rui Long and Kai Shen and Liang Xiang},
      year={2025},
      eprint={2504.02605},
      archivePrefix={arXiv},
      primaryClass={cs.SE},
      url={https://arxiv.org/abs/2504.02605}, 
}

@article{twist2025llms,
  title={LLMs Love Python: A Study of LLMs' Bias for Programming Languages and Libraries},
  author={Twist, Lukas and Zhang, Jie M and Harman, Mark and Syme, Don and Noppen, Joost and Nauck, Detlef},
  journal={arXiv preprint arXiv:2503.17181},
  year={2025}
}

@article{liu2025code,
  title={Code Copycat Conundrum: Demystifying Repetition in LLM-based Code Generation},
  author={Liu, Mingwei and Li, Juntao and Wang, Ying and Du, Xueying and Ou, Zuoyu and Chen, Qiuyuan and An, Bingxu and Wei, Zhao and Xu, Yong and Zou, Fangming and others},
  journal={arXiv preprint arXiv:2504.12608},
  year={2025}
}

@article{zhang2025unveiling,
  title={Unveiling Provider Bias in Large Language Models for Code Generation},
  author={Zhang, Xiaoyu and Zhai, Juan and Ma, Shiqing and Bao, Qingshuang and Jiang, Weipeng and Shen, Chao and Liu, Yang},
  journal={arXiv preprint arXiv:2501.07849},
  year={2025}
}

@article{pan2025code,
  title={Do Code LLMs Understand Design Patterns?},
  author={Pan, Zhenyu and Song, Xuefeng and Wang, Yunkun and Cao, Rongyu and Li, Binhua and Li, Yongbin and Liu, Han},
  journal={arXiv preprint arXiv:2501.04835},
  year={2025}
}

@misc{gu2024cruxevalbenchmarkcodereasoning,
      title={CRUXEval: A Benchmark for Code Reasoning, Understanding and Execution}, 
      author={Alex Gu and Baptiste Rozière and Hugh Leather and Armando Solar-Lezama and Gabriel Synnaeve and Sida I. Wang},
      year={2024},
      eprint={2401.03065},
      archivePrefix={arXiv},
      primaryClass={cs.SE},
      url={https://arxiv.org/abs/2401.03065}, 
}

@misc{liu2024codemindframeworkchallengelarge,
      title={CodeMind: A Framework to Challenge Large Language Models for Code Reasoning}, 
      author={Changshu Liu and Shizhuo Dylan Zhang and Ali Reza Ibrahimzada and Reyhaneh Jabbarvand},
      year={2024},
      eprint={2402.09664},
      archivePrefix={arXiv},
      primaryClass={cs.SE},
      url={https://arxiv.org/abs/2402.09664}, 
}

@misc{ahmed2025otter,
      title={Otter: Generating Tests from Issues to Validate SWE Patches}, 
      author={Toufique Ahmed and Jatin Ganhotra and Rangeet Pan and Avraham Shinnar and Saurabh Sinha and Martin Hirzel},
      year={2025},
      eprint={2502.05368},
      archivePrefix={arXiv},
      primaryClass={cs.SE},
      url={https://arxiv.org/abs/2502.05368}, 
}

@article{cuadron2025danger,
  title={The danger of overthinking: Examining the reasoning-action dilemma in agentic tasks},
  author={Cuadron, Alejandro and Li, Dacheng and Ma, Wenjie and Wang, Xingyao and Wang, Yichuan and Zhuang, Siyuan and Liu, Shu and Schroeder, Luis Gaspar and Xia, Tian and Mao, Huanzhi and others},
  journal={arXiv preprint arXiv:2502.08235},
  year={2025}
}

@article{deshpande2025trail,
  title={TRAIL: Trace Reasoning and Agentic Issue Localization},
  author={Deshpande, Darshan and Gangal, Varun and Mehta, Hersh and Krishnan, Jitin and Kannappan, Anand and Qian, Rebecca},
  journal={arXiv preprint arXiv:2505.08638},
  year={2025}
}

@inproceedings{mialon2023gaia,
  title={Gaia: a benchmark for general ai assistants},
  author={Mialon, Gr{\'e}goire and Fourrier, Cl{\'e}mentine and Wolf, Thomas and LeCun, Yann and Scialom, Thomas},
  booktitle={The Twelfth International Conference on Learning Representations},
  year={2023}
}

@article{ceka2026understanding,
  title={Understanding Automated Program Repair Agents Through the Lens of Traceability: An Empirical Study},
  author={Ceka, Ira and Mitchell, Hailie and Pujar, Saurabh and Buratti, Luca and Ramji, Shyam and Yang, Junfeng and Kaiser, Gail and Ray, Baishakhi},
  journal={arXiv preprint arXiv:2506.08311},
  year={2026}
}

@article{mei2025survey,
  title={A survey of context engineering for large language models},
  author={Mei, Lingrui and Yao, Jiayu and Ge, Yuyao and Wang, Yiwei and Bi, Baolong and Cai, Yujun and Liu, Jiazhi and Li, Mingyu and Li, Zhong-Zhi and Zhang, Duzhen and others},
  journal={arXiv preprint arXiv:2507.13334},
  year={2025}
}

@article{rashid2025swe,
  title={SWE-PolyBench: A multi-language benchmark for repository level evaluation of coding agents},
  author={Rashid, Muhammad Shihab and Bock, Christian and Zhuang, Yuan and Buchholz, Alexander and Esler, Tim and Valentin, Simon and Franceschi, Luca and Wistuba, Martin and Sivaprasad, Prabhu Teja and Kim, Woo Jung and others},
  journal={arXiv preprint arXiv:2504.08703},
  year={2025}
}

@misc{openai_gpt4omini_blog,
  author       = {OpenAI},
  title        = {{GPT-4o mini: advancing cost-efficient intelligence}},
  howpublished = {\url{https://openai.com/index/gpt-4o-mini-advancing-cost-efficient-intelligence/}},
  year         = {2024},
  month        = jul,
}

@article{guo2024deepseek,
  title={DeepSeek-Coder: When the Large Language Model Meets Programming--The Rise of Code Intelligence},
  author={Guo, Daya and Zhu, Qihao and Yang, Dejian and Xie, Zhenda and Dong, Kai and Zhang, Wentao and Chen, Guanting and Bi, Xiao and Wu, Yu and Li, YK and others},
  journal={arXiv preprint arXiv:2401.14196},
  year={2024}
}

@misc{anthropic2024claude35sonnet,
  author       = {Anthropic},
  title        = {Introducing Claude 3.5 Sonnet},
  howpublished = {\url{https://www.anthropic.com/news/claude-3-5-sonnet}},
  year         = {2024},
  month        = jun,
}

@article{achiam2023gpt,
  title={Gpt-4 technical report},
  author={Achiam, Josh and Adler, Steven and Agarwal, Sandhini and Ahmad, Lama and Akkaya, Ilge and Aleman, Florencia Leoni and Almeida, Diogo and Altenschmidt, Janko and Altman, Sam and Anadkat, Shyamal and others},
  journal={arXiv preprint arXiv:2303.08774},
  year={2023}
}
\bibliographystyle{tmlr}

\appendix
\appendix
\section{Appendix}
\label{sec:appendix}
\subsection{Paper Roadmap}
1. Introduction \dotfill{} 2\\
\hspace*{1.5em} Motivation, scope, core reasoning families, and survey contributions.\\
2. Survey Methodology \dotfill{} 3\\
\hspace*{1.5em} Literature search strategy, inclusion criteria, citation tracing, and scope limits.\\
3. Related Surveys \dotfill{} 4\\
\hspace*{1.5em} Prior surveys on reasoning, code LLMs, evaluation, agents, and the gap filled here.\\
4. Code Reasoning: Techniques \dotfill{} 5\\
\hspace*{1.5em} Taxonomy of reasoning methods for code tasks.\\
\hspace*{1.5em} 4.1 Code Chain-of-Thought \dotfill{} 6\\
\hspace*{3em} Plan-based methods: PlanSearch, Self-Planning, ClarifyGPT.\\
\hspace*{3em} Structure-based methods: SCoT, CGO, MoT, CodeChain, VisualCoder.\\
\hspace*{3em} Training with CoT: UniCoder, COTTON, ChainCoder, SemCoder, MSCoT.\\
\hspace*{1.5em} 4.2 Self-refinement \dotfill{} 7\\
\hspace*{3em} Execution reflection: Self-Debugging, CodeCoT, AlphaCodium, Revisiting Self-Debugging, $\mu$Fix.\\
\hspace*{3em} Training with feedback: LEVER, CYCLE, LEDEX.\\
\hspace*{3em} Automated test generation: UTGEN, AceCoder, DSTC, ASTER.\\
\hspace*{1.5em} 4.3 Inference Scaling \dotfill{} 8\\
\hspace*{3em} Sampling methods: AlphaCode, REx, S*, CodeTree, ToC.\\
\hspace*{3em} Search methods: Tree-of-Thoughts, GToT, ORPS, SWE-Search.\\
\hspace*{1.5em} 4.4 SWE Agents \dotfill{} 8\\
\hspace*{3em} Workflow and optimization: Agentless, AutoCodeRover, SWE-Agent, CodeAct, OpenHands,\\
\hspace*{3em} MASAI, CodeR, PairCoder, HyperAgent, AgileCoder.\\
\hspace*{3em} Training with trajectories: Lingma SWE-GPT, SWE-Gym, SWE-Fixer, SWE-RL.\\
5. Code Reasoning: Tasks \dotfill{} 10\\
\hspace*{1.5em} Benchmarks and evaluation settings used to study code reasoning.\\
\hspace*{1.5em} 5.1 Code Generation \dotfill{} 10\\
\hspace*{3em} HumanEval, HumanEvalPack, MBPP, APPS, CodeContests, LiveCodeBench,\\
\hspace*{3em} BigCodeBench, CRUXEval, ConvCodeBench, Spider.\\
\hspace*{1.5em} 5.2 Test Generation \dotfill{} 10\\
\hspace*{3em} TestEval, SWT-Bench, Otter.\\
\hspace*{1.5em} 5.3 Issue Resolution \dotfill{} 10\\
\hspace*{3em} SWE-Bench, SWE-Bench Multimodal, Multi-SWE-Bench, SWE-PolyBench, M$^3$ToolEval.\\
\hspace*{1.5em} 5.4 Reasoning and Understanding \dotfill{} 11\\
\hspace*{3em} ReEval, ExeRScope, CodeMMLU, CodeMind.\\
6. Comparison and Discussion \dotfill{} 11\\
\hspace*{1.5em} Cross-paper synthesis of results, including observations on plan-based vs. structure-based CoT,\\
\hspace*{1.5em} modularity, self-refinement, inference scaling, and agentic systems.\\
7. Gaps and Future Directions \dotfill{} 15\\
\hspace*{1.5em} Open problems in structure-aware reasoning, self-refinement, inference scaling, and agent benchmarks.\\
8. Limitations \dotfill{} 18\\
\hspace*{1.5em} Scope limits, paper selection limits, taxonomy subjectivity, and threats to cross-paper comparison validity.\\
9. Conclusion \dotfill{} 19\\
\hspace*{1.5em} Main takeaways and implications for future work.\\
Acknowledgments \dotfill{} 19

\subsection{Benchmarks}
\label{ap:bench}

\textit{HumanEval (HE)}~\cite{chen2021evaluatinglargelanguagemodels} is a set of 164 hand-written programming problems. Each problem includes a function signature, docstring, body, and several unit tests, with an average of 7.7 tests per problem. A multi-language version of HE is also available in HumanEval-XL \cite{peng2024humanevalxlmultilingualcodegeneration}.

\textit{MBPP}~\cite{austin2021programsynthesislargelanguage} (The Most Basic Programming Problems) benchmark has 1k crowd-sourced Python programming problems and was designed to be solvable by entry level programmers. Each problem consists of a task description, code solution and three automated test cases. 
\textit{EvalPlus}~\cite{evalplus-neurips-2023} augments a given evaluation dataset with large amounts of new test cases
created by an automatic test input generator, powered by both LLM- and
mutation-based strategies. EvalPlus includes \textit{MBPP+, HumanEval+}, and  \textit{EvalPerf}.

\textit{APPS}~\cite{hendrycks2021measuringcodingchallengecompetence} is another benchmark for
code generation with 10k samples that measures the ability of models to take an arbitrary natural language specification and generate satisfactory Python code.
More recent extensions of some of the above benchmarks such as \textit{HumanEval-ET, MBPP-ET}, and \textit{APPS-ET} were introduced by \cite{codescore-2025}, where the amount of correct test cases were extended for each benchmark \(100+\) on average according to the reference code.

\textit{CodeContests}~\cite{codecontests-science.abq1158} is a code generation dataset with problems curated from competitive programming platforms such as Codeforces, requiring solutions to challenging code generation problems. This dataset has solutions to the given problems in Python, Java, and C++, with an English description of the code problems. 

\subsection{Code Evaluation}
\label{ap:codeeval}

To address the poor correlation with human evaluation of exact or fuzzy match metrics, ICE-Score was recently proposed as an evaluation metric that instructs LLMs for code assessments \cite{zhuo2024ice}. The ICE-Score evaluation showed superior correlations with functional correctness and human preferences, without the need for test oracles or references. The efficacy of ICE-Score was measured w.r.t. human preference and execution success for four programming languages.

Additionally, CodeScore \cite{codescore-2025} is another code evaluation metric that was recently proposed to measure the functional correctness of generated codes on three input formats (Ref-only, NL-only, and Ref\&NL). CodeScore can be obtained through the UniCE framework that assists models in learning code execution and predicting an estimate of execution
PassRatio.

\subsection{Metrics}
Functional correctness of generated code by LLMs is mainly measured by passing tests. One of the basic metrics to measure the correctness of code is the percentage of tasks in a given benchmark where the generated code successfully passes all tests. 
\cite{chen2021evaluatinglargelanguagemodels} shows that exact or fuzzy match metrics (e.g., BLEU) are not adequate or reliable indicators of functional correctness of code, by showing that functionally different programs generated by a model often have higher BLEU scores than functionally equivalent ones.

The metric \(pass@k\) is the probability of generating at least one solution passing all test cases successfully in \(k\) trials. 
The \textit{AvgPassRatio} measures the degree of correctness of generated code on evaluation test cases, it considers whether the generated code is completely correct on evaluation test cases or not.
Another metric is the percentage of problems solved using \(n\) submissions from \(k\) samples per problem, denoted as \(n@k\).

\subsection{Comparison Margins}
{
\small
\setlength{\tabcolsep}{2pt}
\setlength{\LTleft}{-2cm}
\setlength{\LTcapwidth}{1.2\textwidth}
\begin{longtable}{lllllr}
  \toprule
  \bfseries Method A (Category) & \bfseries Method B (Category) & \bfseries Winner (Category) & \bfseries Bench. & \bfseries Model & \bfseries Margin \\
  \midrule
  \endfirsthead
  \toprule
  \bfseries Method A (Category) & \bfseries Method B (Category) & \bfseries Winner (Category) & \bfseries Bench. & \bfseries Model & \bfseries Margin \\
  \midrule
  \endhead
  \midrule
  \multicolumn{6}{r}{\textit{Continued on next page}} \\
  \endfoot
  \bottomrule
  \caption{Margins of difference for comparisons discussed in Sec.~\ref{sec:comparison}. Each row shows the two approaches being compared (with their technique category), the winner, the benchmark, the model, and the margin (in percentage points). Metrics are pass@1 for code generation benchmarks and resolved rate for SWE-Bench. All results are sourced from each approach's original paper.}
  \label{tab:margins} \\
  \multicolumn{6}{p{1.2\textwidth}}{\small$^{\clubsuit}$ PlanSearch is categorized as Code CoT (Tab.~\ref{tab:llm-hybrid}) but is a notable outlier among CoT methods, exhibiting large margins over other CoT approaches; given these margins, the role of inference scaling as a secondary component in PlanSearch warrants consideration.} \\
  \endlastfoot
  % Observation 1
  \multicolumn{6}{l}{\textbf{Observation 1: Structure CoT vs.\ Plan CoT}} \\
  \midrule
  \rowcolor{gray!10} CGO (Struct.\ CoT) & Self-Planning (Plan CoT) & CGO (Struct.\ CoT) & MBPP-S & gpt-3.5-turbo & +3.4 \\
  CGO (Struct.\ CoT) & Self-Planning (Plan CoT) & CGO (Struct.\ CoT) & MBPP+ & gpt-3.5-turbo & +5.6 \\
  \rowcolor{gray!10} CGO (Struct.\ CoT) & Self-Planning (Plan CoT) & CGO (Struct.\ CoT) & HE & gpt-3.5-turbo & +1.9 \\
  CGO (Struct.\ CoT) & Self-Planning (Plan CoT) & CGO (Struct.\ CoT) & HE+ & gpt-3.5-turbo & +1.2 \\
  \rowcolor{gray!10} CGO (Struct.\ CoT) & ClarifyGPT (Plan CoT) & CGO (Struct.\ CoT) & MBPP-S & gpt-3.5-turbo & +11.9 \\
  CGO (Struct.\ CoT) & ClarifyGPT (Plan CoT) & CGO (Struct.\ CoT) & HE & gpt-3.5-turbo & +0.2 \\
  \rowcolor{gray!10} CGO (Struct.\ CoT) & Self-Planning (Plan CoT) & CGO (Struct.\ CoT) & MBPP-S & Llama-3-8B & +0.2 \\
  CGO (Struct.\ CoT) & Self-Planning (Plan CoT) & CGO (Struct.\ CoT) & MBPP+ & Llama-3-8B & +1.0 \\
  \rowcolor{gray!10} SCoT (Struct.\ CoT) & Self-Planning (Plan CoT) & SCoT (Struct.\ CoT) & MBPP & gpt-4o-mini & +11.8 \\
  SCoT (Struct.\ CoT) & Self-Planning (Plan CoT) & SCoT (Struct.\ CoT) & MBPP+ & gpt-4o-mini & +9.0 \\
  \rowcolor{gray!10} SCoT (Struct.\ CoT) & Self-Planning (Plan CoT) & SCoT (Struct.\ CoT) & APPS & gpt-3.5-turbo & +0.7 \\
  \midrule
  % Observation 2
  \multicolumn{6}{l}{\textbf{Observation 2: Modular CoT vs.\ Structure/Plan CoT}} \\
  \midrule
  \rowcolor{gray!10} MoT (Mod.\ CoT) & SCoT (Struct.\ CoT) & MoT (Mod.\ CoT) & MBPP & DeepSeek-R1 & +17.0 \\
  MoT (Mod.\ CoT) & Self-Planning (Plan CoT) & MoT (Mod.\ CoT) & MBPP & DeepSeek-R1 & +6.5 \\
  \rowcolor{gray!10} MoT (Mod.\ CoT) & SCoT (Struct.\ CoT) & MoT (Mod.\ CoT) & HE & DeepSeek-R1 & +10.3 \\
  MoT (Mod.\ CoT) & Self-Planning (Plan CoT) & MoT (Mod.\ CoT) & HE & DeepSeek-R1 & +9.7 \\
  \rowcolor{gray!10} MoT (Mod.\ CoT) & SCoT (Struct.\ CoT) & MoT (Mod.\ CoT) & MBPP & gpt-4o-mini & +10.0 \\
  MoT (Mod.\ CoT) & SCoT (Struct.\ CoT) & MoT (Mod.\ CoT) & MBPP+ & gpt-4o-mini & +6.7 \\
  \rowcolor{gray!10} CodeChain (Mod.\ CoT) & SCoT (Struct.\ CoT) & CodeChain (Mod.\ CoT) & APPS & gpt-3.5-turbo & +4.4 \\
  CodeChain (Mod.\ CoT) & Self-Planning (Plan CoT) & CodeChain (Mod.\ CoT) & APPS & gpt-3.5-turbo & +5.1 \\
  \midrule
  % Observation 3
  \multicolumn{6}{l}{\textbf{Observation 3: Self-refinement vs.\ Code CoT}} \\
  \midrule
  \rowcolor{gray!10} $\mu$-Fix (Self-refine.) & CGO (Struct.\ CoT) & $\mu$-Fix (Self-refine.) & HE & gpt-3.5-turbo & +15.6 \\
  $\mu$-Fix (Self-refine.) & SCoT (Struct.\ CoT) & $\mu$-Fix (Self-refine.) & HE & gpt-3.5-turbo & +29.6 \\
  \rowcolor{gray!10} $\mu$-Fix (Self-refine.) & Self-Planning (Plan CoT) & $\mu$-Fix (Self-refine.) & HE & gpt-3.5-turbo & +17.5 \\
  $\mu$-Fix (Self-refine.) & ClarifyGPT (Plan CoT) & $\mu$-Fix (Self-refine.) & HE & gpt-3.5-turbo & +15.8 \\
  \rowcolor{gray!10} Self-Debug (Self-refine.) & ClarifyGPT (Plan CoT) & Self-Debug (Self-refine.) & MBPP-ET & gpt-3.5-turbo & +4.8 \\
  Self-Debug (Self-refine.) & SCoT (Struct.\ CoT) & Self-Debug (Self-refine.) & MBPP & gpt-3.5-turbo & +27.2 \\
  \rowcolor{gray!10} Rev.\ Self-Debug (Self-refine.) & PlanSearch (Code CoT) & Rev.\ Self-Debug (Self-refine.) & HE+ & Claude-3.5 & +7.4 \\
  $\mu$-Fix (Self-refine.) & CodeChain (Mod.\ CoT) & $\mu$-Fix (Self-refine.) & APPS & gpt-3.5-turbo & +9.3 \\
  \rowcolor{gray!10} $\mu$-Fix (Self-refine.) & SCoT (Struct.\ CoT) & $\mu$-Fix (Self-refine.) & APPS & gpt-3.5-turbo & +13.7 \\
  $\mu$-Fix (Self-refine.) & Self-Planning (Plan CoT) & $\mu$-Fix (Self-refine.) & APPS & gpt-3.5-turbo & +14.4 \\
  \midrule
  % Observation 4
  \multicolumn{6}{l}{\textbf{Observation 4: Inference Scaling vs.\ Code CoT}} \\
  \midrule
  \rowcolor{gray!10} ORPS (Inf.\ Scaling) & MoT (Mod.\ CoT) & ORPS (Inf.\ Scaling) & MBPP & gpt-4o-mini & +21.8 \\
  ORPS (Inf.\ Scaling) & SCoT (Struct.\ CoT) & ORPS (Inf.\ Scaling) & MBPP & gpt-4o-mini & +31.8 \\
  \rowcolor{gray!10} ORPS (Inf.\ Scaling) & Self-Planning (Plan CoT) & ORPS (Inf.\ Scaling) & MBPP & gpt-4o-mini & +43.6 \\
  REx (Inf.\ Scaling) & CodeChain (Mod.\ CoT) & REx (Inf.\ Scaling) & APPS & gpt-4 & +8.5 \\
  \rowcolor{gray!10} S* (Inf.\ Scaling) & PlanSearch (Code CoT) & S* (Inf.\ Scaling) & LCB & o1-mini & +15.8 \\
  S* (Inf.\ Scaling) & PlanSearch (Code CoT) & S* (Inf.\ Scaling) & LCB & gpt-4o-mini & +22.3 \\
  \rowcolor{gray!10} PlanSearch (Code CoT$^{\clubsuit}$) & MoT (Mod.\ CoT) & PlanSearch (Code CoT$^{\clubsuit}$) & MBPP+ & gpt-4o-mini & +15.4 \\
  PlanSearch (Code CoT$^{\clubsuit}$) & SCoT (Struct.\ CoT) & PlanSearch (Code CoT$^{\clubsuit}$) & MBPP+ & gpt-4o-mini & +22.1 \\
  \rowcolor{gray!10} PlanSearch (Code CoT$^{\clubsuit}$) & Self-Planning (Plan CoT) & PlanSearch (Code CoT$^{\clubsuit}$) & MBPP+ & gpt-4o-mini & +31.1 \\
  CodeTree (Inf.\ Scaling) & MoT (Mod.\ CoT) & CodeTree (Inf.\ Scaling) & MBPP+ & gpt-4o-mini & +18.9 \\
  \rowcolor{gray!10} CodeTree (Inf.\ Scaling) & SCoT (Struct.\ CoT) & CodeTree (Inf.\ Scaling) & MBPP+ & gpt-4o-mini & +25.6 \\
  CodeTree (Inf.\ Scaling) & Self-Planning (Plan CoT) & CodeTree (Inf.\ Scaling) & MBPP+ & gpt-4o-mini & +34.6 \\
  \midrule
  % Observation 5
  \multicolumn{6}{l}{\textbf{Observation 5: SWE Agents vs.\ other approaches}} \\
  \midrule
  \rowcolor{gray!10} PairCoder (SWE Agent) & ClarifyGPT (Plan CoT) & PairCoder (SWE Agent) & HE & gpt-4 & +6.1 \\
  AgileCoder (SWE Agent) & ClarifyGPT (Plan CoT) & AgileCoder (SWE Agent) & HE & gpt-4 & +3.1 \\
  \rowcolor{gray!10} PairCoder (SWE Agent) & ClarifyGPT (Plan CoT) & PairCoder (SWE Agent) & MBPP-S & gpt-4 & +12.5 \\
  PairCoder (SWE Agent) & CGO (Struct.\ CoT) & PairCoder (SWE Agent) & MBPP+ & gpt-3.5-turbo & +4.0 \\
  \rowcolor{gray!10} PairCoder (SWE Agent) & Self-Planning (Plan CoT) & PairCoder (SWE Agent) & MBPP+ & gpt-3.5-turbo & +9.6 \\
  PairCoder (SWE Agent) & SCoT (Struct.\ CoT) & PairCoder (SWE Agent) & MBPP & gpt-3.5-turbo & +33.6 \\
  \rowcolor{gray!10} AgileCoder (SWE Agent) & SCoT (Struct.\ CoT) & AgileCoder (SWE Agent) & MBPP & gpt-3.5-turbo & +33.9 \\
  PairCoder (SWE Agent) & Self-Debug (Self-refine.) & PairCoder (SWE Agent) & MBPP & gpt-3.5-turbo & +6.4 \\
  \rowcolor{gray!10} AgileCoder (SWE Agent) & Self-Debug (Self-refine.) & AgileCoder (SWE Agent) & MBPP & gpt-3.5-turbo & +6.7 \\
  PairCoder (SWE Agent) & $\mu$-Fix (Self-refine.) & PairCoder (SWE Agent) & HE & DeepSeek-Coder & +1.9 \\
  \rowcolor{gray!10} PairCoder (SWE Agent) & Self-Debug (Self-refine.) & PairCoder (SWE Agent) & HE & DeepSeek-Coder & +8.0 \\
  PairCoder (SWE Agent) & UniCoder (Struct.\ CoT) & PairCoder (SWE Agent) & HE & DeepSeek-Coder & +14.8 \\
  \midrule
  % Observation 6
  \multicolumn{6}{l}{\textbf{Observation 6: Agents + Inference Scaling vs.\ other approaches}} \\
  \midrule
  \rowcolor{gray!10} CodeTree (Agent+Inf.\ Sc.) & MoT (Mod.\ CoT) & CodeTree (Agent+Inf.\ Sc.) & HE+ & gpt-4o-mini & +1.3 \\
  CodeTree (Agent+Inf.\ Sc.) & SCoT (Struct.\ CoT) & CodeTree (Agent+Inf.\ Sc.) & HE+ & gpt-4o-mini & +6.1 \\
  \rowcolor{gray!10} CodeTree (Agent+Inf.\ Sc.) & Self-Planning (Plan CoT) & CodeTree (Agent+Inf.\ Sc.) & HE+ & gpt-4o-mini & +4.9 \\
  CodeTree (Agent+Inf.\ Sc.) & PlanSearch (Code CoT) & CodeTree (Agent+Inf.\ Sc.) & HE+ & gpt-4o-mini & +1.1 \\
  \rowcolor{gray!10} CodeTree (Agent+Inf.\ Sc.) & MoT (Mod.\ CoT) & CodeTree (Agent+Inf.\ Sc.) & MBPP+ & gpt-4o-mini & +18.9 \\
  CodeTree (Agent+Inf.\ Sc.) & SCoT (Struct.\ CoT) & CodeTree (Agent+Inf.\ Sc.) & MBPP+ & gpt-4o-mini & +25.6 \\
  \rowcolor{gray!10} CodeTree (Agent+Inf.\ Sc.) & Self-Planning (Plan CoT) & CodeTree (Agent+Inf.\ Sc.) & MBPP+ & gpt-4o-mini & +34.6 \\
  CodeTree (Agent+Inf.\ Sc.) & PlanSearch (Code CoT) & CodeTree (Agent+Inf.\ Sc.) & MBPP+ & gpt-4o-mini & +3.5 \\
  \rowcolor{gray!10} SWE-Search (Agent+Inf.\ Sc.) & MASAI (SWE Agent) & SWE-Search (Agent+Inf.\ Sc.) & SWE-Lite & gpt-4o & +2.7 \\
  SWE-Search (Agent+Inf.\ Sc.) & Agentless (SWE Agent) & SWE-Search (Agent+Inf.\ Sc.) & SWE-Lite & gpt-4o & +6.7 \\
  \rowcolor{gray!10} SWE-Search (Agent+Inf.\ Sc.) & AutoCodeRover (SWE Agent) & SWE-Search (Agent+Inf.\ Sc.) & SWE-Lite & gpt-4o & +8.3 \\
  SWE-Search (Agent+Inf.\ Sc.) & OpenHands (SWE Agent) & SWE-Search (Agent+Inf.\ Sc.) & SWE-Lite & gpt-4o & +9.0 \\
  \rowcolor{gray!10} SWE-Search (Agent+Inf.\ Sc.) & SWE-Agent (SWE Agent) & SWE-Search (Agent+Inf.\ Sc.) & SWE-Lite & gpt-4o & +12.7 \\
\end{longtable}
}

\clearpage
\subsection{Survey Coverage Justifications}
{
\small
\setlength{\tabcolsep}{3pt}
\setlength{\LTleft}{-2cm}
\setlength{\LTright}{-2cm}
\setlength{\LTcapwidth}{\textwidth}
\begin{longtable}{p{3.5cm}p{2.5cm}p{12cm}}
  \toprule
  \bfseries Survey & \bfseries Dimension & \bfseries Justification \\
  \midrule
  \endfirsthead
  \toprule
  \bfseries Survey & \bfseries Dimension & \bfseries Justification \\
  \midrule
  \endhead
  \midrule
  \multicolumn{3}{r}{\textit{Continued on next page}} \\
  \endfoot
  \bottomrule
  \caption{Justifications for coverage ratings in Tab.~\ref{tab:surveys}. Each entry explains why a survey received its rating (\xmark~= not covered, \checkmark~= partial, \cmark~= comprehensive) for a given dimension, evaluated specifically for code and SWE tasks.}
  \label{tab:survey-justification}
  \endlastfoot

  \multirow{2}{*}{\citealp{dong2022survey}}
    & Reasoning \checkmark & References CoT reasoning but does not expand on it; reasoning is not the paper's focus. \\
    & All others \xmark & Surveys in-context learning broadly; does not cover SWE tasks, agents, taxonomy, or benchmarks for code. \\
  \midrule

  \multirow{2}{*}{\citealp{qiao2022reasoning}}
    & Reasoning \cmark & Extensively studies CoT and other prompting approaches with a detailed taxonomy of reasoning methods. \\
    & All others \xmark & Does not include software engineering tasks; taxonomies and benchmarks are for general reasoning (e.g., arithmetic, commonsense), not code. \\
  \midrule

  \multirow{2}{*}{\citealp{huang2022towards}}
    & Reasoning \cmark & Extensively studies CoT and other prompting approaches covering deductive, inductive, and analogical reasoning. \\
    & All others \xmark & Does not include software engineering tasks; benchmarks are for general reasoning, not code. \\
  \midrule

  \multirow{2}{*}{\citealp{zan2022large}}
    & SWE Tasks \checkmark & Surveys 27 LLMs for natural language to code generation; covers code generation only, which is partial SWE coverage. \\
    & All others \xmark & Does not discuss reasoning techniques. Catalogs code generation benchmarks but not in the context of evaluating reasoning approaches. \\
  \midrule

  \multirow{2}{*}{\citealp{chu2023navigate}}
    & Reasoning \checkmark & Covers CoT extensively and defines XoT (Tree-of-Thought, Program-of-Thought, etc.), but limited to the CoT family; does not cover execution-based refinement, inference scaling, or agentic reasoning. \\
    & SWE Tasks \checkmark & Mentions Program-of-Thought and code generation, but focuses on reasoning benchmarks rather than SWE benchmarks. \\
  \midrule

  \multirow{2}{*}{\citealp{jiang2024survey}}
    & Reasoning \checkmark & Extensive code generation survey covering many research topics; touches reasoning but it is not the primary focus. \\
    & SWE Tasks \checkmark & Covers code generation extensively but not broader SWE tasks (e.g., issue resolution, test generation). \\
  \midrule

  \multirow{3}{*}{\cite{sun2024survey}}
    & Reasoning \cmark & Wide-ranging survey covering 50 models with comprehensive coverage of techniques. \\
    & SWE Tasks \cmark & Covers 20 different code-related task categories. \\
    & Benchmarks \checkmark & Covers benchmarks but not with the depth of a dedicated benchmark survey. \\
  \midrule

  \multirow{2}{*}{\citealp{plaat2024reasoning}}
    & Reasoning \cmark & Classifies in-context reasoning into prompting, evaluating, and control (inference scaling and search) based strategies. \\
    & All others \xmark & Does not focus on coding tasks. Any code or agent mentions are incidental to general reasoning. Taxonomy is for general reasoning steps, not code-specific reasoning techniques. \\
  \midrule

  \multirow{2}{*}{\cite{wang2024enhancing}}
    & SWE Tasks \checkmark & Focuses exclusively on reinforcement learning in code generation; covers code generation only. \\
    & Benchmarks \checkmark & Covers code generation benchmarks within the RL context. \\
  \midrule

  \multirow{2}{*}{\cite{chen2024survey}}
    & SWE Tasks \cmark & Comprehensive coverage of evaluation techniques across coding tasks. \\
    & Benchmarks \checkmark & Covers benchmarks as part of evaluation, but evaluation methodology is the focus. \\
  \midrule

  \multirow{3}{*}{\cite{yehudai2025survey}}
    & SWE Tasks \cmark & Comprehensive coverage of SWE tasks through agent evaluation. \\
    & Agents \cmark & Dedicated focus on LLM-agents including SWE Agents. \\
    & Benchmarks \cmark & Comprehensive benchmark coverage for agent evaluation. \\
  \midrule

  \multirow{2}{*}{\citealp{xu2025towards}}
    & Reasoning \cmark & Discusses reinforcement learning based reasoning techniques comprehensively. \\
    & All others \xmark & Does not discuss code-specific reasoning strategies, SWE tasks, agents, or code benchmarks. \\
  \midrule

  \multirow{2}{*}{\citealp{huynh2025large}}
    & Reasoning \checkmark & Surveys many topics in the space; touches reasoning but not deeply. \\
    & SWE Tasks \checkmark & Covers SWE topics including challenges and applications, but not comprehensively. \\
  \midrule

  \multirow{5}{*}{\citealp{yang2025code}}
    & Reasoning \checkmark & Covers reasoning through code-assisted methods (e.g., PoT), but lacks full coverage of broader reasoning paradigms (e.g., RL, inference scaling). \\
    & SWE Tasks \checkmark & Includes code generation, debugging, and agents as part of the code-reasoning interplay, but SWE tasks are not the primary organizational lens. \\
    & Agents \checkmark & Discusses agents in the context of code-reasoning interaction, but no dedicated or in-depth agent taxonomy. \\
    & Taxonomy \checkmark & Provides a taxonomy of code-reasoning interplay, not a general taxonomy of reasoning or SWE systems. \\
    & Benchmarks \checkmark & Covers both code and reasoning benchmarks, but without deep or systematic evaluation. \\
    \midrule
  \multirow{5}{*}{\citealp{mei2025survey}}
    & Reasoning \cmark & Comprehensive survey on context engineering that covers reasoning within its framework. \\
    & SWE Tasks \checkmark & Covers many aspects of SWE reasoning and tasks, but has a much more general focus and is unable to cover many aspects of code-specific reasoning. \\
    & Agents \checkmark & Covers agents within context engineering, but not with SWE-agent specific depth. \\
    & Taxonomy \checkmark & Provides a taxonomy for context engineering, not for code reasoning specifically. \\
    & Benchmarks \checkmark & Mentions SWE-Bench and other benchmarks but without detailed analysis. \\

\end{longtable}
}

\clearpage
\subsection{Results Tables}

%\newcommand{\ignore}[1]{}
% Define colors
\definecolor{lightblue}{RGB}{224,242,255}
\definecolor{lightorange}{RGB}{255,236,204}
\definecolor{lightgreen}{RGB}{217,240,211}
\definecolor{lightred}{RGB}{255,204,204}

\ignore{
    \begin{table*}[t]
      \centering
      \small
      \begin{tabular}{llrrr}
        \toprule
        \bfseries Approach & \bfseries Model & \bfseries APPS Introductory & \bfseries APPS Interview & \bfseries APPS Competition \\
        \midrule
        \multirow[t]{2}{*}{\colorbox{lightblue}{CodeChain}}
          & gpt-4     & 71.1 & 55.0 & 23.3 \\
          & gpt-3.5-turbo-16k   & 54.5 & 28.1 & 12.4 \\
        \midrule
        \multirow[t]{1}{*}{ChainCoder1B}
          & --              & 17.5 & 7.4 & 5.5 \\
          \midrule
        \multirow[t]{1}{*}{\colorbox{lightgreen}{AlphaCode} 1B}
          & --              & 14.4 & 5.6 & 4.6 \\
        \bottomrule
      \end{tabular}
      \caption{Performance across the APPS benchmark \cite{apps2021}, including \textbf{APPS Introductory}, \textbf{Interview}, and \textbf{Competition} subsets. Performance by pass@1 (\%) for CodeChain. For \textbf{ChainCoder 1B}  and \textbf{AlphaCode 1B}  performance is by n@k where n is 5 and k is 1,000.}
      \label{tab:apps-results}
    \end{table*}
}

\begin{table*}[ht!]
  \centering
  \hspace*{-2cm}
  \small
  \begin{tabular}{llrrrrr}
    \toprule
    \bfseries Approach & \bfseries Model & \bfseries \makecell{APPS \\ Introductory} & \bfseries \makecell{APPS \\ Interview} & \bfseries \makecell{APPS \\ Competition}  & \bfseries APPS-ET   & \bfseries APPS \\
    \midrule
    \multirow[t]{2}{*}{\colorbox{lightblue}{CodeChain \cite{le2023codechain}}}
      & gpt-4                 & 71.1 & 55.0 & 23.3 & -- & 61.5\\
      & gpt-3.5-turbo-16k     & 54.5 & 28.1 & 12.4 & -- & 26.4\\
      & WizardCoder     & 26.3 & 7.5 & 3.8 & -- & 10.5\\
    \midrule
    \multirow[t]{1}{*}{\colorbox{lightblue}{ChainCoder $\lozenge$ \cite{zheng2023chaincoder}}}
      & ChainCoder-1B                    & 17.5 & 7.4  & 5.5  & -- & --\\
    \midrule
    \multirow[t]{1}{*}{\colorbox{lightgreen}{AlphaCode $\lozenge$ \cite{li2022competition}}}
      & AlphaCode-1B                    & 14.4 & 5.6  & 4.6  & -- & --\\
    \midrule
    \multirow[t]{2}{*}{\colorbox{lightblue}{Self-Planning \cite{jiang2024self}}}
      & gpt-3.5-turbo         & --   & --   & --   & 8.3 & 21.3\\
      & DeepSeekCoder         & --   & --   & --   & 1.0 & 4.0 \\
    \midrule
    \multirow[t]{2}{*}{\colorbox{lightblue}{SCoT \citep{li2025structured}}}
      & gpt-3.5-turbo         & --   & --   & --   & 7.7 & 22.0 \\
      & DeepSeek-Coder-6.7B-Instr         & --   & --   & --   & 1.3 & 4.3 \\
    \midrule
    \multirow[t]{2}{*}{\colorbox{lightorange}{Self-Debugging \citep{chen2023teaching}}}
      & gpt-3.5-turbo         & --   & --   & --   & 6.2 & 18.7\\
      & DeepSeek-Coder-6.7B-Instr         & --   & --   & --   & 1.3 & 4.7 \\
    \midrule
    \multirow[t]{2}{*}{\colorbox{lightorange}{CYCLE \citep{ding2024cycle}}}
      & CYCLE-350M         & --   & --   & --   & -- & 8.7\\
      & CYCLE-1B         & --   & --   & --   & -- & 10.9 \\
      & CYCLE-2.7B         & --   & --   & --   & -- & 11.6 \\
      & CYCLE-3B         & --   & --   & --   & -- & 11.3 \\
    \midrule
    \multirow[t]{2}{*}{\colorbox{lightorange}{$\mu$-Fix \citep{tian-fixing-spec-cg-2025}}}
      & gpt-3.5-turbo         & --   & --   & --   & 10.3 & 35.7\\
      & DeepSeek-Coder-6.7B-Instr         & --   & --   & --   & 5.0 & 14.0 \\
    \midrule
    \multirow[t]{1}{*}{\colorbox{lightgreen}{REx \cite{tang2024code}}}
      & gpt-4                    & -- & --  & --  & -- & $\sim 70$\\
    \bottomrule
  \end{tabular}
  \caption{
Performance across the APPS benchmark \cite{apps2021}, including the \textbf{APPS Introductory}, \textbf{Interview}, \textbf{Competition}, \textbf{APPS-ET}, and \textbf{APPS} overall sets. Default performance is reported as $pass@1$ (\%). Approaches marked with \(\lozenge\) use the $n@k$ metric, where \(n = 5\) and \(k = 1{,}000\).
}
  \label{tab:apps-results}
\end{table*}

%LCB and CodeContests table

\begin{table*}[t]
  \centering
  \small
  \begin{tabular}{llrrr}
    \toprule
    \bfseries Approach & \bfseries Model & \bfseries LCB & \bfseries CodeContests & \bfseries M\textsuperscript{3}ToolEval \\
    \midrule
    \multirow[t]{2}{*}{\colorbox{lightgreen}{S* \citep{li2025s}}}
      &  Qwen-2.5-Coder-Instruct 32B  & 70.1 & 21.8 & - \\
      & gpt-4o-mini   &  61.3 & 23.0 & - \\
      & R1-Distill-32B   &  85.7 & - & - \\
      & o1-mini  &  85.3 & 48.5 & - \\
    \midrule
    \multirow[t]{2}{*}{\colorbox{lightblue}{PlanSearch \cite{wang2024planning}}}
      & DeepSeek-Coder-V2  &  41.4 & -- & - \\
      & gpt-4o-mini   &  39.0 & -- & - \\
      & gpt-4o   &   41.3 & -- & - \\
      & Claude-Sonnet-3.5  &  40.3 & - & - \\
      & o1-mini  &  69.5 & -- & - \\
      \midrule
    \multirow[t]{1}{*}{\colorbox{lightblue}{CodeChain \dag{} \cite{le2023codechain}} }
      & gpt-3.5            & - & 14.1 & - \\
    \midrule
    \multirow[t]{1}{*}{\colorbox{lightblue}{ChainCoder \ddag{} \cite{zheng2023chaincoder}} }
      & ChainCoder-1B            & - & $\sim15$ & - \\
    \midrule
    \multirow[t]{1}{*}{\colorbox{lightgreen}{AlphaCode \ddag{}
 \cite{li2022competition}} }
      & AlphaCode-9B              & - & 14.3 & - \\
      & AlphaCode-41B              & - & 15.6 & - \\
    \midrule
        \multirow[t]{1}{*}{\colorbox{lightred}{PairCoder 
 \cite{zhang2024pair}} }
      & gpt-3.5-turbo             & - & 15.2 & - \\
      & DeepSeek-Coder              & - & 14.6 & - \\
    \midrule
            \multirow[t]{1}{*}{\colorbox{lightred}{CodeTree
 \cite{zhang2024pair}} }
      & gpt-4o-mini            & - & 26.4 & - \\
      & gpt-4o              & - & 43.0 & - \\
      & Llama-3.1-8B              & - & 12.1 & - \\
    \midrule
        \multirow[t]{1}{*}{\colorbox{lightorange}{AlphaCodium \dag{} \cite{ridnik2024code}} }
      & DeepSeek-33B              & - & 24.0 & - \\
      & gpt-3.5              & - & 17.0 & - \\
      & gpt-4              & - & 29.0 & - \\
    \midrule
                \multirow[t]{1}{*}{\colorbox{lightred}{CodeAct
 \cite{wang2024executable}} }
      & gpt-4            & -- & -- & 74.4 \\
    \midrule
                \multirow[t]{1}{*}{\colorbox{lightred}{Tree-of-Code
 \cite{ni2024tree}} }
      & Mix-modal            & -- & -- & 81.6 \\
    \bottomrule
  \end{tabular}
\caption{Performance across the \textbf{LiveCodeBench (LCB)},  \textbf{CodeContests (test set)}, and \textbf{M\textsuperscript{3}ToolEval}. Default results are reported as \(pass@1\). Approaches marked with \(\dag\) indicate \(pass@5\), while those marked with \(\ddag\) use the \(n@k\) of \(10@1k\) rate. S* results reflect performance on LCB v2.}
  \label{tab:lcb-results}
\end{table*}

\begin{table*}[t]
  \centering
  \small
  \begin{tabular}{llrrr}
    \toprule
    \bfseries Approach & \bfseries Model & \makecell{\bfseries SWE-Bench \\ \bfseries Verified} & \makecell{\bfseries SWE-Bench \\ \bfseries Lite} & \bfseries SWE-Bench \\
    \midrule
    \multirow[t]{2}{*}{\colorbox{lightred}{Agentless \cite{xia2024agentless}}}
      & gpt-4o  & 33.2 & 24.3 & - \\
      & o1-preview  & 41.3 & - & - \\
      & DeepSeek-V3  & 42.0 & - & - \\
      & DeepSeek-R1  & 49.2 & - & - \\
      & Claude-3.5-Sonnet  & 53.0 & - & - \\
    \midrule
    \multirow[t]{2}{*}{\colorbox{lightred}{AutoCodeRover \cite{zhang2024autocoderover}}}
      &  Qwen2-72B-Instruct  & - & 9.3 & - \\
      & gpt-4o  & 28.8 & 22.7 & - \\
      & gpt-4  & - & 19.0 & - \\
    \midrule
        \multirow[t]{2}{*}{\colorbox{lightred}{MASAI \cite{arora2024masai}}}
      & gpt-4o  & -- & 28.3 & - \\
    \midrule
    \multirow[t]{2}{*}{\colorbox{lightred}{SWE-Agent \cite{yang2024sweagent}}}
      &  Claude-3.5-Sonnet & 33.6 & 23.0 & - \\
      & gpt-4o  & 23.2 & 18.3 & - \\
    \midrule
    \multirow[t]{2}{*}{\colorbox{lightred}{SWE-Gym \cite{pan2024training}}}
      &  Qwen-2.5-Coder-Instruct 32B & 20.6 & 15.3 & - \\
      & SWE-Gym-32B  & 32.0 & 26.0 & - \\
    \midrule
        \multirow[t]{2}{*}{\colorbox{lightred}{SWE-Search \cite{antoniades2024swe}}}
      & gpt-4o & - & 31.0 & - \\
      & gpt-4o-mini & - & 17.0 & - \\
      & Qwen-2.5-72b-Instruct & - & 24.7 & - \\
      & Deepseek-V2.5 & - & 21.0 & - \\
      & Llama-3.1-70b-Instruct & - & 17.7 & - \\
    \midrule
    \multirow[t]{2}{*}{\colorbox{lightred}{Lingma \cite{ma2024lingma}}}
      &  Lingma SWE-GPT 72B  & 30.2 & 22.0 & - \\
      &  Lingma SWE-GPT 7B  & 18.2 & 12.0 & - \\
    \midrule

    \multirow[t]{2}{*}{\colorbox{lightred}{SWE-Fixer \cite{xie2025swe}}}
      &  SWE-Fixer-72B  & 32.8 & 24.7 & - \\
    \midrule
        \multirow[t]{2}{*}{\colorbox{lightred}{HyperAgent \cite{phan2024hyperagent}}}
      &  - & 33.0 & 26.0 & - \\
    \midrule
    \multirow[t]{2}{*}{\colorbox{lightred}{SWE-RL \cite{wei2025swe}}}
      &  Llama3-SWE-RL-70B & 41.0 & - & - \\
    \midrule
    \multirow[t]{2}{*}{\colorbox{lightred}{CodeR \cite{chen2024coder}}}
      & gpt-4 & -- & 28.3 & - \\
    \midrule
        \multirow[t]{2}{*}{\colorbox{lightred}{CodeTree \cite{li2024codetree}}}
      & gpt-4o-mini & -- & -- & 27.6 \\
    \midrule
    \multirow[t]{2}{*}{\colorbox{lightred}{OpenHands \cite{wang2024openhands}}}
      & gpt-4o-mini & -- & 7.0 & - \\
      & gpt-4o & -- & 22.0 & - \\
      & Claude-3.5-Sonnet & -- & 26.0 & - \\

    \bottomrule
  \end{tabular}
\caption{Performance on \textbf{SWE-Bench Verified}, and \textbf{SWE-Bench Lite}, and \textbf{SWE-Bench}. Performance is measured by resolved rate.}
  \label{tab:swe-bench}
\end{table*}

\begin{table*}[t]
  \centering
  \small
  \hspace*{-1.5cm}
  \begin{tabular}{llrrrr}
    \toprule
    \bfseries Approach & \bfseries Model & \bfseries MBPP+ & \bfseries MBPP & \bfseries MBPP‑ET & \bfseries MBPP‑S \\
    \midrule
    \multirow[t]{5}{*}{\colorbox{lightblue}{PlanSearch \cite{wang2024planning} }}
      & gpt-4o-mini           & 73.5 & --   & -- & -- \\
      & gpt-4o                & 77.2 & --   & -- & -- \\
      & DeepSeekCoder‑V2      & 76.3 & --   & -- & -- \\
      & Claude‑3.5‑sonnet     & 77.1 & --   & -- & -- \\
    \midrule
    \multirow[t]{2}{*}{\colorbox{lightblue}{ClarifyGPT \cite{mu2023clarifygpt} }}
      & gpt-3.5-turbo         & --   & --   & 55.6 & 74.1 \\
      & gpt-4           & --   & -- & 58.5 & 78.7 \\
    \midrule
    \multirow[t]{4}{*}{\colorbox{lightblue}{Self-Planning \cite{jiang2024self}}}
     & Codex           & -- & --   & 41.9 & 55.7 \\
      & gpt-4o-mini           & 42.4 & 52.1   & 48.2 & -- \\
      & DeepSeek-R1           & 55.4 & 68.4   & 65.5 & -- \\
      & gpt-3.5-turbo         & 68.1 & -- & -- & 82.6 \\
      & Llama‑3 8B Instr.     & 56.9 & -- & -- & 67.9\\
    \midrule
    \multirow[t]{4}{*}{\colorbox{lightblue}{SCoT \citep{li2025structured}}}
      & gpt-3.5-turbo         & --   & 47.0   & -- & -- \\
      & Codex                 & --   & 38.3   & -- & -- \\
      & gpt-4o-mini           & 51.4 & 63.9 & 55.6 & -- \\
      & DeepSeek-R1           & 46.9 & 57.9 & 61.3 & -- \\
    \midrule
    \multirow[t]{2}{*}{\colorbox{lightblue}{MoT \cite{pan2025modularization} }}
      & DeepSeek-R1           & 60.4 & 74.9   & 68.0 & -- \\
      & gpt-4o-mini           & 58.1 & 73.9   & 58.9 & -- \\
    \midrule
    \multirow[t]{2}{*}{\colorbox{lightblue}{CGO \cite{yeo2025chain} }}
      & gpt-3.5-turbo         & 73.7 & -- & -- & 86.0 \\
      & Llama‑3 8B Instr.     & 57.9 & -- & -- & 68.1 \\
    \midrule
    \multirow[t]{2}{*}{\colorbox{lightblue}{UniCoder \cite{sun2024unicoder} }}
      & Deepseek-Coder         & --   & 64.3 & -- & -- \\
      & CodeLlama-7B             & --   & 65.2   & -- & -- \\
    \midrule
    \multirow[t]{4}{*}{ \colorbox{lightorange}{Self-Debugging \cite{chen2023teaching}}}
      & Codex        & --   & 70.8 & -- & -- \\
      & gpt-3.5-turbo             & --   & 74.2   & 60.4 & -- \\
      & gpt-4             & --   & 80.6   & -- & -- \\
      & StarCoder             & --   & 53.2   & -- & -- \\
      & DeepSeek-Coder-6.7B-Instruct        & --   & --   & 56.9 & -- \\
    \midrule
    \multirow[t]{4}{*}{\colorbox{lightorange}{LeDex \cite{jiang2025ledextrainingllmsbetter}}}
      & StarCoder-15B        & 54.3   & 58.2 & -- & -- \\
      & CodeLlama-7B             & 52.9   & 58.1   & -- & -- \\
      & CodeLlama-13B            & 57.9   & 61.9   & -- & -- \\
    \midrule
    \multirow[t]{4}{*}{\colorbox{lightorange}{Revisiting Self-Debugging \cite{chen2025revisit}}}
      & gpt-4o        & 76.5   & 91.5 & -- & -- \\
      & Claude-3.5-sonnet             & 77.0   & 92.6   & -- & -- \\
      & Llama-3-70B-Instr.             & 71.2   & 84.4   & -- & -- \\
      & Qwen-2.5-Coder-7B-Instr             & 70.6   & 84.7   & -- & -- \\
    \midrule
    \multirow[t]{4}{*}{\colorbox{lightgreen}{ORPS \cite{yu2024outcome}}}
      & Llama-3.1-8B-Instruct        & - &  90.4 & -- & -- \\
      & DeepSeek-Coder-7B-Instruct-v1.5  & -   &  93.0   & -- & -- \\
      & Qwen-2.5-Coder-7B-Instruct           & -  &  94.9   & -- & -- \\
      & Qwen-2.5-Coder-14B-Instruct            & -  & 95.3  & -- & -- \\
      & gpt-4o-mini           & -  & 95.7  & -- & -- \\
    \bottomrule
  \end{tabular}
\caption{
Performance on the \textbf{MBPP\,+}, \textbf{MBPP}, \textbf{MBPP-ET}, and \textbf{MBPP-sanitized} benchmarks. All results are reported as $pass@1$.
}
  \label{tab:mbpp-family}
\end{table*}

\begin{table*}[t]
  \centering
  \small
  \begin{tabular}{llrrrr}
    \toprule
    \bfseries Approach & \bfseries Model & \bfseries MBPP+ & \bfseries MBPP & \bfseries MBPP‑ET & \bfseries MBPP‑S \\
    \midrule
    \multirow[t]{4}{*}{\colorbox{lightred}{CodeTree \cite{li2024codetree}}}
      & gpt-4o-mini        & 77.0   & 96.8 & -- & -- \\
      & gpt-4o             & 80.7   & 98.7   & -- & -- \\
      & Llama-3.1-8B-Instr.             & 73.3   & 90.5   & -- & -- \\
      \midrule
       \multirow[t]{4}{*}{\colorbox{lightred}{AgileCoder \citep{nguyen2024agilecoder}}}
      & gpt-3.5-turbo       & --   &  80.9 & -- & -- \\
      & claude-3-haiku             & --   & 84.3   & -- & -- \\
      \midrule
    \multirow[t]{4}{*}{\colorbox{lightred}{PairCoder \citep{zhang2024pair}}}
      & gpt-3.5-turbo       & 77.7   &  80.6 & -- & -- \\
      & DeepSeek-Coder            & 75.7   & 78.8   & -- & -- \\
      & gpt-4            & --   & --   & -- & 91.2 \\
      \midrule
    \multirow[t]{4}{*}{\colorbox{lightorange}{CYCLE \citep{ding2024cycle}}}
      & CYCLE-350M      & --   & -- & -- & 32.6 \\
      & CYCLE-1B             & --   & --   & -- & 35.8 \\
      & CYCLE-2.7B             & --   & --   & -- & 48.5 \\
      & CYCLE-3B             & --   & --   & -- & 51.3 \\
      \midrule
    \multirow[t]{4}{*}{\colorbox{lightorange}{$\mu$-Fix \citep{tian-fixing-spec-cg-2025}}}
      & gpt-3.5-turbo       & --  & -- & 69.1 & -- \\
      & DeepSeek-Coder-6.7B-Instruct              & --   & --   & 63.3 & -- \\
    \midrule
    \multirow[t]{4}{*}{\colorbox{lightorange}{SemCoder \citep{ding2024semcoder}}}
      & SemCoder-S-6.7B      & 68.5  & 79.6 & -- & -- \\
      & SemCoder-6.7B           & 65.3   & 79.9   & -- & -- \\
    \bottomrule
  \end{tabular}
\caption{
Performance on the \textbf{MBPP\,+}, \textbf{MBPP}, \textbf{MBPP-ET}, and \textbf{MBPP-sanitized} benchmarks (continued). All results are reported as $pass@1$.
}
  \label{tab:mbpp-family-continued}
\end{table*}

\begin{table*}[t]
  \centering
  \small
  \setlength{\tabcolsep}{3pt}
  \renewcommand{\arraystretch}{1}

  \begin{tabular}{llrrrrr}
    \toprule
    \bfseries \makecell{Approach} & \bfseries \makecell{Model} &
    \bfseries \makecell{HE+} & \bfseries \makecell{HE} &
    \bfseries \makecell{HE-XL} & \bfseries \makecell{HE-X} &
    \bfseries \makecell{HE-ET} \\
    \midrule
    % ----------------------- PlanSearch -----------------------
    \multirow[t]{5}{*}{\colorbox{lightblue}{PlanSearch \cite{wang2024planning} }}
      & gpt-4o-mini          & 83.7 & --   & -- & -- & -- \\
      & gpt-4o               & 86.4 & --   & -- & -- & -- \\
      & DeepSeekCoder-V2     & 82.8 & --   & -- & -- & -- \\
      & Claude-3.5-sonnet    & 81.6 & --   & -- & -- & -- \\
    \midrule
    % ----------------------- ClarifyGPT -----------------------
    \multirow[t]{2}{*}{\colorbox{lightblue}{ClarifyGPT \cite{mu2023clarifygpt} }}
      & gpt-3.5-turbo        &  --  & 74.4 & -- & -- & 64.8 \\
      & gpt-4          &  --  & 87.8 & -- & -- & 78.1 \\
    \midrule
    % ----------------------- Self-Planning --------------------
    \multirow[t]{5}{*}{\colorbox{lightblue}{Self-Planning \cite{jiang2024self}}}
      & Codex    &  --  & 60.3 & -- & 60.3 & 46.2 \\
      & gpt-4o-mini          & 79.9 & 87.2 & -- &  --  & 87.1 \\
      & DeepSeek-R1          & 79.3 & 85.4 & -- &  --  & 85.3 \\
      & gpt-3.5-turbo        & 67.3 & 72.7 & -- &  --  &  --  \\
      & LLaMA-3 8B Instr.    & 52.8 & 60.1 & -- &  --  &  --  \\
    \midrule
    % ----------------------- SCoT -----------------------------
    \multirow[t]{6}{*}{\colorbox{lightblue}{SCoT \citep{li2025structured}}}
      & gpt-3.5-turbo        &  --  & 60.6 & -- & -- & -- \\
      & Codex                &  --  & 49.8 & -- & -- & -- \\
      & gpt-4o-mini          & 78.7 &  86.6 & -- & -- & 86.0 \\
      & DeepSeek-R1          & 79.3 &  84.8  & -- & -- & -- \\
      & DeepSeekCoder        &  --  &  --  & 69.3 & -- & -- \\
      & Qwen-2.5-Coder       &  --  &  --  & 74.4 & -- & -- \\
    \midrule
    % ----------------------- MoT ------------------------------
    \multirow[t]{2}{*}{\colorbox{lightblue}{MoT \cite{pan2025modularization} }}
      & DeepSeek-R1          & 88.4 & 95.1 & -- & -- & 94.5 \\
      & gpt-4o-mini          & 83.5 & 92.1 & -- & -- & 91.5 \\
    \midrule
        % ----------------------- MsoT ------------------------------
    \multirow[t]{2}{*}{\colorbox{lightblue}{MSCoT \cite{pan2025modularization} }}
      & DeepSeek-Coder          & -- & -- & 66.0 & -- & -- \\
      & Qwen2.5-Coder          & -- & -- & 72.3 & -- & -- \\
    \midrule
    % ----------------------- CGO ------------------------------
    \multirow[t]{2}{*}{\colorbox{lightblue}{CGO \cite{yeo2025chain} }}
      & gpt-3.5-turbo        & 68.5 & 74.6 & -- & -- & -- \\
      & LLaMA-3 8B Instr.    & 56.2 & 62.4 & -- & -- & -- \\
    \midrule
    % ----------------------- UniCoder -------------------------
    \multirow[t]{2}{*}{\colorbox{lightblue}{UniCoder \cite{sun2024unicoder} }}
      & DeepSeek-Coder       &  --  & 70.6 & -- & -- & -- \\
      & CodeLlama-7B           &  --  & 65.4 & -- & -- & -- \\
    \midrule
    % ----------------------- COTTON ---------------------------
    \multirow[t]{4}{*}{\colorbox{lightblue}{COTTON \cite{yang2024chain} }}
      & gpt-3.5-turbo        & 76.2 & 74.4 & -- & -- & -- \\
      & DeepSeekCoder        &  --  &  --  & 61.8 & -- & -- \\
      & Qwen-2.5-Coder       &  --  &  --  & 68.7 & -- & -- \\
    \bottomrule
  \end{tabular}

    \caption{
    Performance on the \textbf{HumanEval\,+}, \textbf{HumanEval}, \textbf{HumanEval-XL}, \textbf{HumanEval-X}, and \textbf{HumanEval-ET} benchmarks. All results are reported as $pass@1$.
    }
  \label{tab:he-family}
\end{table*}

\begin{table*}[t]
  \centering
  \small
  \setlength{\tabcolsep}{3pt}
  \renewcommand{\arraystretch}{1}

  \begin{tabular}{llrrrrr}
    \toprule
    \bfseries \makecell{Approach} & \bfseries \makecell{Model} &
    \bfseries \makecell{HE+} & \bfseries \makecell{HE} &
    \bfseries \makecell{HE-XL} & \bfseries \makecell{HE-X} &
    \bfseries \makecell{HE-ET} \\
    \midrule
    % ----------------------- Agile Coder ----------------------
    \multirow[t]{2}{*}{%
      \makecell[l]{%
        \colorbox{lightred}{Agile Coder \cite{nguyen2024agilecoder}}
      }%
    }
      & gpt-3.5-turbo        &  --  & 70.5 & -- & -- & -- \\
      & claude-3-haiku       &  --  & 79.3 & -- & -- & -- \\
      & gpt-4      &  --  & 90.9 & -- & -- & -- \\
    \midrule
        \multirow[t]{2}{*}{%
      \makecell[l]{%
        \colorbox{lightred}{CodeAct \cite{wang2024executable}}
      }%
    }
      & CodeActAgent(LLaMA-2-7B)       &  --  & 18.1 & -- & -- & -- \\
      & CodeActAgent(Mistral-7B)       &  --  & 34.7 & -- & -- & -- \\
      \midrule
    \multirow[t]{2}{*}{%
      \makecell[l]{%
        \colorbox{lightred}{PairCoder \cite{zhang2024pair}}
      }%
    }
      & gpt-3.5-turbo       &  77.4  & 87.8 & -- & -- & -- \\
      & DeepSeek-Coder       &  76.2  & 85.4 & -- & -- & -- \\
      & gpt-4      &  --  & 93.9 & -- & -- & -- \\
    \midrule
        \multirow[t]{2}{*}{%
      \makecell[l]{%
        \colorbox{lightred}{CodeTree \cite{li2024codetree}}
      }%
    }
      & gpt-4o-mini       &  84.8  & 94.5 & -- & -- & -- \\
      & gpt-4o       &  86.0  & 94.5 & -- & -- & -- \\
      & Llama-3.1-8B      &  72.0  & 82.3 & -- & -- & -- \\
    \midrule
        \multirow[t]{2}{*}{%
      \makecell[l]{%
        \colorbox{lightorange}{$\mu$-Fix \cite{tian-fixing-spec-cg-2025}}
      }%
    }
      & gpt-3.5-turbo        &  80.5  & 90.2 & -- & -- & 79.9 \\
      & DeepSeek-Coder-6.7B-Instr       &  78.7  & 83.5 & -- & -- & 75.0 \\
    \midrule
        \multirow[t]{2}{*}{%
      \makecell[l]{%
        \colorbox{lightorange}{Self-Debugging \cite{chen2023teaching}}
      }%
    }
      & gpt-3.5-turbo        &  71.3  & 77.4 & -- & -- & -- \\
      & DeepSeek-Coder-6.7B-Instr       &  73.2  & 77.4 & -- & -- & -- \\
    \midrule
            \multirow[t]{2}{*}{%
      \makecell[l]{%
        \colorbox{lightorange}{LeDex \cite{jiang2025ledextrainingllmsbetter}}
      }%
    }
      & StarCoder-15B       &  46.3  & 52.3 & -- & -- & -- \\
      & CodeLlama-7B       &  50.0  & 55.8 & -- & -- & -- \\
      & CodeLlama-13B       &  56.7  & 61.7 & -- & -- & -- \\
    \midrule
                \multirow[t]{2}{*}{%
      \makecell[l]{%
        \colorbox{lightorange}{CYCLE \cite{ding2024cycle}}
      }%
    }
      & CYCLE-350M      &  -- & 20.7 & -- & -- & -- \\
      & CYCLE-1B       &  --  & 22.0 & -- & -- & -- \\
      & CYCLE-2.7B       &  --  & 29.3 & -- & -- & -- \\
      & CYCLE-3B      &  --  & 29.9 & -- & -- & -- \\
    \midrule
    % ------------- Revisiting Self-Debugging ------------------
    \multirow[t]{4}{*}{\colorbox{lightorange}{Revisiting Self-Debugging \cite{chen2025revisit}}}
      & gpt-4o               & 87.8 & 92.1 & -- & -- & -- \\
      & Claude-3.5-Sonnet    & 89.0 & 94.5 & -- & -- & -- \\
      & Llama-3-70B Instr.   & 73.8 & 79.9 & -- & -- & -- \\
      & Qwen-2.5-Coder       & 81.7 & 86.0 & -- & -- & -- \\
    \midrule
        \multirow[t]{4}{*}{\colorbox{lightorange}{SemCoder \cite{ding2024semcoder}}}
      & SemCoder-S-6.7B              & 74.4 & 79.3 & -- & -- & -- \\
    & SemCoder-6.7B              & 68.9 & 73.2 & -- & -- & -- \\
    \bottomrule
  \end{tabular}

    \caption{
    Performance on the \textbf{HumanEval\,+}, \textbf{HumanEval}, \textbf{HumanEval-XL}, \textbf{HumanEval-X}, and \textbf{HumanEval-ET} benchmarks (continued). All results are reported as $pass@1$.
    }
  \label{tab:he-family-continued}
\end{table*}

\end{document}